\documentclass[aps, prx, floatfix,
  reprint, 
  superscriptaddress, 
  longbibliography, noeprint,
]{revtex4-2} 
\usepackage[T1]{fontenc}
\usepackage[utf8]{inputenc}
\usepackage{microtype}
\usepackage{graphicx, tikz, xcolor} 
\usepackage{array}

\usepackage{amsmath, amssymb}
\usepackage{mathtools} 
\usepackage{csquotes}  

\usepackage{physics2}                 
\usephysicsmodule{ab}                 
\usephysicsmodule{ab.braket}          
\usepackage{fixdif, derivative}       
\usepackage{bm}

\usepackage[pdftex,unicode,pdfusetitle]{hyperref}
\hypersetup{
    colorlinks=true,
    citecolor=blue,
    linkcolor=magenta,
    urlcolor=blue,
    bookmarksnumbered=true,
    pdfborder={0 0 0},
    bookmarkstype=toc
}
\graphicspath{ {./Figure/} }

\usepackage{amsthm} 
\newtheorem{theorem}{Theorem}
\newtheorem{proposition}{Proposition}
\newtheorem{lemma}{Lemma}
\newtheorem{definition}{Definition}

\DeclareMathOperator{\tr}{tr}
\DeclareMathOperator{\Texp}{Texp}
\DeclareMathOperator{\Log}{ln}

\usepackage{xparse}

\NewDocumentCommand{\Appendix}{m}{%
    Appendix\IfBlankF{#1}{~\ref{Appendix#1}}%
}

\NewDocumentCommand{\abs}{s m}{%
  \IfBooleanF{#1}{\ab}|{#2}|%
}
\NewDocumentCommand{\order}{s m}{%
  \operatorname{\mathcal{O}}\IfBooleanF{#1}{\ab}(#2)%
}
\NewDocumentCommand{\norm}{s o m}{%
    \IfBooleanF{#1}{\ab}\|{#3}\|\IfNoValueF{#2}{_{#2}}%
}
\NewDocumentCommand{\Norm}{s m}{%
    \IfBooleanF{#1}{\ab}\|{#2}\|_{\mathrm{f}}%
}

\NewDocumentCommand{\dyad}{s m g}{%
    \IfBooleanTF{#1}{\ketbra*}{\ketbra}|{#2}><{\IfNoValueTF{#3}{#2}{#3}}|%
}
\NewDocumentCommand{\expval}{s m g}{%
    \IfBooleanTF{#1}{\braket*}{\braket}<\IfNoValueTF{#3}{{#2}}{{#3}|{#2}|{#3}}>%
}
\NewDocumentCommand{\comm}{s m m}{%
    \IfBooleanF{#1}{\ab}[{#2},{#3}]%
}
\NewDocumentCommand{\acomm}{s m m}{%
    \IfBooleanF{#1}{\ab}\{{#2},{#3}\}%
}

\begin{document}
\title{
    Optimal Work Extraction from Finite-Time Closed Quantum Dynamics
}
%
\author{Shoki Sugimoto}
\email{shoki.sugimoto@ap.t.u-tokyo.ac.jp}
  \affiliation{Department of Applied Physics, The University of Tokyo, 7-3-1 Hongo, Bunkyo-ku, Tokyo 113-0033, Japan}
  \affiliation{Nonequilibrium Quantum Statistical Mechanics RIKEN Hakubi Research Team, Pioneering Research Institute (PRI), RIKEN, 2-1 Hirosawa, Wako, Saitama 351-0198, Japan}
\author{Takahiro Sagawa}
  \affiliation{Department of Applied Physics, The University of Tokyo, 7-3-1 Hongo, Bunkyo-ku, Tokyo 113-0033, Japan}
  \affiliation{Quantum-Phase Electronics Center (QPEC), The University of Tokyo, 7-3-1 Hongo, Bunkyo-ku, Tokyo 113-8656, Japan}
\author{Ryusuke Hamazaki}
  \affiliation{Nonequilibrium Quantum Statistical Mechanics RIKEN Hakubi Research Team, Pioneering Research Institute (PRI), RIKEN, 2-1 Hirosawa, Wako, Saitama 351-0198, Japan}
  \affiliation{RIKEN Center for Interdisciplinary Theoretical and Mathematical Sciences (iTHEMS), RIKEN, Wako 351-0198, Japan
}
\begin{abstract}
    Extracting useful work from quantum systems is a fundamental problem in quantum thermodynamics.
    In scenarios where rapid protocols are desired---whether due to practical constraints or deliberate design choices---a fundamental trade-off between power and efficiency is yet to be established.
    Here, we investigate the problem of finite-time optimal work extraction from closed quantum systems, subject to a constraint on the magnitude of the control Hamiltonian.
    We first reveal the trade-off relation between power and work under a general setup, showing that these fundamental performance metrics cannot be maximized simultaneously.
    We then identify a solvable class of finite-time optimal work-extraction problems.
    This class includes nontrivial many-body models such as the Heisenberg model and the $\mathrm{SU}(n)$-Hubbard model.
    The key assumption is that the control Hamiltonian is optimized over a Lie algebra preserved by the uncontrolled dynamics.
    Within this class, the optimal work-extraction problem admits an exact reduction to a nonlinear self-consistent equation, circumventing extensive search over time-dependent control paths.
    The resulting optimal protocol turns out to be particularly simple:
    it suffices to use a time-independent control Hamiltonian in the interaction picture, determined by that equation.
    By exploiting the Lie-algebraic structure of the controllable terms, our approach is applicable to quantum many-body systems through efficient numerical computation.
    Our results highlight the necessity of rapid protocols to achieve the maximum power and provide an exact route to finite-time optimal work extraction in many-body quantum systems.
\end{abstract}
\maketitle

\section{Introduction}
Work extraction is a fundamental problem in thermodynamics, central to understanding how nonequilibrium resources can be transformed into useful work.
According to Planck’s formulation of the second law of thermodynamics, no work can be extracted from a closed system in thermal equilibrium through any adiabatic cycle.
In quantum mechanics, this principle is captured by the passivity of thermal equilibrium states~\cite{Lenard_JStatPhys1978f, Pusz_CommunicationsInMathematicalPhysics1978d, Skrzypczyk2015-zt, Mitsuhashi_PhysicalReviewX2022a}.
By contrast, the extractable work from non-passive states has been actively studied in quantum thermodynamics for decades~\cite{allahverdyan2004maximal, Alicki_PhysRevEStatNonlinSoftMatterPhys2013v, Perarnau-Llobet_PhysicalReviewX2015t, Francica2020-qv, Touil_JPhysAMathTheor2022x, Hokkyo_PhysRevLett2025j, Paulson_JPhysPhotonics2025l}, partially motivated by the experimental advancements in quantum batteries~\cite{Hu_QuantumSciTechnol2022r,Quach2022-au, Joshi_PhysRevACollPark2022t, Burkard_RevModPhys2023p, Campaioli_RevModPhys2024z}.
Without control limitations, the maximum extractable work is known as ergotropy~\cite{allahverdyan2004maximal, Sone_EntropyBasel2021j, Campisi_arXivcond-matStat-mech2026r}.

In real experimental setups, however, work extraction is often constrained by finite-time limitations due to factors such as decoherence, system stability, experimental control limitations, or the need for rapid processing.
In this context, determining the maximum work extractable within a fixed operational time is a highly nontrivial task, given that the magnitude of the control Hamiltonian, which is relevant for the control timescale, is constrained.
There are several efforts~\cite{Garcia-Pintos2020-ix, Mohan_PhysRevACollPark2022w, Shrimali_PhysRevACollPark2024q, Hamazaki_CommunPhys2024i} to estimate the extractable work for finite-time closed quantum dynamics, on the basis of the exact bounds (i.e., speed limits) on the dynamics~\cite{Mandelstam_Tamm1945, Margolus_PhysicaD1998x, Deffner_JPhysA:MathTheor2017h, Gong_IntJModPhysB2022d}. 
However, these studies provide only lower bounds, and it remains unclear whether they are close to the truly achievable maximum extractable work.

This raises the fundamental question of determining the optimal extractable work within a given finite time $T$ and the optimal protocol to achieve it.
While the previous work~\cite{Gyhm_PhysRevACollPark2024l} focuses on determining the minimum time required to fully extract ergotropy, our focus here is on the finite-time optimization of work extraction, where both the extractable work and the optimal control Hamiltonian depend on the imposed time constraint $T$.
Note that, to approach such optimization problems, it is common to use direct optimization methods over time-dependent protocols, such as Krotov’s method~\cite{Bershchanskii1970-sh, Somloi_ChemPhys1993d, Sklarz_PhysRevA2002e, Schirmer2011-mo, Morzhin2019-ws}, GRAPE (Gradient Ascent Pulse Engineering)~\cite{Khaneja_JMagnReson2005a}, among others~\cite{Dominy_JPhysAMathTheor2008h, Caneva_PhysRevA2011q, Zahedinejad_PhysRevA2014s, Machnes_PhysRevLett2018p, Bukov_PhysRevX2018u, Sorensen_PhysRevACollPark2018o}.
However, these methods involve computationally intensive searches over time-dependent control paths and are often intractable.
It is therefore crucial to establish general principles that do not rely on direct numerical optimizations of time-dependent dynamics.

\begin{figure*}[tb]
    \centering
    \includegraphics[width=\linewidth]{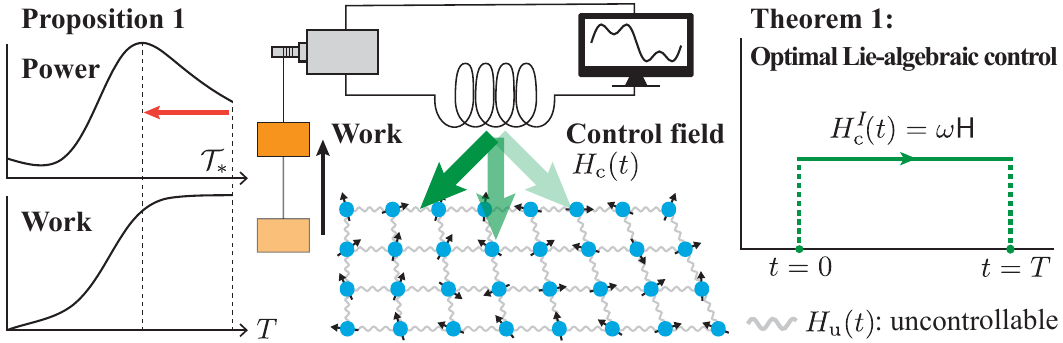}
    \caption{{Schematic of the finite-time work extraction.}
    The dynamics of a closed quantum system are governed by a time-dependent Hamiltonian of the form $H(t) = H_{\mathrm{c}}(t) + H_{\mathrm{u}}(t)$, where $H_{\mathrm{c}}(t)$ is the controllable part subject to optimization~(green arrows), and $H_{\mathrm{u}}(t)$ denotes the uncontrollable part not subject to optimization~(gray curves).
    Under the norm constraint $\Norm*{ H_{\mathrm{c}}(t) } \leq \omega$, the power attains its maximum strictly before the time $\mathcal{T}_{\ast}$ required for the maximum work extraction, indicating the trade-off relation between work and power. 
    This relation highlights the importance of considering fast control processes with $T < \mathcal{T}_{\ast}$ to enhance the power~(Proposition~\ref{thmM_TradeOff}).
    We derive the optimal work extraction protocol under Lie-algebraic control, where $H_{\mathrm{c}}(t)$ is optimized over a subspace $\mathcal{V}$ closed under the commutation relation, and $H_{\mathrm{u}}(t)$ leaves $\mathcal{V}$ invariant, i.e., $i \comm*{ H_{\mathrm{u}}(t) }{ \mathcal{V} } \subseteq \mathcal{V}$~(Theorem~\ref{thmM_MainResult}).
    The optimal protocol is remarkably simple and consists of three steps:
    (i) turn on the control via a quench to $H_{\mathrm{c}}(t=+0) = \omega \mathsf{H}$;
    (ii) steer the system under the control Hamiltonian $H_{\mathrm{c}}^{I}(t) = \omega \mathsf{H}$, which is \textit{time-independent} in the interaction picture;
    and (iii) switch off the control at the final time $t=T$.
    Here, $\mathsf{H}$ is a time-independent, rescaled Hamiltonian that satisfies the closed-form self-consistent equation~\eqref{eqM_SelfConsistentEq}.
    }
    \label{figM_OptimalProtocol}
\end{figure*}
In this article, we develop a general theory of finite-time optimal work extraction in closed quantum systems controlled by a Hamiltonian whose magnitude is constrained (see Fig.~\ref{figM_OptimalProtocol}).
We first prove the trade-off relation between the power and the work, which are two fundamental performance metrics for control schemes~(Proposition~\ref{thmM_TradeOff}).
Specifically, we show that these two quantities cannot be maximized simultaneously.
This fact justifies the importance of the finite-time control setting we consider:
to achieve high power, we should adopt rapid control protocols while sacrificing the work extracted from a single resource.

We then introduce a new framework of Lie-algebraic control, which allows us to solve the time dependence of the optimal protocol (Theorem~\ref{thmM_MainResult}). 
This can be regarded as a solvable class of finite-time optimal control problems, analogous to solvable (i.e., integrable) models in condensed-matter physics.
We find that, among the infinitely many possible control protocols, an optimal control protocol is remarkably simple:
it is driven by a \emph{time-independent} control Hamiltonian in the interaction picture with respect to the uncontrollable part.
The form of the optimal Hamiltonian is determined by a single self-consistent equation, which is dramatically simplified from the original optimization problem.
Importantly, the self-consistent equation admits efficient numerical solutions, from which the optimal extractable work also follows.
As a direct consequence, we obtain an exact and saturable quantum speed limit for work extraction, in contrast to conventional bounds that are generally not attainable.

The key assumption underlying this framework is that the uncontrollable Hamiltonian preserves the subspace over which the controllable Hamiltonian is optimized, a condition that naturally arises in physically relevant models such as Heisenberg models with controlled magnetic fields~\cite{Le_PhysRevACollPark2018z} and $\mathrm{SU}(n)$-Hubbard models with controllable fermion flavors~\cite{Katsura_PhysRevA2013h}.

As a demonstration, we analytically solve the self-consistent equation in the case of $\mathfrak{su}(2)$ control, including the fully controlled two-level systems and magnetic control of the Heisenberg model.
We find an exact expression for the optimal extractable work, which exhibits a cosine dependence on $T$.
Even for more complex control algebras, such as $\mathfrak{su}(n)$, we show that the equation is numerically tractable using gradient descent, and demonstrate it for the fully controlled three-level systems.
Crucially, the approach extends to many-body systems, e.g., $\mathrm{SU}(n)$-Hubbard models, where the number of controllable degrees of freedom is independent of the system size.
In such cases, the Lie-algebraic structure allows the self-consistent equation to be solved within a reduced representation, so that each iteration step in the gradient method requires no access to the full many-body Hilbert space.

Moreover, in the case where the controllable parts span all Hermitian or all traceless Hermitian operators, our framework naturally extends to the optimization of expectation values of general observables at final time.
This enables applications to a broad class of tasks, such as fidelity maximization with respect to a target state.

This paper is organized as follows. 
In Sec.~\ref{SectionII}, we set up the problem of finite-time work extraction under the norm constraint. 
In Sec.~\ref{SectionIII}, we establish a trade-off relation between power and work under the norm constraint (Proposition~\ref{thmM_TradeOff}), whose proof is given in Appendix~\ref{AppendixA}. 
In Sec.~\ref{SectionIV}, we introduce the framework of Lie-algebraic control and derive both the optimal work extraction protocol and the self-consistent equation~\eqref{eqM_SelfConsistentEq} for its generator, presented as Theorem~\ref{thmM_MainResult}; the proof is deferred to Appendix~\ref{AppendixC}. 
We then provide an analytical solution to the self-consistent equation in the simplest nontrivial case of $\mathfrak{su}(2)$ control, as well as a numerical solution for more complex control algebras. 
In Sec.~\ref{SectionV}, we demonstrate that numerical solutions are also feasible for many-body systems without diagonalizing candidate control operators in the full Hilbert space, by employing the representation theory of Lie algebras. 
In Sec.~\ref{SectionVI}, we discuss an extension of the framework beyond work extraction to the optimization of expectation values of general observables, with applications such as fidelity maximization. 
Finally, in Sec.~\ref{SectionVII}, we conclude the paper with a summary and outlook.

\section{Finite-time work extraction under constraints}
\label{SectionII}
We consider the problem of extracting the maximum possible work from a closed quantum system within a finite time $T$.
Such finite-time constraints are physically motivated: in realistic setting, operations must be completed within a finite duration due to experimental or environmental limitations.
Moreover, for a given task, faster protocols are typically more desirable, making time an essential resource to be accounted for.
Specifically, let us consider the following setting: 
prepare an initial state $\rho^\mathrm{i}$ with an initial Hamiltonian $H(0)=H^\mathrm{i}$; 
control the system by a time-dependent Hamiltonian $H(t)$ for $t \in (0,T)$; 
and calculate the extracted work $W$ on the condition that the final Hamiltonian is set to be $H(T)=H^\mathrm{f}$. 
In this setting, $W$ is given by~\cite{Lenard_JStatPhys1978f,allahverdyan2004maximal}
\begin{equation}
    W(U) \coloneqq -\tr[ H^{\mathrm{f}} U \rho^{\mathrm{i}} U^{\dagger} ] +\tr[H^{\mathrm{i}} \rho^{\mathrm{i}}].
    \label{defM_Work}
\end{equation}
Here, the time-evolution operator is given by $U(t) = \Texp[ -i \int_{0}^{t} H(s) \d{s} ]$, where $\Texp$ denotes the time-ordered exponential.
We remark that, within this setting, extracting work is equivalent to lowering the system’s energy, and can therefore also be regarded as a cooling problem in a finite time.
In the following, we consider only the finite-dimensional Hilbert space $\mathcal{H}$, whose dimension is denoted as $D$.
The space of all Hermitian operators acting on $\mathcal{H}$ is then denoted as $\mathcal{B}(\mathcal{H})$.
We also set $\hbar = 1$ throughout the paper.

If there are no constraints on the controllability of the Hamiltonian, i.e., if $H(t)$ can take any operators in $\mathcal{B}(\mathcal{H})$, the maximal extractable work becomes the ergotropy~\cite{allahverdyan2004maximal}.
To see this, we first recall that the ergotropy is obtained by transforming the initial state $\rho^\mathrm{i}$ into the passive state $\rho_\mathrm{erg}$ via a unitary operation $\mathcal{U}_\mathrm{erg}$, i.e., $\rho_\mathrm{erg}=\mathcal{U}_\mathrm{erg}\rho^\mathrm{i}\mathcal{U}_\mathrm{erg}^\dag$.
Here, the passive state $\rho_\mathrm{erg}$ is defined as follows:
by performing spectral decompositions of the initial state and the final Hamiltonian as 
$\rho^\mathrm{i}=\sum_n q_n \ketbra*|q_n>< q_n|$ ($q_1\geq q_2\geq\cdots\geq q_D$) and $H^\mathrm{f}=\sum_n E_n \ketbra*|E_n><E_n|$ ($E_1\leq E_2\leq\cdots\leq E_D$), respectively, $\rho_\mathrm{erg}$ is given by
\begin{align}\label{passive}
    \rho_\mathrm{erg}=\sum_n q_n \ketbra*|E_n><E_n|
\end{align}
Accordingly, the ergotropy is obtained as $\mathcal{W}_\mathrm{erg}=W(\mathcal{U}_\mathrm{erg})$.
Then, if there is no constraint on the Hamiltonian $H(t)$, we can choose, for instance,
\begin{align}
    H_\mathrm{erg}(t) = \frac{i}{T}\Log \mathcal{U}_\mathrm{erg}
    \label{eq_ErgotropyControl}
\end{align}
for $t \in (0,T)$, with which we indeed obtain $\mathcal{W}_\mathrm{erg}$ within time $T$.
Note that we allow a sudden quench of the Hamiltonian, i.e., $H(t)$ may not be continuous at time $t=0$ and $T$.
This construction shows that, without constraints, the ergotropy can always be achieved within arbitrarily short times. 
However, the Hamiltonian in Eq.~\eqref{eq_ErgotropyControl} diverges in the limit $T \to 0$, implying that arbitrarily strong control fields would be required. 
Since such unbounded fields are physically unrealizable, this construction does not represent a feasible protocol when $T$ is not large enough. 
These considerations motivate us to consider more realistic control settings, in which the Hamiltonian is not freely tunable as we address in the following paragraphs.

In realistic control settings, it is usually infeasible to dynamically set all components of the system Hamiltonian at our will, which significantly influences the optimization problem~\cite{Deffner_JPhysBAtMolOptPhys2014b}.
For example, while certain components of the Hamiltonian (e.g., external fields or tunable local potentials) can be dynamically controlled, other components (e.g., confining potentials, background fields, or intrinsic couplings in many-body systems) are not dynamically controllable.
To reflect this limitation, we decompose the time-dependent Hamiltonian as 
\begin{align}
    H(t) = H_{\mathrm{c}}(t) + H_\mathrm{u}(t),
\end{align}
where $H_{\mathrm{c}}(t)$ is the controllable part subject to optimization, and $H_\mathrm{u}(t)$ is the uncontrollable part not subject to optimization.
This decomposition becomes particularly relevant when considering work extraction from many-body systems, where full control over all degrees of freedom is formidable.
Mathematically, $H_{\mathrm{c}}(t)$ is constrained to take values in a subspace $\mathcal{V}\subset\mathcal{B}(\mathcal{H})$.
In the following, we assume that $\mathcal{V}$ is time-independent, which is often the case for the practical control scenarios (including the examples discussed later).
The extent to which $H_{\mathrm{u}}(t)$ must be known for implementing an optimal protocol in our Lie-algebraic control framework will be discussed after Theorem~\ref{thmM_MainResult}.
Here, we emphasize that \enquote{uncontrollable} means not dynamically tunable, rather than unknown, consistent with standard practice in quantum control and optimization theory~\cite{Sklarz_PhysRevA2002e, Khaneja_JMagnReson2005a, Dirr_GAMM-Mitteilungen2008h, DAlessandro_Other2021w, Andrei_Other2022y}.
The formulation in this section is general and underlies Proposition~\ref{thmM_TradeOff}; the additional Lie-algebraic conditions required for the exact finite-time solution will be introduced separately in Sec.~\ref{SectionIV}.

Furthermore, as a key constraint considered in our work, we impose that the magnitude of the controllable Hamiltonian is bounded from above, which is represented as
\begin{align}
\Norm{ H_{\mathrm{c}}(t) } \leq \omega,
\end{align}
where $\Norm{X} \coloneqq \sqrt{\tr(X^{\dagger} X) / \tr I}$ denotes the normalized Frobenius norm~\cite{Carlini_PhysRevLett2006r, Allan_Quantum2021q}.
This constraint reflects physical limitations on available control resources, such as the intrinsic difficulty of generating strong control fields in experimental setups, which leads to bounds on the amplitude of externally applied fields.
Moreover, the norm constraint plays a fundamental role in the optimization problem within the finite time $T$, since $\omega^{-1}$ is relevant for the intrinsic timescale for the control operation.
Note that we choose the normalized Frobenius norm because it is analytically tractable and scales naturally in many-body systems.
The normalization by $\sqrt{\tr I}$ ensures that typical values of the norm $\Norm{ H_{\mathrm{c}}(t) }$ avoid exponential growth in many-body systems, thereby facilitating natural comparison across systems of different sizes.
For instance, in the case of a global magnetic field control $H_{\mathrm{c}}(t) = \bm{B}(t) \cdot \sum_{j=1}^{N} \bm{S}_j$, 
the normalized norm becomes $\Norm{ H_{\mathrm{c}}(t) } = \sqrt{N} \abs*{\bm{B}(t)}$, which scales only polynomially with the system size.

With the above setting, we now formalize optimal extractable work within the finite operational time $T$ under the constrained control of $\{H_{\mathrm{c}}(t)\}_{t \in (0,T)}$.
The control Hamiltonian acts only during $t \in (0,T)$ and is switched off at the initial and final times, $H_{\mathrm{c}}(0) = H_{\mathrm{c}}(T) = 0$.
We fix the initial and final Hamiltonians, denoted by $H^{\mathrm{i}}$ and $H^{\mathrm{f}}$, respectively.
The uncontrollable Hamiltonian $\{H_\mathrm{u}(t)\}_{t \in [0,T]}$ is fixed, not subject to optimization, and satisfies the boundary conditions $H_{\mathrm{u}}(0) = H^{\mathrm{i}}$ and $H_{\mathrm{u}}(T) = H^{\mathrm{f}}$.
We also fix the initial state, denoted by $\rho^{\mathrm{i}}$.
Then, we define
\begin{equation}
    \label{optimize}
    \mathcal{W}(T;\omega) \coloneqq \max_{\{H_{\mathrm{c}}(t)\}_{t \in (0,T)},\, \Norm*{H_{\mathrm{c}}(t)} \leq \omega} W(U(T)),
\end{equation}
as the optimal extractable work.
Note that, while we explicitly write down the norm constraint to highlight its importance, we also implicitly assume that $\{H_{\mathrm{c}}(t)\}_{t \in (0,T)}$ is constrained within $\mathcal{V}$.
For brevity, we write $\mathcal{W}(T) \equiv \mathcal{W}(T;\omega)$ whenever the $\omega$-dependence is irrelevant.
Since the second term in Eq.~\eqref{defM_Work} just gives a constant and does not affect the optimization in Eq.~\eqref{optimize}, we replace it with $\tr[H^{\mathrm{f}}\rho^{\mathrm{i}}]$ without loss of generality, which ensures $W(U\!=\!I) = 0$ and $\mathcal{W}(0)=0$ with $I$ being the identity operator.
Note that $\mathcal{W}(T)$ is not necessarily monotonic as a function of operational time $T$, owing to the effect of the uncontrollable dynamics.


\section{Trade-off between power and work}
\label{SectionIII}
As our first main result, we show that a rapid protocol is necessary to achieve the maximum power of work extraction, which demonstrates the importance of our finite-time optimization setting.
To be specific, we consider two fundamental performance metrics of the quantum battery, work $\mathcal{W}(T)$ and power $\mathcal{P}(T)$~\cite{Binder_NewJPhys2015j,Campaioli_RevModPhys2024z}, the latter of which is defined as the optimized extractable work per time,
\begin{align}
    \mathcal{P}(T) \coloneqq \frac{\mathcal{W}(T)}{T}.
\end{align}
As for work, we denote its maximum amount when there is no time constraint as
\begin{align}
    \mathcal{W}_{\ast} \coloneqq \max_T{\mathcal{W}(T)} \eqqcolon \mathcal{W}(\mathcal{T}_{\ast}).
\end{align}
Here, $\mathcal{T}_{\ast}$ is the minimal time for $\mathcal{W}(T)$ to achieve $\mathcal{W}_\ast$.
This is typically finite, although it can be infinite depending on the setup ($H_{\mathrm{u}}(t)$ and $\mathcal{V}$).

A crucial question is whether we can simultaneously maximize these two performance metrics, the power and the work.
Since the maximum work extraction $\mathcal{W}_{\ast} = \mathcal{W}(T)$ is already obtained by $T=\mathcal{T}_{\ast}$, this question is equivalent to asking whether $\mathcal{P}(T)$ reaches its maximum at $T=\mathcal{T}_{\ast}$.
If that were the case, the best protocol for given independent batteries would be trivial: 
just extract the maximal work $\mathcal{W}_{\ast}$ with time $\mathcal{T}_{\ast}$ for one battery and repeat the protocol for the other ones, and then the total time to extract a certain amount of work would also be minimized.

However, we can generally show the following no-go theorem for the simultaneous maximization of the power and the work, which highlights the importance of considering fast control processes within $T<\mathcal{T}_{\ast}$~(see \Appendix{A} for a proof).
\begin{proposition}[Trade-off between power and work]
    \label{thmM_TradeOff}
    The power $\mathcal{P}(T)$ becomes maximum strictly before the time $T=\mathcal{T}_{\ast}$.
    Consequently, maxima of the power $\mathcal{P}(T) $ and the work $\mathcal{W}(T)$ cannot be achieved simultaneously.
\end{proposition}

Note that this trade-off differs in nature from that in classical heat engines, where the maximum conversion efficiency between heat and work (Carnot efficiency) can only be reached in the quasistatic limit, resulting in vanishing power~\cite{Shiraishi_PhysRevLett2016v}.
In quantum systems with finite Hilbert-space dimension, the maximum work can often be achieved within finite time, meaning that maximum work is compatible with nonzero power.
That being said, our trade-off relation reveals a fundamental need to balance power and work in the design of work extraction protocols.

As detailed in \Appendix{A}, the essential and nontrivial step in the proof (in the case of finite $\mathcal{T}_{\ast}$) is to show that $\mathcal{W}(T)$ is differentiable at time $T=\mathcal{T}_{\ast}$ once and 
\begin{align}
    \dot{\mathcal{W}}(\mathcal{T}_{\ast})=0.
\end{align}
Note that such differentiability is far from obvious because $\mathcal{W}(T)$ is defined as the optimal value of a maximization problem over control protocols.
As such, it can lose higher-order differentiability when the optimal protocol does not change smoothly as $T$ varies.
In fact, even twice differentiability of $\mathcal{W}(T)$ at $T = \mathcal{T}_{\ast}$ is not guaranteed.
This already fails in the analytically solvable case of the controlled Heisenberg model discussed around Eq.~\eqref{wctheisenberg}, where the left-hand second derivative of $\mathcal{W}(T)$ at $T=\mathcal{T}_{\ast}$ is strictly negative, but the right-hand second derivative vanishes.
Once we obtain $\dot{\mathcal{W}}(\mathcal{T}_{\ast})=0$, the proposition follows from 
\begin{align}
\dot{\mathcal{P}}(\mathcal{T}_{\ast}) = -\frac{\mathcal{W}(\mathcal{T}_{\ast})}{ \mathcal{T}_{\ast}^2} < 0,
\end{align}
which implies that the power $\mathcal{P}$ achieves its maximum strictly before $T = \mathcal{T}_{\ast}$.

Proposition~\ref{thmM_TradeOff} clarifies the role of the finite-time optimization problem studied in the rest of this paper.
Since the maximum power is attained strictly before $\mathcal{T}_{\ast}$, the regime relevant for high-power operation is necessarily a finite-time regime with $T < \mathcal{T}_{\ast}$, where the extractable work has not yet saturated.
This motivates the need to determine the optimal work extraction protocol at a given finite operational time, rather than only characterizing the unconstrained maximum-work limit.
In Sec.~\ref{SectionIV}, we show that this finite-time optimization problem becomes tractable under Lie-algebraic control.

\section{Lie-algebraic control}
\label{SectionIV}
\subsection{Setup}
\label{SectionIV.A}
While Proposition~\ref{thmM_TradeOff} offers a conceptually important no-go theorem that generally holds under the norm constraint, it does not provide explicit values for extractable work and power.
Here, we identify a solvable class of finite-time optimal work-extraction problems, for which the generally complicated and often intractable optimization problem in Eq.~\eqref{optimize} reduces to a fundamental self-consistent equation (see Eq.~\eqref{eqM_SelfConsistentEq}).
This class is defined by the following two Lie-algebraic assumptions:

\begin{enumerate}
\renewcommand{\labelenumi}{(\roman{enumi})}
\item The controllable part $H_{\mathrm{c}}(t)$ is optimized over a subspace $\mathcal{V}$ of $\mathcal{B}(\mathcal{H})$ closed under the commutator, i.e., $i\comm*{X}{Y} \in \mathcal{V}$ for all $X, Y \in\mathcal{V}$.
\item The uncontrollable part $H_{\mathrm{u}}(t)$ leaves $\mathcal{V}$ invariant, i.e., $i \comm*{ H_{\mathrm{u}}(t) }{ X } \in \mathcal{V}$ for all $X \in \mathcal{V}$ and all $t \in [0,T]$.
\end{enumerate}
Mathematically, $H_{\mathrm{c}}(t)$ is optimized over a Lie algebra $\mathcal{V}$~(we adopt a convention that a compact Lie algebra consists of Hermitian operators), and $H_{\mathrm{u}}(t)$ lies in its normalizer algebra.
In particular, $H_{\mathrm{u}}(t)$ may contain components within $\mathcal{V}$ itself, such as fixed background fields along the controllable directions and $\mathcal{V}$-components of confining potentials, which are not subject to optimization.
In contrast, the other components of $H_{\mathrm{u}}(t)$ commute with all of $\mathcal{V}$ and therefore have the symmetry of the Lie group $e^{i \mathcal{V}}$.

Since the above conditions are formalized in a mathematical way, we illustrate them with physical examples.
One extreme case of such control setups is the control over all Hermitian operators under norm constraint, which corresponds to $\mathcal{V} \cong \mathfrak{u}(D)$.
Another example is the control over all traceless operators under norm constraint, which corresponds to $\mathcal{V} \cong \mathfrak{su}(D)$.
Here, $\mathfrak{u}(D)$ and $\mathfrak{su}(D)$ denote the Lie algebras of the unitary and special unitary groups, respectively.
We can even consider many-body systems, such as Heisenberg-type models with controlled magnetic fields:
\begin{align}
    H_{\mathrm{c}}(t) &= \bm{B}(t)\cdot\bm{S},
    \nonumber \\
    H_{\mathrm{u}}(t) &= \sum_{x,y} J_{xy}(t) \bm{S}_{x}\cdot \bm{S}_{y} + \bm{B}_{0}(t)\cdot \bm{S},
    \label{eqM_Heisenberg}
\end{align}
where $\bm{S} \coloneqq \sum_{x} \bm{S}_x$ is the total spin operator.
Here, $\bm{B}(t)$ is a controllable magnetic field subject to the norm constraint, while $\bm{B}_{0}(t)$ represents a background field that is not subject to optimization.
This exemplifies the part of $H_{\mathrm{u}}(t)$ lying within $\mathcal{V}$ and reflects experimental situations where a background field coexists with the controllable drive.
In this case, we confirm that $\mathcal{V} \cong \mathfrak{su}(2)$, independent of the system size.

Under assumption~(ii), the time-evolution operator factorizes as $U(t) = U_{\mathrm{u}}(t) U_{\mathrm{c}}(t)$, where $U_{\mathrm{u}}(t)$ is the unitary generated by $H_{\mathrm{u}}(t)$, and
\begin{align}
    U_{\mathrm{c}}(t) 
    &\coloneqq \Texp\ab[-i\int_{0}^{t} U_{\mathrm{u}}(s)^{\dagger} H_{\mathrm{c}}(s) U_{\mathrm{u}}(s) \d{s}]
    \nonumber \\
    &\eqqcolon \Texp\ab[-i\int_{0}^{t} H_{\mathrm{c}}^{I}(s) \d{s}] \in e^{i \mathcal{V}}.
\end{align}
Here, the fact that $U_{\mathrm{c}}(t) \in e^{i \mathcal{V}}$ follows from assumption~(ii) of Lie-algebraic control, which implies that the control Hamiltonian in the interaction picture, $H_{\mathrm{c}}^{I}(t) \coloneqq U_{\mathrm{u}}(t)^{\dagger} H_{\mathrm{c}}(t) U_{\mathrm{u}}(t)$, belongs to $\mathcal{V}$.
When $H_{\mathrm{u}}(t)$ commutes with $\mathcal{V}$ for all $t \in [0,T]$, the interaction picture coincides with the original picture, i.e., $H_{\mathrm{c}}^{I}(t) \equiv H_{\mathrm{c}}(t)$, and hence an explicit transformation to the interaction picture is unnecessary.
A representative example is the Heisenberg model in Eq.~\eqref{eqM_Heisenberg} with vanishing background field during the control period, $\bm{B}_0(t) \equiv 0$ for $t \in (0,T)$.

We decompose both the initial state and the final Hamiltonian as $\rho^{\mathrm{i}} = \rho_{\mathcal{V}}^{\mathrm{i}} + \rho_{\perp}^{\mathrm{i}}$ and $H^{\mathrm{f}} = H_{\mathcal{V}}^{\mathrm{f}} + H_{\perp}^{\mathrm{f}}$, where the subscript $\mathcal{V}$ denotes the projection onto $\mathcal{V}$ (see Eq.~\eqref{rhocortho}), and the subscript $\perp$ indicates the orthogonal complement.
Under assumption~(ii), this decomposition implies that the work separates into the controllable and uncontrollable parts as $W(U) = W_{\mathrm{c}}(U_{\mathrm{c}}, U_{\mathrm{u}}) + W(U_{\mathrm{u}})$, where~(see \Appendix{B})
\begin{multline}\label{workdec}
   W_{\mathrm{c}}(U_{\mathrm{c}}, U_{\mathrm{u}}) \\
   \coloneqq -\tr[ (U_{\mathrm{u}}^{\dagger} H_{\mathcal{V}}^{\mathrm{f}} U_{\mathrm{u}})\, U_{\mathrm{c}} \rho^{\mathrm{i}} U_{\mathrm{c}}^{\dagger} ] + \tr[ (U_{\mathrm{u}}^{\dagger} H_{\mathcal{V}}^{\mathrm{f}} U_{\mathrm{u}}) \rho^{\mathrm{i}} ].
\end{multline}
Here, $W_{\mathrm{c}}(U_{\mathrm{c}}, U_{\mathrm{u}})$ represents the additional work, relative to the uncontrolled dynamics, arising from the control.
It coincides with the work~\eqref{defM_Work} evaluated with the final Hamiltonian $U_{\mathrm{u}}^{\dagger} H_{\mathcal{V}}^{\mathrm{f}} U_{\mathrm{u}}$, the initial state $\rho^{\mathrm{i}}$, and the time-evolution operator $U_{\mathrm{c}}$.
Since the uncontrollable part $H_{\mathrm{u}}(t)$ is fixed and not subject to optimization, we focus solely on the contribution from the controllable part $W_{\mathrm{c}}$, which is denoted as $W_{\mathrm{c}}(U_{\mathrm{c}}) \equiv W_{\mathrm{c}}(U_{\mathrm{c}}, U_{\mathrm{u}})$ for brevity.
Its optimized value is denoted as 
\begin{align}
    \mathcal{W}_\mathrm{c}(T) \coloneqq \mathcal{W}(T) - W(U_\mathrm{u}(T)).
\end{align}
We also denote the final Hamiltonian in the interaction picture by $H_{\mathcal{V}}^{\mathrm{f},I}\!(T) \coloneqq U_{\mathrm{u}}(T)^{\dagger} H_{\mathcal{V}}^{\mathrm{f}} U_{\mathrm{u}}(T)$.
Again, when $H_{\mathrm{u}}(t)$ commutes with $\mathcal{V}$ for all $t \in [0,T]$, this reduces to $H_{\mathcal{V}}^{\mathrm{f},I}\!(T) \equiv H_{\mathcal{V}}^{\mathrm{f}}$, and an explicit transformation to the interaction picture is unnecessary.

In the interval of $T$ such that $H_{\mathcal{V}}^{\mathrm{f},I}\!(T)$ is independent of time, i.e., $H_{\mathcal{V}}^{\mathrm{f},I}\!(T) \equiv H_{\mathcal{V}}^{\mathrm{f}}$, the optimized contribution $\mathcal{W}_\mathrm{c}(T)$ from the controllable part is nondecreasing.
This is in contrast with the optimal total work $\mathcal{W}(T)$, which is not necessarily monotonic owing to the uncontrollable part $W(U_\mathrm{u}(T))$.
The monotonicity is because $W_{\mathrm{c}}(U_{\mathrm{c}}, U_{\mathrm{u}}(T))$ does not depend on $U_{\mathrm{u}}(T)$ in this case, and any unitary $U_{\mathrm{c}}$ implementable within an operational time $T$ remains implementable within any larger operational time.
Indeed, if $H_{\mathrm{c}}^{I}(t)$ implements the unitary $U_{\mathrm{c}}(T)$ within time $T$, then the control Hamiltonian
\begin{equation}\label{nondec}
    \tilde{H}_{\mathrm{c}}^{I}(t)
    =
    \begin{cases}
        H_{\mathrm{c}}^{I}(t) & (0 < t < T);\\
        0 & (T \leq t < T')
    \end{cases}
\end{equation}
implements the same unitary within any larger $T'\, (> T)$.
We note that the time-independence condition, $H_{\mathcal{V}}^{\mathrm{f},I}(T) \equiv H_{\mathcal{V}}^{\mathrm{f}}$, can be satisfied beyond the commuting case described above.
For example, this equality also holds in the Heisenberg-type model in Eq.~\eqref{eqM_Heisenberg} with a static background field $\bm{B}_{0}(t) \equiv \bm{B}_{0}$ and a final Hamiltonian $H_{\mathcal{V}}^{\mathrm{f}} = \bm{B}_{0}\cdot \bm{S}$.

To assess whether the norm constraint is active, i.e., whether optimal control Hamiltonians saturate the constraint, it is useful to quantify how far the optimal unitaries are from the identity.
We therefore define
\begin{equation}
    \label{eqM_OptimalGeodesicDistance}
    \ell_{T} \coloneqq \min_{U_{\mathrm{c}} \in \displaystyle\operatorname*{argmax}_{V_{\mathrm{c}} \in e^{i\mathcal{V}}} 
    W_{\mathrm{c}}(V_{\mathrm{c}}, U_{\mathrm{u}}(T))} \Norm*{ \Log U_{\mathrm{c}} },
\end{equation}
where $\Log$ denotes the inverse of the exponential map in the vicinity of the identity.
This $\ell_{T}$ is the shortest distance on $e^{i\mathcal{V}}$ between the identity and the unitaries that maximize the controllable work~\cite{Lee_Other2019c,Lee_Other1997d}.
For each fixed operational time $T$, the effective final operator $H_{\mathcal{V}}^{\mathrm{f},I}(T)$ is already fixed, and the optimization in Eq.~\eqref{eqM_OptimalGeodesicDistance} is therefore a static optimization problem on the compact Lie group $e^{i\mathcal V}$.
This problem is independent of the finite-time constrained optimization, and a detailed Lie-theoretic treatment will be reported elsewhere.
Accordingly, $\ell_T$ can be determined before applying Theorem~\ref{thmM_MainResult} below.
For brevity, we simply write $\ell$ if $\ell_T$ does not depend on $T$.

Once $\ell_T$ is determined from the static optimization problem in Eq.~\eqref{eqM_OptimalGeodesicDistance}, one first checks whether $\omega T > \ell_{T}$.
If so, the norm constraint is not active, i.e., there exists an optimal control Hamiltonian that does not saturate the constraint.
To see this, let $\tilde{U}_{\mathrm{c}}$ be an optimal time-evolution operator achieving the minimum in Eq.~\eqref{eqM_OptimalGeodesicDistance}, $\Norm*{\Log \tilde{U}_{\mathrm{c}}} = \ell_{T}$.
Then, it can be implemented by the following control Hamiltonian that does not saturate the constraint:
\begin{equation}
    H_{\mathrm{c}}^{I}(t) = \frac{ i \Log \tilde{U}_{\mathrm{c}} }{ T },\quad \text{with} \quad \Norm*{ H_{\mathrm{c}}^{I}(t) } = \frac{\ell_{T}}{T} < \omega.
    \label{eqM_MaximumWorkExtractionProtocol}
\end{equation}
Only in the complementary regime $\omega T \le \ell_T$ does the finite-time constrained optimization remain nontrivial.

As an example, consider the case where $H_\mathrm{c}(t)$ can take arbitrary operators and $H_\mathrm{u} \equiv 0$.
In this case, the final Hamiltonian is independent of $T$, i.e., $H_{\mathcal{V}}^{\mathrm{f},I}(T) \equiv H_{\mathcal{V}}^{\mathrm{f}}$.
Consequently, $\mathcal{W}_\mathrm{c}(T)=\mathcal{W}(T)$ is a nondecreasing function of $T$, and the distance $\ell_T$ defined in Eq.~\eqref{eqM_OptimalGeodesicDistance} is independent of $T$. 
As discussed around Eq.~\eqref{passive}, the maximum work (ergotropy) is achieved by a unitary $U_\mathrm{c}=\mathcal{U}_\mathrm{erg}$, which is represented as $\mathcal{U}_\mathrm{erg}(\{\phi_n\}) = \sum_{n} e^{i \phi_{n}} \ketbra*|E_{n}><q_{n}|$ with arbitrary real phases $\phi_{n}$.
Then, the distance $\ell \equiv \ell_T$ is given by
\begin{align}
    \ell=\min_{\{\phi_n\}}\Norm*{\Log \mathcal{U}_\mathrm{erg}(\{\phi_n\})} \eqqcolon \Norm*{\Log \mathcal{U}_\mathrm{erg}(\{\tilde{\phi}_n\}) },
\end{align}
where $\tilde{U}_\mathrm{c}=\mathcal{U}_\mathrm{erg}(\{\tilde{\phi}_n\})$.
By definition of $\ell$ in Eq.~\eqref{eqM_OptimalGeodesicDistance}, we cannot achieve the ergotropy $\mathcal{W}_\mathrm{erg}$ for $T < \ell/\omega$. 
In contrast, we can extract $\mathcal{W}_\mathrm{erg}$ for $T \geq \ell/\omega$ with the control Hamiltonian defined in Eq.~\eqref{eqM_MaximumWorkExtractionProtocol}.

\subsection{Optimal protocol under the Lie-algebraic control}
After determining $\ell_T$ from the static optimization problem in Eq.~\eqref{eqM_OptimalGeodesicDistance}, we now turn to the complementary regime $\omega T \le \ell_T$, in which the norm constraint is active.
In this nontrivial case, we present our main theorem for maximum work extraction within finite time, which dramatically simplifies the optimization problem in Eq.~\eqref{optimize} into the problem of solving a time-independent nonlinear equation.
\begin{theorem} \label{thmM_MainResult}
    When $0 < \omega T \leq \ell_{T}$,
    the optimal work extraction protocol proceeds in the following three steps~(Fig.~\ref{figM_OptimalProtocol}):
    (i) turn on the control via a quench to $H_{\mathrm{c}}(+0) = \omega \mathsf{H}$;
    (ii) steer the system under the control Hamiltonian 
    \begin{equation}
        H_{\mathrm{c}}^{I}(t) = \omega \mathsf{H}\quad\text{for all $t \in (0,T)$,}
        \label{eqM_OptimalControlHamiltonian}
    \end{equation}
    which is time-independent in the interaction picture;
    and (iii) switch off the control at the final time $t=T$.
    Here, $\mathsf{H}$ is the time-independent rescaled Hamiltonian that satisfies the following closed nonlinear equation
    \begin{equation}
        C \mathsf{H} = -i \comm*{ H_{\mathcal{V}}^{\mathrm{f},I}\!(T) }{ e^{ -i\omega T \mathsf{H} } \rho^{\mathrm{i}}_\mathcal{V} e^{ +i\omega T \mathsf{H} } },\quad\!\! \Norm*{ \mathsf{H} } = 1,
        \label{eqM_SelfConsistentEq}
    \end{equation}
    for some scalar $C \geq 0$.

    Moreover, the optimal work is explicitly given as
    \begin{equation}
        \mathcal{W}_\mathrm{c}(T) = W_\mathrm{c}\ab(U_{\mathrm{c}} = e^{-i\omega T \mathsf{H}}) = 
        T D \int_{0}^{\omega} C(T;\omega') \d{\omega'},
        \label{workEq}
    \end{equation}
    where $C(T;\omega')$ denotes the scalar $C$ determined by Eq.~\eqref{eqM_SelfConsistentEq} for the norm constraint $\Norm*{H_{\mathrm{c}}(t)} \leq \omega'$ (instead of $\omega$), and $D$ denotes the Hilbert space dimension.
    In this sense, the scalar $C$ quantifies the power gain achievable by increasing the admissible control strength.
    When $H_{\mathcal{V}}^{\mathrm{f},I}\!(T) \equiv H_{\mathcal{V}}^{\mathrm{f}}$, $C$ is also proportional to $\pdv{\mathcal{W}_{\mathrm{c}}}/{T}$ almost everywhere because the operational time $T$ enters Eq.~\eqref{eqM_SelfConsistentEq} only through the product $\omega T$.
\end{theorem}
Unlike typical optimal control problems where the optimal Hamiltonian exhibits highly complex time dependence~\cite{MooreTibbetts_PhysRevA2012w, Deffner_JPhysBAtMolOptPhys2014b, Bukov_PhysRevX2018u, Bukov_PhysRevACollPark2018u, Brady_PhysRevLett2021u, Dionis_PhysRevACollPark2023i, Mazzoncini_PhysRevACollPark2023g, Li_SciRep2023z, Chen_PhysRevRes2025b}, the optimal protocol here is remarkably simple: 
the rescaled Hamiltonian $\mathsf{H}$ is independent of time $t \in (0,T)$.
As discussed in later subsections, the optimal Hamiltonian determined from Eq.~\eqref{eqM_SelfConsistentEq} is analytically obtained for $\mathfrak{su}(2)$ control and numerically solved for a more general Lie algebra.

Another relevant aspect is the knowledge of the uncontrollable Hamiltonian required to implement the optimal protocol~\eqref{eqM_OptimalControlHamiltonian}.
In typical control problems, full knowledge of $H_{\mathrm{u}}(t)$ is required to determine an optimal protocol~\cite{Sklarz_PhysRevA2002e, Khaneja_JMagnReson2005a, Dirr_GAMM-Mitteilungen2008h, DAlessandro_Other2021w, Andrei_Other2022y}, 
whereas in our Lie-algebraic control framework only the $\mathcal{V}$-component of $H_{\mathrm{u}}(t)$ needs to be known.
To make this explicit, we decompose $H_{\mathrm{u}}(t)$ orthogonally as $H_{\mathrm{u},\mathcal{V}}(t)+H_{\mathrm{u},\perp}(t)$.
Then, only $H_{\mathrm{u},\mathcal{V}}(t)$ must be specified, while no precise knowledge of $H_{\mathrm{u},\perp}(t)$ is required (other than that $H_{\mathrm{u}}(t)$ itself satisfies assumption~(ii) of Lie-algebraic control).
This is because we can show that the orthogonal component $H_{\mathrm{u}, \perp}(t)$ commutes with all of $\mathcal{V}$ and therefore does not affect the dynamics in $\mathcal{V}$~(see \Appendix{B}).
Consequently, the optimal control Hamiltonian can be written as
\begin{equation*}
    H_{\mathrm{c}}(t) = U_{\mathrm{u}}(t) H_{\mathrm{c}}^{I}(t) U_{\mathrm{u}}^{\dagger}(t) = U_{\mathrm{u},\mathcal{V}}(t) H_{\mathrm{c}}^{I}(t) U_{\mathrm{u},\mathcal{V}}^{\dagger}(t),
\end{equation*}
where $U_{\mathrm{u},\mathcal{V}}(t) \coloneqq \Texp\ab[-i \int_{0}^{t} H_{\mathrm{u},\mathcal{V}}(s) \d{s}]$ is the unitary generated by the $\mathcal{V}$-component alone.
Therefore, precise knowledge of $H_{\mathrm{u},\perp}(t)$ is unnecessary, making the optimal protocol in Eq.~\eqref{eqM_OptimalControlHamiltonian} robust against uncertainty or noise in this component and enhancing its practical relevance.

We remark that the required knowledge of $H_{\mathrm{u},\mathcal{V}}(t)$ can be obtained in different ways. 
For instance, it may be accessed either by direct measurement of the $\mathcal{V}$-component, or inferred from symmetry considerations that imply $H_{\mathrm{u},\mathcal{V}}(t)$ effectively vanishes (more precisely, $\comm*{ H_{\mathrm{u},\mathcal{V}}(t) }{ \mathcal{V} } \equiv 0$). 
In the latter case, no explicit transformation to the interaction picture is needed, since it then coincides with the original picture.

The above robustness with respect to $H_{\mathrm{u},\perp}(t)$ relies on assumption (ii) of Lie-algebraic control.
A natural question is then how the optimal work extraction is affected when this assumption is weakly violated.
Suppose that the Hamiltonian contains an additional perturbation \(\varepsilon K(t)\) with \(\Norm*{K(t)} \le 1\), so that
\begin{equation}
    H(t) = [H_{\mathrm{u}}(t) + \varepsilon K(t)] + H_{\mathrm{c}}(t),
\end{equation}
where possibly \(i\comm*{K(t)}{\mathcal{V}} \not\subset \mathcal{V}\).
For a fixed admissible control protocol \(H_{\mathrm{c}}(t)\), the corresponding
time-evolution operator is written as
\begin{equation}
\begin{aligned}
    U(T) &= U_{\mathrm{u}}(T) U_{\mathrm{c}}(T) \nonumber \\
    &\quad \times
    \ab(
        I -i\varepsilon \int_{0}^{T}
        U_{\mathrm{c}}(t)^{\dagger} K^{I}(t) U_{\mathrm{c}}(t)\,\d{t}
    )
    + \order{\varepsilon^{2}}.
\end{aligned}
\end{equation}
Therefore, the extracted work is modified only at order \(\varepsilon T\):
\begin{equation}
    W(U(T)) = W(U_{\mathrm u}(T)U_{\mathrm c}(T)) + \order{\varepsilon T}.
\end{equation}

To see the implication for the optimal value, let
\begin{equation}
    W_{\varepsilon}[H_{\mathrm c}]
    \coloneqq 
    W\ab(U_{\varepsilon}(T;H_{\mathrm c})),
    \quad
    W_{0}[H_{\mathrm c}]
    \coloneqq 
    W\ab(U_{0}(T;H_{\mathrm c})),
\end{equation}
where \(U_{\varepsilon}(T;H_{\mathrm c})\) and \(U_{0}(T;H_{\mathrm c})\) denote the
final-time unitaries generated by
\([H_{\mathrm u}(t)+\varepsilon K(t)] + H_{\mathrm c}(t)\) and
\(H_{\mathrm u}(t)+H_{\mathrm c}(t)\), respectively.
Let \(H_{\mathrm{c}}^{(0)}(t)\) and \(H_{\mathrm{c}}^{(\varepsilon)}(t)\) be optimal
control protocols for the unperturbed and perturbed problems, respectively.
Since the above expansion applies to any fixed admissible protocol, we have
\begin{equation}
    W_{\varepsilon}[H_{\mathrm{c}}^{(\varepsilon)}]
    - W_{0}[H_{\mathrm{c}}^{(0)}]
    \le
    W_{\varepsilon}[H_{\mathrm{c}}^{(\varepsilon)}]
    - W_{0}[H_{\mathrm{c}}^{(\varepsilon)}]
    = \order{\varepsilon T},
\end{equation}
where we used the optimality of \(H_{\mathrm{c}}^{(0)}(t)\) for the unperturbed problem.
Similarly,
\begin{equation}
    W_{\varepsilon}[H_{\mathrm{c}}^{(\varepsilon)}]
    - W_{0}[H_{\mathrm{c}}^{(0)}]
    \ge
    W_{\varepsilon}[H_{\mathrm{c}}^{(0)}]
    - W_{0}[H_{\mathrm{c}}^{(0)}]
    = \order{\varepsilon T},
\end{equation}
where we used the optimality of $H_{\mathrm{c}}^{(\varepsilon)}(t)$ for the perturbed problem.
Therefore, the optimal extractable work itself is modified only perturbatively:
\begin{equation}
    W_{\varepsilon}[H_{\mathrm{c}}^{(\varepsilon)}]
    - W_{0}[H_{\mathrm{c}}^{(0)}] = \order{\varepsilon T}.
\end{equation}
Moreover, the protocol optimal in the unperturbed case remains
\(\order{\varepsilon T}\)-optimal in the presence of weak perturbations:
\begin{align}
    W_{\varepsilon}[H_{\mathrm{c}}^{(0)}]
    &= W_{\varepsilon}[H_{\mathrm{c}}^{(\varepsilon)}]
    \begin{multlined}[t]
        + \ab(W_{\varepsilon}[H_{\mathrm{c}}^{(0)}] - W_{0}[H_{\mathrm{c}}^{(0)}]) \\
        + \ab(W_{0}[H_{\mathrm{c}}^{(0)}] - W_{\varepsilon}[H_{\mathrm{c}}^{(\varepsilon)}])
    \end{multlined}
    \nonumber \\
    &= W_{\varepsilon}[H_{\mathrm{c}}^{(\varepsilon)}] + \order{\varepsilon T}.
\end{align}
On the other hand, the optimal protocol itself may vary discontinuously with $\varepsilon$ in the presence of near-degeneracies or flat directions of the control landscape. 
A systematic analysis of such protocol-level robustness is left for future work.

As shown in \Appendix{C}, the proof of Theorem~\ref{thmM_MainResult} employs the Lagrange multiplier method with inequality constraints~\cite{Andrei_Other2022y} to maximize the extracted work under the norm constraint.
The stationary condition of the Lagrangian, together with the Lie-algebraic assumption, leads to the nonlinear equation given in \eqref{eqM_SelfConsistentEq}.
The relation between $\mathcal{W}_\mathrm{c}$ and $C$ in Eq.~\eqref{workEq} is regarded as the so-called envelope theorem~\cite{Milgrom_Econometrica2002f} in the optimization theory: 
this is proven by carefully showing the differentiability of the optimal work $\mathcal{W}_\mathrm{c}(T;\omega)$ as a function of $\omega$ and obtaining $\pdv{\mathcal{W}_\mathrm{c}}/{\omega} = T D C$ from the direct calculation.

\subsection{Analytical solution in $\mathfrak{su}(2)$ control}
\label{SectionIV.C}
To illustrate the general theory in the simplest nontrivial setting that allows for analytical treatment, we now consider the case where the control algebra is $\mathcal{V} \cong \mathfrak{su}(2)$, i.e., 
\begin{align}
    \mathcal{V} = \operatorname{span}_{\mathbb{R}}\{S_{1}, S_{2}, S_{3}\}
\end{align}
with $\comm*{S_{\alpha}}{S_{\beta}} = i\sum_{\gamma} \varepsilon_{\alpha\beta\gamma} S_{\gamma}$.
An example of such a setup is a two-level system where all of the traceless control operations are possible under the norm constraint.
In this case, $S_\alpha$ is given by the spin-1/2 operators. 
Another nontrivial example is the Heisenberg-type models with controlled magnetic fields introduced in Eq.~\eqref{eqM_Heisenberg}.
In this case, $S_\alpha$ corresponds to the three components of $\bm{S}$.

\begin{figure}[tb]
    \centering
    \includegraphics[width=\linewidth]{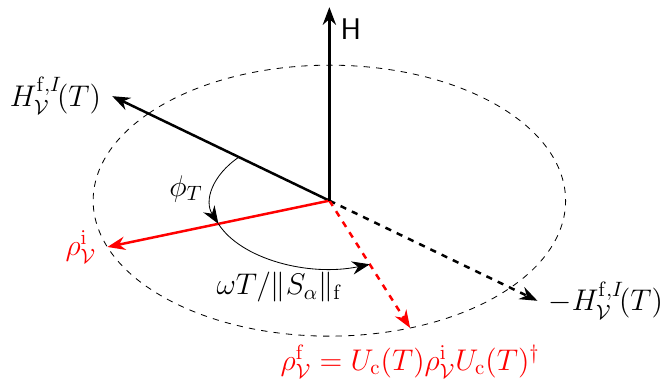}
    \caption{{Optimal work-extraction protocol within time $T$ under $\mathfrak{su}(2)$ control.}
    We analytically find that the solution~\eqref{eq_SU2OptimalHamiltonian} is to rotate the projected density operator from $\rho^{\mathrm{i}}_\mathcal{V}$ to $\rho^{\mathrm{f}}_\mathrm{c} \coloneqq U_{\mathrm{c}}(T) \rho^{\mathrm{i}}_\mathcal{V} U_{\mathrm{c}}(T)^{\dagger}$ along the shortest possible path in the Bloch sphere so that $\rho^{\mathrm{f}}_\mathrm{c}$ becomes maximally aligned with $-H_{\mathcal{V}}^{\mathrm{f},I}\!(T)$.
    Here, we describe operators as vectors in the three-dimensional space of controllable operators.
    }
    \label{figM_OptimalProtocol_su(2)}
\end{figure}

For the $\mathfrak{su}(2)$ control, we can obtain the optimal Hamiltonian $\mathsf{H}$ analytically.
To see this, we first observe from the self-consistent equation~\eqref{eqM_SelfConsistentEq} that the optimal rescaled Hamiltonian must be orthogonal to the final Hamiltonian in the Hilbert--Schmidt inner product, i.e., $\tr[ \mathsf{H}\, H_{\mathcal{V}}^{\mathrm{f},I}\!(T) ] = 0$.
Moreover, since $\mathsf{H}$ commutes with the time-evolution operator $e^{ -i \omega T \mathsf{H} }$, 
Eq.~\eqref{eqM_SelfConsistentEq} is written as 
\begin{align}
    C \mathsf{H} = -i \comm*{ e^{ +i\omega T \mathsf{H} } H_{\mathcal{V}}^{\mathrm{f},I}\!(T)e^{ -i\omega T \mathsf{H} }  }{ \rho^{\mathrm{i}}_\mathcal{V} }.
\end{align}
This equation further implies that $\mathsf{H}$ is orthogonal to $\rho^{\mathrm{i}}_\mathcal{V}$, $\tr[\mathsf{H}\, \rho^{\mathrm{i}}_\mathcal{V}]=0$.
Since the controllable space of operators is three-dimensional, these two orthogonality relations determine the optimal Hamiltonian $\mathsf{H}$ up to an overall sign: it must be parallel or antiparallel to $i\comm*{ H_{\mathcal{V}}^{\mathrm{f},I}\!(T) }{ \rho^{\mathrm{i}}_\mathcal{V} }$~(except when they commute) in the operator space.
This sign is fixed by the nonnegativity of the coefficient $C\geq 0$ in Eq.~\eqref{eqM_SelfConsistentEq},
which yields the unique solution (see \Appendix{D})
\begin{align}
    \mathsf{H} = \frac{-i\comm*{ H_{\mathcal{V}}^{\mathrm{f},I}\!(T) }{ \rho^{\mathrm{i}}_\mathcal{V} } }{ \Norm*{ \comm*{ H_{\mathcal{V}}^{\mathrm{f},I}\!(T) }{ \rho^{\mathrm{i}}_\mathcal{V} } }}.
    \label{eq_SU2OptimalHamiltonian}
\end{align}
Geometrically, this solution corresponds to rotating $\rho^{\mathrm{i}}_\mathcal{V}$ along the shortest Bloch-sphere path so that the final state aligns as closely as possible with $-H_{\mathcal{V}}^{\mathrm{f},I}\!(T)$~(Fig.~\ref{figM_OptimalProtocol_su(2)}).

With this explicit solution, the optimal extractable work $\mathcal{W}_\mathrm{c}(T)$ under $\mathfrak{su}(2)$ control admits a closed-form expression.
Let $\phi_{T} \in [0,\pi]$ be the angle between $H_{\mathcal{V}}^{\mathrm{f},I}\!(T)$ and $\rho^{\mathrm{i}}_\mathcal{V}$ in the operator space, defined by
\begin{align}
    \phi_{T}\coloneqq \arccos\ab( \frac{ \tr[H_{\mathcal{V}}^{\mathrm{f},I}\!(T) \rho_{\mathcal{V}}^{\mathrm{i}}] }{ \sqrt{ \tr[(H_{\mathcal{V}}^{\mathrm{f}})^2] \tr[(\rho_{\mathcal{V}}^{\mathrm{i}})^2] } } ).
\end{align}
Then, the extractable work for operational times $T \leq \ell_{T}/\omega$ is obtained as
\begin{multline}
    \mathcal{W}_\mathrm{c}(T) = D\, \Norm*{ H_{\mathcal{V}}^{\mathrm{f}} } \Norm*{ \rho_{\mathcal{V}}^{\mathrm{i}} } \\ 
    \times \ab[ \cos \phi_{T} - \cos\ab( \frac{\omega T }{ \Norm*{S_{\alpha}} }+ \phi_{T}) ],
    \label{eq_SolutionForSU2}
\end{multline}
where $S_{\alpha}$ is any of the orthonormal generators $\{S_{1}, S_{2}, S_{3}\}$ of $\mathcal{V} \cong \mathfrak{su}(2)$.

Importantly, the above results are applicable to both few-level and many-body systems.
For the case with the two-level system (i.e., full control of traceless operators under norm constraint), we can apply the above results with 
$D=2,\ H_{\mathcal{V}}^{\mathrm{f}} = H^\mathrm{f} - \mathrm{tr}[H^\mathrm{f}]/2,\ \rho^{\mathrm{i}}_\mathcal{V} = \rho^\mathrm{i} - I/2$, and $\Norm*{S_\alpha} = 1/2$. 
For $\mathfrak{su}(2)$ control of the Heisenberg model given in Eq.~\eqref{eqM_Heisenberg} with static background field, $\bm{B}_{0}(t) = \bm{B}_{0}$ for all $t \in [0,T]$.
The final Hamiltonian is given by $H_{\mathcal{V}}^{\mathrm{f}} = H_{\mathrm{u}}(T) = \bm{B}_{0} \cdot \bm{S}$.
In this case, the uncontrollable contribution $W(U_{\mathrm{u}}(T))$ vanishes under the convention introduced below Eq.~\eqref{optimize}, so that we have $\mathcal{W}_{\mathrm{c}}(T) = \mathcal{W}(T)$.
Under these settings, a direct calculation shows
\begin{align}
    \label{HHeisenberg}
    \mathsf{H} = \frac{(\bm{B}_{0} \times \expval*{\bm{S}}_{\mathrm{i}})\cdot{\bm{S}}}{ \Norm{ (\bm{B}_{0} \times \expval*{\bm{S}}_{\mathrm{i}})\cdot{\bm{S}} } }
\end{align}
and
\begin{align}
    \label{HcHeisenberg}
    H_{\mathrm{c}}(t) 
    &= e^{ -i t \bm{B}_{0} \cdot \bm{S} }\, \omega \mathsf{H}\, e^{ i t \bm{B}_{0} \cdot \bm{S} } \nonumber \\
    &= \omega \mathsf{H} \cos(\abs*{\bm{B}_{0}}\, t) - i\comm{\frac{ \bm{B}_{0} }{ \abs*{\bm{B}_{0}} }\cdot\bm{S}}{\omega \mathsf{H}} \sin(\abs*{\bm{B}_{0}}\, t).
\end{align}
The corresponding optimal extractable work is
\begin{align}
    \label{wctheisenberg}
    \mathcal{W}(T) =
    \abs*{\bm{B}_{0}} \cdot \abs*{\langle\bm{{S}}\rangle_\mathrm{i}} \ab[ \cos\phi-\cos\ab(\frac{\omega T}{\Norm*{S_{\alpha}}}+\phi) ],
\end{align}
where $\expval*{\bm{S}}_{\mathrm{i}} \coloneqq \mathrm{tr}[\bm{S}{\rho}^\mathrm{i}]$, and $\phi \coloneqq \arccos\ab( \frac{\bm{B}_{0}\cdot\langle{\bm{S}}\rangle_\mathrm{i}}{ \abs*{\bm{B}_{0}} \cdot \abs*{\langle{\bm{S}}\rangle_\mathrm{i}} })$ is independent of $T$.
Since $\Norm*{S_{\alpha}} = \mathcal{O}(V^{1/2})$, $\omega=\mathcal{O}(V^{1/2})$, and $\langle{\bm{S}}\rangle_\mathrm{i}=\mathcal{O}(V)$, we find that the extensive work extraction $\mathcal{W}(T)=\mathcal{O}(V)$ is possible with operational times $T=\mathcal{O}(V^0)$.
In particular, the minimum time for the maximum work extraction is given by 
\begin{equation}
    \mathcal{T}_{\ast} = \frac{\ell}{\omega} = \frac{\Norm*{S_{\alpha}} (\pi - \phi)}{\omega},
\end{equation}
where we note that $\ell \equiv \ell_{T}$ does not depend on $T$ in this case.
The maximum work is given by
\begin{equation}
    \mathcal{W}_{\ast} = \abs*{\bm{B}_{0}} \cdot \abs*{\langle\bm{{S}}\rangle_\mathrm{i}} ( \cos\phi + 1 ).
\end{equation}

As seen from the above results, our method shows how to design the time dependence of the optimal control $H_{\mathrm{c}}(t)$ using the time-independent operator $\mathsf{H}$ given in Eq.~\eqref{HHeisenberg}.
Specifically, as demonstrated in Eq.~\eqref{HcHeisenberg}, an optimal protocol can be implemented by dynamically adjusting the direction of the control field, even in the presence of a (possibly large) background field $\bm{B}_0$.

From another perspective, our result also provides a quantitative bound on finite-time cooling of this many-body system, where the objective is to drive the system close to its ground state.
Specifically, if the system with Heisenberg interactions (Eq.\eqref{eqM_Heisenberg}) can only be controlled by global (i.e., translation-invariant) magnetic fields, then the energy reduction achievable within time $T$ under such protocols is strictly bounded above by $\mathcal{W}(T)$ in Eq.\eqref{wctheisenberg}. 
Moreover, the optimal cooling protocol is realized using the time-independent control operator in Eq.~\eqref{HHeisenberg}.

\subsection{Numerical solution for higher-dimensional $\mathcal{V}$}
\label{SectionIV.D}
While Eq.~\eqref{eqM_SelfConsistentEq} becomes analytically intractable for $\dim \mathcal{V} > 3$, its solution can still be obtained via a gradient-based numerical method.
For this purpose, we introduce a cost function 
\begin{align}
    g(X) \coloneqq \min_{C\geq 0} \Norm*{CX - F(X)}^2\quad (X \in \mathcal{V}),
\end{align}
where a candidate control operator $X$ satisfies $\Norm*{X} = 1$, and $F(X) \coloneqq -i\comm*{ H_{\mathcal{V}}^{\mathrm{f},I}\!(T) }{ e^{-i\omega T X}\, \rho_{\mathcal{V}}^{\mathrm{i}}\, e^{+i\omega T X} }$ is the right-hand side of Eq.~\eqref{eqM_SelfConsistentEq}.
This cost function quantifies how well a candidate control operator $X$ satisfies the self-consistent equation~\eqref{eqM_SelfConsistentEq}.
A zero value of $g(X)$ implies that $X$ solves the equation for some $C \geq 0$, and hence serves as a candidate for the optimal control operator $\mathsf{H}$.

The minimization over $C$ for the above function $g(X)$ can be carried out analytically, and we obtain
\begin{equation}
    g(X) = \Norm*{ F(X) }^{2} - \ab( \frac{1}{D} \tr[X F(X)]^{+})^2,
    \label{eq_CostFunction}
\end{equation}
where $x^{+} \coloneqq \max\{0,x\}$ for $x \in \mathbb{R}$.
Importantly, the expression for the gradient of $g(X)$ can be obtained analytically, enabling a gradient-based search for a solution of the self-consistent equation~(see \Appendix{E}).
With the gradient at hand, we implement the standard steepest descent algorithm to numerically solve the self-consistent equation.
We emphasize that this numerical algorithm is much more straightforward than the direct numerical optimization of time-dependent dynamics in the original problem.

\begin{figure}
    \centering
    \includegraphics[width=\linewidth]{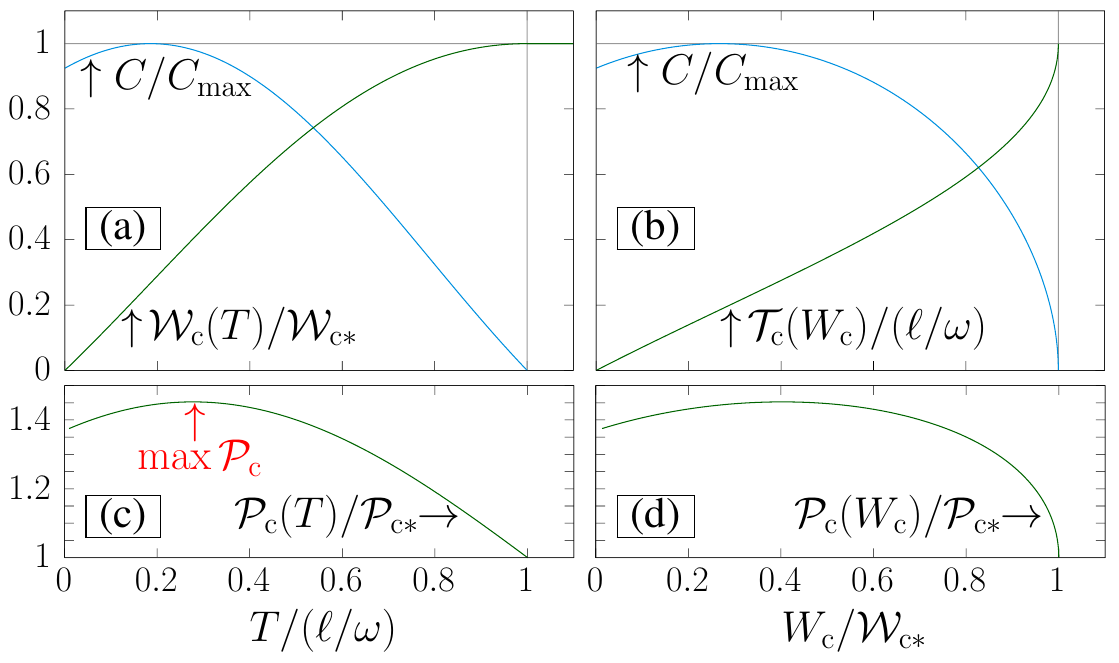}
    \caption{{Optimal work extraction from a random three-level system under the full control ($\mathcal{V} \cong \mathfrak{su}(3)$).}
    (a) Optimal amount of extracted work as a function of $T$~(the green curve). 
    The blue curve shows the constant $C$ in Eq.~\eqref{eqM_SelfConsistentEq}, which is proportional to the derivative of the optimal work.
    It remains finite for $T \in (0, \ell/\omega)$ and vanishes as $T \to \ell/\omega$.
    (b) Minimum time $\mathcal{T}_{\mathrm{c}}(W_{\mathrm{c}})$ required to extract a given amount of work $W_{\mathrm{c}}$, which is obtained by inverting $W_{\mathrm{c}} = \mathcal{W}_{\mathrm{c}}(T)$.
    (c) Optimal power $\mathcal{P}_{\mathrm{c}}(T) \coloneqq \mathcal{W}_{\mathrm{c}}(T) / T$ of work extraction within time $T$, normalized by the reference power $\mathcal{P}_{\mathrm{c}\ast} \coloneqq \mathcal{W}_{\mathrm{c}\ast} / (\ell/\omega)$.
    Its maximum is attained strictly before the minimum time for the maximum work extraction.
    (d) Optimal power $\mathcal{P}_{\mathrm{c}}$ as a function of the extracted work $W_{\mathrm{c}}$.
    It decreases rapidly to $\mathcal{P}_{\mathrm{c}\ast}$ as $W_{\mathrm{c}} \to \mathcal{W}_{\mathrm{c}\ast}$.
    }
    \label{fig_NumericalResultForSU3}
\end{figure}

Figure~\ref{fig_NumericalResultForSU3}(a) shows the optimal work $\mathcal{W}_{\mathrm{c}}(T)$ obtained from the above numerical method, for full control of a random three-level system.
Here we assume control over all traceless Hermitian operators, corresponding to $\mathcal{V} = \mathfrak{su}(3)$.
The uncontrollable Hamiltonian is fixed as $H_{\mathrm{u}}(t) = H^{\mathrm{f}}$ for all $t \in [0,T]$.
Consequently, $H_{\mathcal{V}}^{\mathrm{f},I}(T) = H_{\mathcal{V}}^{\mathrm{f}}$ for any $T$, 
and $\ell_{T} \equiv \ell$ is independent of $T$.
Both the optimal work $\mathcal{W}_{\mathrm{c}}(T)$ and the optimal operator $\mathsf{H}$ exhibit a nontrivial dependence on $T$. 
This is more complicated than $\mathfrak{su}(2)$ control, where the condition $H_{\mathcal{V}}^{\mathrm{f},I}(T) \equiv H_{\mathcal{V}}^{\mathrm{f}}$ results in a simple cosine curve for the optimal extractable work $\mathcal{W}_{\mathrm{c}}(T)$, and the optimal operator $\mathsf{H}$ becomes independent of $T$.
Consistent with the discussions around Eqs.~\eqref{nondec}--\eqref{eqM_MaximumWorkExtractionProtocol}, $\mathcal{W}_{\mathrm{c}}(T)$ is nondecreasing in $T$ and attains its maximum at $T= \ell/\omega$.
We also show the normalized value of $C$, which is proportional to $\pdv{\mathcal{W}_\mathrm{c}}/{T}$ as discussed in Theorem~\ref{thmM_MainResult}.
Note that $C$ becomes zero at $T = \ell/\omega$, because $\mathcal{W}_\mathrm{c}(T) $ attains its maximum there.

By inverting the function $\mathcal{W}_{\mathrm{c}}(T)$, we obtain the minimum time $\mathcal{T}_{\mathrm{c}}(W_{\mathrm{c}})$ required to extract a given amount of work $W_{\mathrm{c}}$, as shown in Fig.~\ref{fig_NumericalResultForSU3}(b).
This yields the exact speed limit for finite-time work extraction from closed quantum systems: $T \geq \mathcal{T}_{\mathrm{c}}(W_{\mathrm{c}})$, \textit{where equality is attainable}.
This is in contrast with previous attempts~\cite{Garcia-Pintos2020-ix, Mohan_PhysRevACollPark2022w, Shrimali_PhysRevACollPark2024q, Deffner_JPhysA:MathTheor2017h, Hamazaki_CommunPhys2024i} to evaluate the time for work extraction via quantum speed limits, for which the attainability of equality remained unclear.
Note that the speed limit becomes steeper as $T$ approaches $\ell/\omega$.
Indeed, we have $\odv{\mathcal{T}_\mathrm{c}}/{W_c}=(\odv{\mathcal{W}_\mathrm{c}}/{T})^{-1}\rightarrow\infty$, since $\odv{\mathcal{W}_\mathrm{c}}/{T}\propto C\rightarrow 0$ in this case.

Furthermore, our results reveal a nontrivial relationship between power, operational time, and extractable work, as shown in Fig.~\ref{fig_NumericalResultForSU3}(c)(d).
As given in Proposition~\ref{thmM_TradeOff}, the power attains its maximum at some time strictly before the minimum time required for the maximum work extraction (note that Proposition~\ref{thmM_TradeOff} similarly applies to $\mathcal{W}_{\mathrm{c}}(T)$ and $\mathcal{P}_{\mathrm{c}}(T)$, in place of $\mathcal{W}(T)$ and $\mathcal{P}(T)$).
This implies that extracting a modest amount of energy 
from each system and rapidly switching to fresh systems yields higher power than fully extracting the available energy $\mathcal{W}_{\mathrm{c}\ast}$ from each system.


\section{Efficient numerical solution for many-body systems}
\label{SectionV}
The numerical method introduced above is general and applicable to any finite-dimensional system under Lie-algebraic control.
However, for physically relevant many-body systems, 
the cost of computing the gradient $\nabla g(X)$, required at each iteration step, scales exponentially in system size as $\order*{D^3}$.
This fact makes it infeasible to employ the naive gradient-based method in many-body systems. 

However, the number of controllable degrees of freedom, given by $\dim \mathcal{V}$ in our case of Lie-algebraic control, typically remains constant with system size;
we can exploit this fact to devise an efficient numerical solution of the self-consistent equation even for many-body systems.
Crucially, all relevant operations in the optimization problem, such as commutators, matrix exponentials, and the Hilbert--Schmidt inner product, can be evaluated entirely within a much smaller representation of $\mathcal{V}$ whose dimension is independent of system size.
That is, all computations involved in the cost function~\eqref{eq_CostFunction} and its gradient can be performed within this reduced representation.

\subsection{Case study for $\mathrm{SU}(n)$-Hubbard models}
While the reduction of the representation generally follows from several facts from the theory of Lie algebras, we first illustrate how it applies to a concrete physical model.
As an example, we consider an $\mathrm{SU}(n)$-Hubbard model with controllable fermion flavors defined on a lattice $\Lambda$ of size $V \coloneqq \abs*{\Lambda}$.
The total particle number is denoted by $N$, and the Hilbert space dimension is given by $D = \binom{nV}{N}$, which is exponentially large in system size.
As we will see below, the reduction can be carried out analytically in this case, with the dimension of the reduced representation being $n$, independently of $D$.
The Hamiltonian is given by $H(t) = H_{\mathrm{c}}(t) + H_{\mathrm{u}}(t)$ with
\begin{equation*}
    H_{\mathrm{c}}(t) = \sum_{\alpha\beta} u_{\alpha\beta}(t) \sum_{x \in \Lambda} c_{x\alpha}^{\dagger} c_{x\beta}
    \ \ab(\eqqcolon \sum_{\alpha\beta} u_{\alpha\beta}(t) E_{\alpha\beta}),
\end{equation*}
and
\begin{multline}\label{hubbard}
    H_{\mathrm{u}}(t) = -J(t) \sum_{\expval*{x,y}} \sum_{\alpha=1}^{n} \ab( c_{x\alpha}^{\dagger} c_{y\alpha} + \mathrm{h.c.} ) \\
    + \frac{U(t)}{2} \sum_{\alpha \neq \beta} \sum_{x \in \Lambda} n_{x\alpha} n_{x\beta}
    +\sum_{\alpha} h_{\alpha}(t) \ab( \sum_{x\in\Lambda} n_{x\alpha} ).
\end{multline}
Here, $c_{x\alpha}$ denotes the fermionic annihilation operator of the $\alpha$th flavor at site $x$, 
\begin{equation}
    E_{\alpha\beta} \coloneqq \sum_{x\in\Lambda}c_{x\alpha}^{\dagger} c_{x\beta},
\end{equation}
$\expval*{x,y}$ denotes the nearest-neighboring sites, $n_{x\alpha} \coloneqq c_{x\alpha}^{\dagger} c_{x\alpha}$, $h_{\alpha}(t)$ are flavor-dependent potentials, and $u_{\alpha\beta}(t) \in \mathbb{C}$ is a control field satisfying $u_{\alpha\beta}(t) = u_{\beta\alpha}(t)^{\ast}$ and $\sum_{\alpha=1}^{n} u_{\alpha\alpha}(t) = 0$.
One can also directly verify that $H_{\mathrm{u}}$ leaves $\mathcal{V}$ invariant, ensuring that the system falls within the Lie-algebraic control.
The $\mathcal{V}$-component of the final Hamiltonian is given by $H_{\mathcal{V}}^{\mathrm{f}} = \sum_{\alpha} h_{\alpha}^{\mathrm{f}} E_{\alpha\alpha}$.

We choose as the initial state $\rho^{\mathrm{i}} = \dyad*{\psi}$, where each flavor has a definite particle number, i.e., $E_{\alpha\alpha} \ket*|\psi> = N_{\alpha} \ket*|\psi>$ for all $\alpha$.
Here, we note that the operators $E_{\alpha\alpha}$ represent the total particle number of the $\alpha$th flavor, and $N_{\alpha} \in \mathbb{N}$ are their values in $\ket*|\psi>$ satisfying $N = \sum_{\alpha} N_{\alpha}$.
As shown in \Appendix{F}, the orthogonal projection of $\rho^{\mathrm{i}}$ onto $\mathcal{V}$ is given by
\begin{align}
    \rho_{\mathcal{V}}^{\mathrm{i}}
    &= \frac{N}{C_{V,N}^{(n)}} \sum_{\alpha = 1}^{n} \ab( \frac{N_{\alpha}}{N} - \frac{1}{n} ) E_{\alpha\alpha},
    \label{eq_SUnRhoProjection}
\end{align}
where $C_{V,N}^{(n)} \coloneqq V \binom{nV-2}{N-1}$.

With the above setup,
we now translate the original problem concerning $D\times D$ matrices into that concerning $n\times n$ matrices.
For this purpose, we introduce the standard representation of $\mathcal{V} \cong \mathfrak{su}(n)$, defined as a linear map satisfying 
\begin{align}
\pi(E_{\alpha\beta}) \coloneqq \dyad*{ e_{\alpha} }{ e_{\beta} }.
\end{align}
Here, $\{ \ket|e_{\alpha}> \}_{\alpha=1}^{n}$ is an orthonormal basis of an $n$-dimensional Hilbert space.
Therefore, $\pi$ reduces the dimensionality of the Hilbert space, on which the operators act, from $D$ to $n$.

Let us consider rewriting the work to optimize with the reduced representation.
As we will explain below, we find that Eq.~\eqref{workdec} is written as
\begin{multline}
    W_\mathrm{c}(U_{\mathrm{c}}) = -C_{V,N}^{(n)} \tr\ab[ \pi(H_{\mathcal{V}}^{\mathrm{f},I}\!(T))\, U_{\mathrm{c} \pi} \pi(\rho_{\mathcal{V}}^{\mathrm{i}}) U_{\mathrm{c} \pi}^{\dagger} ] \\ +\tr[H_{\mathcal{V}}^{\mathrm{f}} \rho_{\mathcal{V}}^{\mathrm{i}}],
    \label{eqM_reducedRep_Work}
\end{multline}
where $\pi(H_{\mathcal{V}}^{\mathrm{f},I}\!(T)) = U_{\mathrm{u}\pi}^{\dagger} \pi(H_{\mathcal{V}}^{\mathrm{f}}) U_{\mathrm{u}\pi}$ and
\begin{align}
    \begin{split}
    U_{\mathrm{u} \pi}(T) &\coloneqq \Texp\ab[ -i \int_{0}^{T} \pi\ab(H_{\mathrm{u},\mathcal{V}}(t)) \d{t} ], 
     \\
    U_{\mathrm{c} \pi}(T) &\coloneqq \Texp\ab[ -i \int_{0}^{T} \pi\ab(H_{\mathrm{c}}^{I}(t)) \d{t} ].
    \end{split} \label{eq_ReducedUnitary}
\end{align}
Here, the only $\mathcal{V}$-component $H_{\mathrm{u},\mathcal{V}}$ of the uncontrollable Hamiltonian appears in Eq.~\eqref{eq_ReducedUnitary}, because the orthogonal component $H_{\mathrm{u},\perp}$ commutes with all of $\mathcal{V}$ and does not affect the dynamics in $\mathcal{V}$~(see \Appendix{B}).
Moreover, the norm constraint $\Norm*{ H_{\mathrm{c}}(t) }$ translates into
\begin{equation}
    \Norm*{ \pi(H_{\mathrm{c}}(t)) } = \kappa \Norm*{ H_{\mathrm{c}}(t) } \leq \kappa \omega,
    \label{eq_ReducedNormConstraint}
\end{equation}
where $\kappa \coloneqq (n C^{(n)}_{V,N} / D)^{-1/2}$~(see below).
Therefore, all the inputs (the work and the norm constraint) to the optimization problem are expressed in the reduced $n$-dimensional representation.

We can then apply Theorem~\ref{thmM_MainResult} to this $n$-dimensional optimization problem to obtain an optimal control Hamiltonian $\pi(H_{\mathrm{c}}^{I}(t)) = \kappa \omega \mathsf{H}_{\pi}$, where $\mathsf{H}_{\pi}$ satisfies the self-consistent equation~\eqref{eqM_SelfConsistentEq}:
\begin{equation}
  C_{\pi} \mathsf{H}_{\pi} = -i \comm*{ \pi(H_{\mathcal{V}}^{\mathrm{f},I}\!(T)) }{ e^{ -i \kappa \omega T \mathsf{H}_{\pi} } \pi(\rho_{\mathcal{V}}^{\mathrm{i}}) e^{ i \kappa \omega T \mathsf{H}_{\pi} } }
  \label{eqM_SelfConsistentEq_Eg}
\end{equation}
with $\Norm*{\mathsf{H}_{\pi}} = 1$ and $C_{\pi}$ being some nonnegative scalar.
As discussed earlier, this equation can be solved numerically based on the gradient method at a cost that does not scale with system size.
Once the solution $\mathsf{H}_{\pi}$ is obtained in this $n$-dimensional representation, the optimal control Hamiltonian in the original representation can be reconstructed as $H_{\mathrm{c}}^{I}(t) = \kappa \omega\, \pi^{-1}(\mathsf{H}_{\pi})$.

We note that, in Eq.~\eqref{eq_ReducedNormConstraint}, we have $\Norm*{ \pi(H_{\mathrm{c}}(t)) }^2 = \sum_{\alpha,\beta=1}^{n} \abs*{ u_{\alpha\beta}(t) }^2$, which represents the strength of the control field $\{ u_{\alpha\beta}(t) \}_{\alpha,\beta=1}^{n}$.
Therefore, the natural scaling of $\omega$ is such that $\kappa \omega = \order{V^0}$.
On the other hand, the operational time $T$ appears in Eq.~\eqref{eqM_SelfConsistentEq_Eg} only through the combination $\kappa \omega T$, except for the explicit $T$-dependence in $H_{\mathcal{V}}^{\mathrm{f},I}(T)$.
Taken together, these observations imply that the exact speed limit $\mathcal{T}_{\mathrm{c}}(W_{\mathrm{c}})$ is also of $\order*{V^0}$ for $W_{\mathrm{c}} = \order*{\mathcal{W}_{\mathrm{c}\ast}}$ when $H_{\mathcal{V}}^{\mathrm{f},I}(T)$ does not depend on $T$.

To obtain Eq.~\eqref{eqM_reducedRep_Work}, we first note that 
\begin{align}
    \pi(\comm*{E_{\alpha\beta}}{E_{\mu\nu}}) = \comm*{\pi(E_{\alpha\beta})}{\pi(E_{\mu\nu})}
\end{align}
for any $E_{\alpha\beta}$ and $E_{\mu\nu}$, manifesting that $\pi$ is indeed a representation of $\mathcal{V}$.
From the Baker--Campbell--Hausdorff formula, this also implies that 
\begin{align}
    \pi(e^{-iX} Y e^{iX}) = e^{-i \pi(X)}  \pi(Y) e^{i \pi(X)}
\end{align}
for any $X,Y \in \mathcal{V}$.
Therefore, the matrix exponential $e^{-iX}$ can be evaluated as $e^{-i \pi(X)}$ in the standard representation $\pi$.
In the same vein, the time-ordered exponential in the reduced representation can be evaluated as in Eq.~\eqref{eq_ReducedUnitary}, so that we have 
\begin{align}
    \begin{split}
        \pi(H_{\mathcal{V}}^{\mathrm{f},I}\!(T)) &= U_{\mathrm{u} \pi}(T)^{\dagger}\, \pi(H_{\mathcal{V}}^{\mathrm{f}})\, U_{\mathrm{u} \pi}(T), \\
        \pi(U_{\mathrm{c}}(T) \rho_{\mathcal{V}}^{\mathrm{i}} U_{\mathrm{c}}(T)^{\dagger}) &= U_{\mathrm{c} \pi}(T)\, \pi(\rho_{\mathcal{V}}^{\mathrm{i}})\, U_{\mathrm{c} \pi}(T)^{\dagger}.
    \end{split}
\end{align}
Moreover, we can verify that the Hilbert--Schmidt inner product in the original Hilbert space and the standard representation are proportional to each other for traceless operators: 
there exists a constant $C_{V,N}^{(n)}$ such that 
\begin{align}
    \tr[ X Y ] = C_{V,N}^{(n)} \tr[ \pi(X) \pi(Y) ]
    \label{eq_SUnInnerProducts}
\end{align}
for all $X$ and $Y$ in $\mathcal{V}$~(see \Appendix{F}).
Therefore, the normalized Frobenius norm in the standard representation is related to the original norm by $\Norm*{\pi(X)} = \kappa \Norm*{X}$ with the rescaling factor $\kappa \coloneqq (n C^{(n)}_{V,N} / D)^{-1/2}$.
This also implies that the norm constraint translates to $\Norm*{\pi(H_{\mathrm{c}}(t))} \leq \kappa \omega$ as in Eq.~\eqref{eq_ReducedNormConstraint}.
Finally, the final Hamiltonian $H_{\mathcal{V}}^{\mathrm{f}}$ and the initial state $\rho_{\mathcal{V}}^{\mathrm{i}}$ are trivially mapped by applying $\pi$.
This completes the reformulation of the problem in the reduced representation with $n\times n$ matrices (as in Eq.~\eqref{eqM_reducedRep_Work}), to which the numerical solution discussed earlier can be efficiently applied.

\begin{figure}
    \centering
    \includegraphics[width=\linewidth]{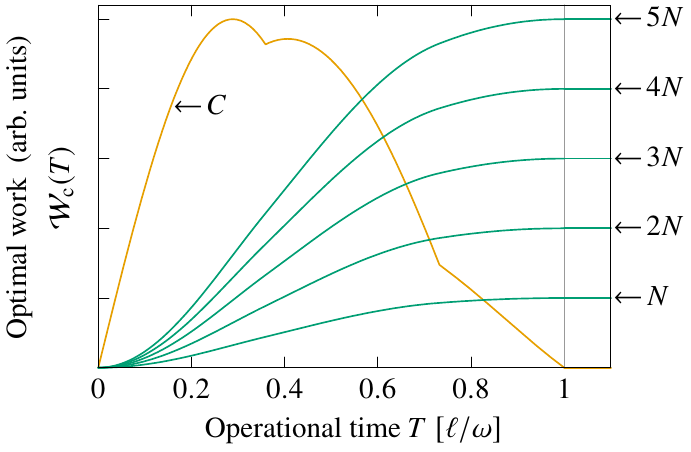}
    \caption{Optimal work extraction from an $\mathrm{SU}(6)$ Hubbard model with static flavor-dependent potentials $h_{\alpha}(t)\equiv h_{\alpha}$.
        The values of $h_{\alpha}$ and the flavor fractions $N_{\alpha}/N$ in Eq.~\eqref{eq_SUnRhoProjection} are randomly sampled as $h_{\alpha} \simeq \{0, 0.11, 0.34, 0.54, 0.84, 1.17\}$ and $N_{\alpha}/N \simeq \{0.022, 0.078, 0.15, 0.24, 0.25, 0.26\}$.
        The initial state is chosen as the one maximizing the energy expectation value. 
        According to Eqs.~\eqref{eq_SUnRhoProjection} and \eqref{eqM_reducedRep_Work}, the optimal work scales extensively with particle number for fixed flavor fractions and is independent of the lattice size $V$  and the parameters $J(t)$ and $U(t)$ in Eq.~\eqref{hubbard}.
        Moreover, $C$ has several cusps as a function of operational time $T$, where a discontinuous transition of the optimal operator $\mathsf{H}$ occurs.
    }
    \label{fig_SU6_WorkScaling}
\end{figure}

Figure~\ref{fig_SU6_WorkScaling} illustrates the scaling of optimal work extraction in an $\mathrm{SU}(6)$ Hubbard model, which has been realized experimentally~\cite{Taie_NatPhys2022d}.
We consider static flavor-dependent potentials $h_{\alpha}(t) \equiv h_{\alpha}$ and randomly choose both the values of $h_{\alpha}$ and the flavor fractions $N_{\alpha}/N$.
The initial state is chosen as the one that maximizes the energy expectation value.
According to Eqs.~\eqref{eq_SUnRhoProjection} and \eqref{eqM_reducedRep_Work}, the optimal work scales extensively with particle number at fixed flavor fractions and is independent of the lattice size $V$ and the parameters $J(t)$ and $U(t)$ in Eq.~\eqref{hubbard}.

Interestingly, $C$ has several cusps as a function of operational time $T$, where we find that the optimal operator $\mathsf{H}$ exhibits a discontinuous change. 
These control phase transitions of the optimal Hamiltonian reflect the complexity of the optimal work extraction problem, while our method offers a way to solve it.

\subsection{Generalization via changing representations}
\label{SectionVB}
Let us now discuss our general method for the reduction of the representation by employing three facts from the representation theory of Lie algebras:
\begin{enumerate}
\renewcommand{\labelenumi}{(\alph{enumi})}
\item
Any subalgebra $\mathcal{V}$ of a compact Lie algebra (in our case, $\mathcal{B}(\mathcal{H}) = \mathfrak{u}(D)$) can be orthogonally decomposed into its semisimple part $\mathcal{V}_{\mathfrak{ss}} \coloneqq \comm*{ \mathcal{V} }{ \mathcal{V} }$ and its center $\mathfrak{z}[\mathcal{V}] \coloneqq \{ X \in \mathcal{V} \colon \comm*{X}{\mathcal{V}} = 0 \}$~\cite{Knapp_Other1996h, Sepanski_Other2006f}.
In the example above, this decomposition is trivial because $\mathcal{V} \cong \mathfrak{su}(n)$ is itself simple.
\item
Any semisimple Lie algebra $\mathcal{V}_{\mathfrak{ss}}$ can be orthogonally decomposed into a direct sum of simple Lie algebras as $\mathcal{V}_{\mathfrak{ss}} = \oplus_{j} \mathcal{V}_{j}$~\cite{Humphreys_Other1972o, Knapp_Other1996h, Fulton1999-ai}.
In the example above, this decomposition is trivial because $\mathcal{V}_{\mathfrak{ss}} \cong \mathfrak{su}(n)$ is already simple.
\item For any representation $\pi_{j}$ of a simple Lie algebra $\mathcal{V}_{j}$, there exists a positive constant $C_{\pi_{j}}$ satisfying
\begin{align}
    C_{\pi_{j}} \tr[\pi_{j}(X) \pi_{j}(Y)] = \tr[XY]
\end{align}
for all $X,Y \in \mathcal{V}_{j}$~\cite{Humphreys_Other1972o}.
In the example above, we confirm this property with $C_{\pi} = C_{V,N}^{(n)}$.
\end{enumerate}
From (a) and (b), the control algebra $\mathcal{V}$ decomposes as
\begin{equation}
    \mathcal{V} = \bigoplus_{j} \mathcal{V}_{j} \oplus \mathfrak{z}[\mathcal{V}],
\end{equation}
where each $\mathcal{V}_{j}$ is simple, and $\mathfrak{z}[\mathcal{V}]$ is the center of $\mathcal{V}$.
Correspondingly, any operator $X \in \mathcal{V}$ decomposes as $X = \sum_{j} X_{j} + X_{\mathfrak{z}}$ with $X_{j} \in \mathcal{V}_{j}$ and $X_{\mathfrak{z}} \in \mathfrak{z}[\mathcal{V}]$.
Combining this fact with (c), the Hilbert--Schmidt inner product can be expressed as 
\begin{align}
\tr[XY] = \sum_{j} C_{\pi_{j}} \tr[ \pi_{j}(X_{j}) \pi_{j}(Y_{j}) ] +\tr[X_{\mathfrak{z}} Y_{\mathfrak{z}}]
\end{align}
for any $X, Y \in \mathcal{V}$.
The last term involving the center must still be evaluated in the full Hilbert space.
However, it can be safely disregarded for the purpose of optimization,
since the center is invariant under any control within $\mathcal{V}$, and conversely, any operator in the center cannot contribute to nontrivial dynamics in the work $W(U(T))$.

As practical representations, one may employ the standard representation by $n\times n$ skew-Hermitian matrices for $\mathfrak{su}(n)$, and $n\times n$ skew-symmetric matrices for $\mathfrak{so}(n)$, and $2n\times 2n$ Hamiltonian matrices for $\mathfrak{sp}(n)$, for example.
The point is that the dimensions of these representations depend only on that of the controllable subspace $\mathcal{V}$ and are independent of $D$.
Given any $X \in \mathcal{V}_{j}$, its representation $\pi_{j}(X)$ can be computed as follows.
Let $\{ \Lambda_{\beta} \}_{\beta}$ be an orthonormal basis of $\mathcal{V}_{j}$.
Then, we have $\pi_{j}(X) = \sum_{\beta} c_\beta \pi_{j}(\Lambda_{\beta})$, where $c_{\beta} \coloneqq \tr[ \Lambda_{\beta} X ] / \tr[ (\Lambda_{\beta})^2 ]$.
The computation of each $c_{\beta}$ costs $\order*{D^2}$ operations, and this needs to be done only once at the initial stage of the numerical calculation.
Without this reduction, each iteration of the gradient descent would require $\order*{D^3}$ operations due to diagonalization in the full Hilbert space, resulting in a total cost of $\order*{N_{\mathrm{step}} D^3}$ with $N_{\mathrm{step}}$ being the number of iterations.
In contrast, once the $\pi_{j}(X)$ are computed, all subsequent calculations proceed within the reduced representation, requiring only $\order*{ \sum_{j} n_j^3}$ operations per iteration, with $n_{j}$ the dimension of the representation $\pi_{j}$.
This significantly reduces the total computational cost to $\order*{D^2 + N_{\mathrm{step}} \sum_{j} n_{j}^3}$.

\section{Extension to optimization of general expectation values}
\label{SectionVI}
While we have so far focused on the work extraction problem, the framework also straightforwardly applies to a broader class of control problems beyond work extraction.
As the simplest example, by reversing the sign of the final Hamiltonian ($H^{\mathrm{f}} \mapsto -H^{\mathrm{f}}$), our framework equally applies to the finite-time charging process of a quantum battery, where we aim to control the system so that its internal energy becomes as large as possible~\cite{Ferraro_PhysRevLett2018f, Mazzoncini_PhysRevACollPark2023g, Rodriguez_NewJPhys2024i, Campaioli_RevModPhys2024z}.

When the control algebra is the space of all Hermitian operators (i.e., $\mathcal{B}(\mathcal{H})$) or all traceless Hermitian operators, the objective function can be generalized to the expectation value of an arbitrary observable $A \in \mathcal{B}(\mathcal{H})$ at the final time.
By replacing $H^{\mathrm{f}}$ with $-A$ in Eq.~\eqref{defM_Work}, the objective function is given by
\begin{equation}
    f(U) \coloneqq \tr[A\, U \rho^{\mathrm{i}} U^{\dagger}] - \tr[A\, \rho^{\mathrm{i}}].
    \label{eqM_GeneralObjectiveExp}
\end{equation}
This extended framework encompasses a variety of physically relevant tasks through appropriate choices of $A$.
Theorem~\ref{thmM_MainResult} applies to the optimization of $f(U)$ as well, in which case the self-consistent equation~\eqref{eqM_SelfConsistentEq} takes the form
\begin{equation}
    C \mathsf{H} = i \comm*{ A^{I}(T) }{ e^{ -i\omega T \mathsf{H} } \rho^{\mathrm{i}} e^{ +i\omega T \mathsf{H} } },\quad\!\! \Norm*{ \mathsf{H} } = 1.
    \label{eqM_GeneralExp_SelfConsistentEq}
\end{equation}
Here, $A_{\mathcal{V}}^{I}(T) = A^{I}(T)$ and $\rho_{\mathcal{V}}^{\mathrm{i}} = \rho^{\mathrm{i}}$ for the full control $\mathcal{B}(\mathcal{H})$, and $A_{\mathcal{V}}^{I}(T) = A^{I}(T) - (\tr A) I/D$ and $\rho_{\mathcal{V}}^{\mathrm{i}} = \rho^{\mathrm{i}} - I/D$ for the control of all traceless Hermitian operators.
The equation~\eqref{eqM_GeneralExp_SelfConsistentEq} applies to both cases.
Again, we stress that this approach provides an achievable speed limit, in contrast with previous speed-limit approaches to obtain bounds for observables~\cite{Garcia-Pintos_PhysRevX2022h, Mohan_PhysRevACollPark2022w, Yadin_PhysRevLett2024s, Hamazaki_PRXQuantum2022r, Shrimali_PhysRevACollPark2024q, Hamazaki_PhysRevRes2024t}.

As an example, consider the task of maximizing the fidelity with respect to a given pure target state $\ket*| \psi_{\mathrm{t}} >$~\cite{Carlini_PhysRevLett2006r}.
In this case, the fidelity is given by 
\begin{equation}
    F(U) = \braket*< \psi_{\mathrm{t}} | U \rho^{\mathrm{i}} U^{\dagger} | \psi_{\mathrm{t}} >,
\end{equation}
which is the expectation value of the projection operator onto the target state.
This corresponds to choosing $A = \dyad{ \psi_{\mathrm{t}} }$ in Eq.~\eqref{eqM_GeneralObjectiveExp} without the second term, which is irrelevant to the optimization.
Now consider the two-level system under the full control of traceless Hermitian operators, where the control algebra is $\mathcal{V} = \mathfrak{su}(2)$ and the uncontrollable Hamiltonian is proportional to the identity, $H_{\mathrm{u}}(t) \propto I$.
The $\mathcal{V}$- and orthogonal components of $A$ and $\rho^{\mathrm{i}}$ are given by $A_{\mathcal{V}} = \dyad*{\psi_{\mathrm{t}}} - I/2$, $\rho_{\mathcal{V}}^{\mathrm{i}} = \rho^{\mathrm{i}} - I/2$, and $A_{\perp} = \rho^{\mathrm{i}}_{\perp} = I/2$.
Substituting these into the analytical solution~\eqref{eq_SolutionForSU2} for $\mathfrak{su}(2)$ control, we obtain the optimal fidelity $\mathcal{F}(T)$ achievable within time $T$ as
\begin{equation}
    \mathcal{F}(T) = -\frac{\sqrt{ 2 \tr[(\rho^{\mathrm{i}})^2] - 1 }}{2} \cos(2\omega T + \phi) + \frac{1}{2},
\end{equation}
where the second term $1/2$ originates from the contribution of the orthogonal components $A_{\perp}$ and $\rho^{\mathrm{i}}_{\perp}$.
Here, $\phi$ is the angle between $A_{\mathcal{V}}$ and $\rho_{\mathcal{V}}^{\mathrm{i}}$ and determines the minimum time required to maximize the fidelity via
\begin{equation}
    \mathcal{T}_{\ast} = \frac{\pi - \phi}{2 \omega} = \frac{1}{2 \omega} \arccos\ab( \frac{ 2 \expval*{ \rho^{\mathrm{i}} }{\psi_{\mathrm{t}}}  - 1 }{ \sqrt{ 2 \tr[(\rho^{\mathrm{i}})^2] - 1 } } ).
    \label{eqM_MinimalTimeForStatePreparation}
\end{equation}

If the initial state is also pure, $\rho^{\mathrm{i}} = \dyad*{ \psi_{\mathrm{i}} }$, the self-consistent equation admits an analytical solution even for general $D$-level systems under full control.
Substituting $A^{I}(T) = \ketbra*|\psi_{\mathrm{t}}><\psi_{\mathrm{t}}|$ and $\rho^{\mathrm{i}} = \ketbra*|\psi_{\mathrm{i}}><\psi_{\mathrm{i}}|$ into Eq.~\eqref{eqM_GeneralExp_SelfConsistentEq}, the optimal control operator $\mathsf{H}$ is shown to be of rank two:
\begin{equation}
    \label{eqM_SelfConsistentEqForPureStatePreparation}
    C \mathsf{H} = i \ab\Big( \ketbra*|\psi_{\mathrm{t}}><\psi_{\mathrm{f}}(\mathsf{H})| - \ketbra*|\psi_{\mathrm{f}}(\mathsf{H})><\psi_{\mathrm{t}}| ),
\end{equation}
where $\ket*|\psi_{\mathrm{f}}(\mathsf{H})> \coloneqq e^{-i\omega T \mathsf{H}} \ket*|\psi_{\mathrm{i}}>$ denotes the final state.
To identify the two-dimensional subspace on which $\mathsf{H}$ acts, we apply this equation to the vector $\ket*|\psi_{\mathrm{f}}(\mathsf{H})> - \ket*|\psi_{\mathrm{t}}> \braket*< \psi_{\mathrm{t}} |\psi_{\mathrm{f}}(\mathsf{H}) >$, obtaining
\begin{multline}
    \mathsf{H} \ab\Big( \ket*|\psi_{\mathrm{f}}(\mathsf{H})> - \ket*|\psi_{\mathrm{t}}> \braket*< \psi_{\mathrm{t}} | \psi_{\mathrm{f}}(\mathsf{H}) > ) 
    \\
    = \frac{i}{C} \ab\Big(1 - \abs*{ \braket*< \psi_{\mathrm{t}} | \psi_{\mathrm{f}}(\mathsf{H}) > }^2) \ket*| \psi_{\mathrm{t}} >.
\end{multline}
This shows that the target state $\ket*| \psi_{\mathrm{t}} >$ lies in the image of $\mathsf{H}$.
We then consider an equivalent form of the self-consistent equation~\eqref{eqM_GeneralExp_SelfConsistentEq}:
\begin{equation}
    C \mathsf{H} = i \comm*{ e^{ +i\omega T \mathsf{H} } A^{I}(T)  e^{ -i\omega T \mathsf{H} } }{ \rho^{\mathrm{i}} },\quad\!\! \Norm*{ \mathsf{H} } = 1,
\end{equation}
which follows from $\comm*{ \mathsf{H} }{ e^{-i\omega T \mathsf{H}} } = 0$.
This form allows us to apply the same reasoning as above with the roles of the target and initial states reversed: the initial state $\ket*|\psi_{\mathrm{i}}>$ also lies in the image of $\mathsf{H}$.
Since $\mathsf{H}$ is a rank-2 traceless Hermitian operator, its image is $\operatorname{span}\{ \ket*|\psi_{\mathrm{i}}>, \ket*|\psi_{\mathrm{t}}> \}$.
In particular, we can expand the final state $\ket*|\psi_{\mathrm{f}}(\mathsf{H}) >$ as
\begin{equation}
    \ket*|\psi_{\mathrm{f}}(\mathsf{H}) > = \alpha(\mathsf{H}) \ket*|\psi_{\mathrm{i}}> + \beta(\mathsf{H}) \ket*|\psi_{\mathrm{t}}>
\end{equation}
for some coefficients $\alpha(\mathsf{H}), \beta(\mathsf{H}) \in \mathbb{C}$.
Substituting this expansion into Eq.~\eqref{eqM_SelfConsistentEqForPureStatePreparation} and using $\Norm*{ \mathsf{H} } = 1$, the solution is obtained as
\begin{equation}
    \mathsf{H} = i \sqrt{ \frac{D}{2} } \frac{ e^{-i\theta} \ketbra*|\psi_{\mathrm{i}}><\psi_{\mathrm{t}}| - e^{i\theta} \ketbra*|\psi_{\mathrm{t}}><\psi_{\mathrm{i}}| }{\sqrt{1 - \abs*{ \braket*< \psi_{\mathrm{t}} | \psi_{\mathrm{i}} > }^2}},
\end{equation}
where the phase $\theta$ is determined to be $\theta = \arg\braket*< \psi_{\mathrm{t}} | \psi_{\mathrm{i}} >$ by taking the trace of Eq.~\eqref{eqM_GeneralExp_SelfConsistentEq}.
(It is arbitrary when $\braket*< \psi_{\mathrm{t}} | \psi_{\mathrm{i}} > = 0$.)

The resulting optimal fidelity is 
\begin{equation}\label{eq:optfid}
    \mathcal{F}(T) = \frac{1}{2} \cos\ab(\sqrt{2 D}\, \omega T - 2 \arccos\abs{ \braket*< \psi_{\mathrm{t}} | \psi_{\mathrm{i}} > }) + \frac{1}{2}.
\end{equation}
This recovers the well-known result for the minimum time to reach the target pure state from an initial pure state~\cite{Carlini_PhysRevLett2006r}, up to the difference in the normalization of the control Hamiltonian:
\begin{equation}
    \mathcal{T}_{\ast} = \frac{1}{\omega \sqrt{D/2}} \arccos\abs{ \braket*< \psi_{\mathrm{t}} | \psi_{\mathrm{i}} > }.
\end{equation}
However, we stress that Eq.~\eqref{eq:optfid} also provides the optimal fidelity for times shorter than $\mathcal{T}_{\ast}$.

Beyond the analytically solvable cases discussed above, the numerical method introduced in Sec.~\ref{SectionIV.D} enables us to obtain optimal solutions for general $D$-level systems with $D \geq 3$ and mixed initial states, where analytical solutions are generally unavailable. 
Therefore, the results in Secs.~\ref{SectionII}--\ref{SectionIV} extend to the optimization of general expectation values, under full control over all Hermitian or all traceless Hermitian operators as considered in this section.
In many-body settings, however, such full control is generally unrealistic, and whether our framework further extends to such many-body settings remains to be explored.

\section{Conclusion and Outlook}
\label{SectionVII}
In this work, we considered finite-time work extraction from closed quantum systems.
We first established the trade-off between power and extractable work under a general setup~(Proposition~\ref{thmM_TradeOff}), showing that they cannot be maximized simultaneously.
This demonstrates the advantage of our finite-time control setting in enhancing the power.
We then derived the optimal protocol for finite-time work extraction from closed quantum systems within the framework of Lie-algebraic control~(Theorem~\ref{thmM_MainResult}).
The optimal control Hamiltonian takes a remarkably simple form: 
it is time-independent in the interaction picture, and it is determined by the closed-form self-consistent equation~\eqref{eqM_SelfConsistentEq}.
We then presented an analytical solution in the simplest nontrivial case of $\mathfrak{su}(2)$ control and applied it to fully controlled two-level systems and the Heisenberg-type model under magnetic-field control with norm constraints.
We also demonstrated that the self-consistent equation remains numerically tractable for larger control algebras, such as $\mathfrak{su}(3)$, where analytical solutions are no longer available.
Thanks to the properties of Lie algebras, the numerical solution to the self-consistent equation~\eqref{eqM_SelfConsistentEq} remains computationally efficient even for quantum many-body systems.

Notably, examples of Lie-algebraic controls include systems with only local and few-body interactions, such as Heisenberg-type models with controlled magnetic fields and $\mathrm{SU}(n)$-Hubbard model with controllable fermion flavors.
In this sense, our theory identifies a physically meaningful class of settings in which a rigorous treatment is available for the optimal work extraction of quantum many-body systems within a finite time.

Note that our problem can equally be regarded as a cooling task, since extracting work from a closed system is equivalent to lowering the system’s energy with respect to the final Hamiltonian. 
From this perspective, our results provide quantitative upper bounds on the cooling capability (i.e., the controllable energy reduction) achievable within finite time for the many-body systems discussed above.
Moreover, by replacing $H^{\mathrm{f}}$ with $-H^{\mathrm{f}}$, our framework also applies to the problem of charging a quantum battery within a finite time, where the objective is to steer the system into a high-energy state with respect to a given Hamiltonian~\cite{Ferraro_PhysRevLett2018f, Mazzoncini_PhysRevACollPark2023g, Rodriguez_NewJPhys2024i, Campaioli_RevModPhys2024z}.

As we have discussed in Sec.~\ref{SectionVI}, our framework readily extends from work extraction to the optimization of general expectation values when the control algebra spans all Hermitian or all traceless Hermitian operators.
In addition to the analytically solvable examples discussed in Sec.~\ref{SectionVI}, the numerical method introduced in Sec.~\ref{SectionIV.D} enables the computation of optimal solutions for finite-time control in more complex systems, such as $D$-level systems with $D \geq 3$ and mixed initial states.

Another important open problem is, as mentioned before, whether the extended framework in Sec.~\ref{SectionVI} further generalizes to many-body settings, where full control over all Hermitian or traceless Hermitian operators is generally unrealistic.
If so, a further question is whether the resulting self-consistent equation can be solved efficiently, analogous to the method established for work extraction in Sec.~\ref{SectionV}.
It would also be worthwhile to explore generalizations of the current framework to situations where the assumptions of Lie-algebraic control are not fully satisfied, as in general many-body systems, and to determine when optimal solutions can still be obtained.

\section*{Acknowledgment}
We thank Masaya Nakagawa, Ken Funo, Yugo Takanashi, Marin Bukov, Nicolò Beato, and Hongzheng Zhao for valuable discussions and comments.
This work was supported by JST ERATO Grant Number JPMJER2302, Japan.
R.H. was supported by JSPS KAKENHI Grant No. JP24K16982.
T.S. is also supported by JST CREST Grant No. JPMJCR20C1 and by Institute of AI and Beyond of the University of Tokyo.


\section*{Data availability}
The data are not publicly available. 
The data are available from the authors upon reasonable request.

\section*{Code availability}
The numerical code used to generate the data plotted in Figs.~\ref{fig_NumericalResultForSU3} and \ref{fig_SU6_WorkScaling} is publicly available at Zenodo: \url{https://doi.org/10.5281/zenodo.17350563}.

\bibliography{references}


\clearpage
\appendix
\section{Proof of Proposition~\ref{thmM_TradeOff}}
\label{AppendixA}
We first consider the case with $\mathcal{T}_{\ast} = \infty$.
Since the work $\mathcal{W}(T)$ is bounded above by the ergotropy $\mathcal{W}_{\mathrm{erg}}$, the power at any time $T$ is bounded above as
\begin{equation}\label{PlecW/T}
    \mathcal{P}(T)= \frac{ \mathcal{W}(T) }{T}\leq \frac{ \mathcal{W}_{\mathrm{erg}} }{T}.
\end{equation}
We choose a reference time $T_{0}$ that satisfies $\mathcal{P}(T_{0}) > 0$ (i.e., nonzero power).
We also define another time $T_1$ through $T_{1} \coloneqq \mathcal{W}_{\mathrm{erg}} / \mathcal{P}(T_{0})$, where Eq.~\eqref{PlecW/T} indicates $T_0\leq T_1$.
Then, for any $T$ longer than $T_1$ (i.e., $T>T_1$), we obtain that the power cannot attain its maximum, since it is strictly smaller than the power at time $T_0$:
\begin{equation}
    \mathcal{P}(T) \leq \frac{ \mathcal{W}_{\mathrm{erg}} }{T} < \frac{\mathcal{W}_{\mathrm{erg}}}{T_1}=\mathcal{P}(T_{0}),\quad (T > T_{1}).
\end{equation}
Therefore, the maximum power is attained somewhere in the interval $[0, T_{1}]$:
\begin{equation}
    \sup_{T \geq 0} \mathcal{P}(T) = \sup_{T \in [0, T_{1}]} \mathcal{P}(T).
\end{equation}
Since the interval $[0, T_{1}]$ is closed, the supremum is attained somewhere in this region:
\begin{equation}
    0 \leq \operatorname*{argmax}_{T\geq 0} \mathcal{P}(T) \leq T_{1} < \mathcal{T}_{\ast} = \infty.
\end{equation}
This completes the proof for the case $\mathcal{T}_{\ast} = \infty$.

The case with $\mathcal{T}_{\ast} < \infty$ requires a separate argument to exclude the possibility that $\operatorname*{argmax}_{T\geq 0} \mathcal{P}(T) = \mathcal{T}_{\ast}$.
We first view the extractable work $W(U(T))$ as a function of the rescaled path $\varphi(\lambda) \coloneqq H_{\mathrm{c}}(\lambda T)$ $(\lambda \in [0,1])$ and $T$, and write
\begin{equation}
    f(\varphi, T) \coloneqq W(U(T)).
\end{equation}
Then, $f(\varphi, T)$ is bounded and continuous with respect to both arguments.

In Eq.~\eqref{optimize}, the control Hamiltonian $H_{\mathrm{c}} \colon [0,T] \to \mathcal{V}$ is optimized under the norm constraint $\Norm*{ H_{\mathrm{c}}(t) } \leq \omega$.
In terms of the rescaled path $\varphi \colon [0,1] \to \mathcal{V}$, we can rewrite Eq.~\eqref{optimize} as
\begin{equation}
    \mathcal{W}(T) = \max_{ \varphi \in \mathsf{P}\mathcal{V}_{\omega} } f(\varphi, T),
\end{equation}
where the set of admissible paths
\begin{equation}
    \mathsf{P}\mathcal{V}_{\omega} \coloneqq \{ \varphi \colon [0,1] \to \mathcal{V}_{\omega} \}
\end{equation}
with 
\begin{equation}
    \mathcal{V}_{\omega} \coloneqq \{ X \in \mathcal{V} \colon \Norm*{X} \leq \omega \},
\end{equation}
is now independent of the operational time $T$.

For each $T$, let $\varphi_T \in \mathsf{P}\mathcal{V}_{\omega}$ denote an optimal control path, i.e.,
\begin{equation}
    \varphi_T = \operatorname*{argmax}_{\varphi\in\mathsf{P}\mathcal{V}_{\omega}} f(\varphi,T).
\end{equation}
Such a maximizer exists because $\mathcal{V}_{\omega}$ is a closed ball of radius $\omega$ in the finite-dimensional space $\mathcal{V}$, which is compact; hence $\mathsf{P}\mathcal{V}_{\omega}$ is also compact.

%

At this stage, we can apply Theorem~1 of Ref.~\cite{Milgrom_Econometrica2002f}, which applies when the admissible set $\mathsf{P}\mathcal{V}_{\omega}$ is independent of $T$ (see below), and yields
\begin{equation}
    \partial_{-}\mathcal{W}(T) \leq \pdv{f}{T}(\varphi_{T}, T) \leq \partial_{+}\mathcal{W}(T).
    \label{eqM_OneSidedDerivatives}
\end{equation}
Here, $\partial_{-/+}$ denotes the left/right derivative.
At the maximum point $T = \mathcal{T}_{\ast}$ of $\mathcal{W}(T)$, we obtain $\mathcal{W}(\mathcal{T}_{\ast} + \delta) \leq \mathcal{W}(\mathcal{T}_{\ast})$ for any $\delta \in \mathbb{R}$ by definition.
This implies that 
\begin{align}
\partial_{+}\mathcal{W}(\mathcal{T}_{\ast}) \leq 0 \leq \partial_{-}\mathcal{W}(\mathcal{T}_{\ast}).
\end{align}
Combining this inequality with Eq.~\eqref{eqM_OneSidedDerivatives}, we conclude that $\mathcal{W}$ is differentiable at $T = \mathcal{T}_{\ast}$ with derivative 
\begin{align}
\dot{\mathcal{W}}(\mathcal{T}_{\ast}) = 0.
\end{align}
Then, the claim of Proposition~\ref{thmM_TradeOff} follows from 
\begin{align}
\dot{\mathcal{P}}(\mathcal{T}_{\ast}) = -\frac{\mathcal{W}(\mathcal{T}_{\ast})}{ \mathcal{T}_{\ast}^2} < 0,
\end{align}
implying that the power $\mathcal{P}$ attains its maximum strictly before $T = \mathcal{T}_{\ast}$.

Since the proof of Eq.~\eqref{eqM_OneSidedDerivatives} is elementary, we include it for completeness.
From the maximality of $\varphi_{T}$ for any given $T$ and the fact that $\mathsf{P}\mathcal{V}_{\omega}$ is independent of $T$, we have
\begin{equation}
    f(\varphi_{T}, T') - f(\varphi_{T}, T) \leq \mathcal{W}(T') - \mathcal{W}(T)
\end{equation}
for any $T'$.
Dividing both sides by $T' - T$ and taking the limit as $T' \to T \pm 0$ yields Eq.~\eqref{eqM_OneSidedDerivatives}.

\section{Decomposition of work into controllable and uncontrollable parts}
\label{AppendixB}
Given an arbitrary orthonormal basis $\{ \Lambda_{j} \}$ of $\mathcal{V}$, the orthogonal projection of an arbitrary operator $X$ onto $\mathcal{V}$ (with respect to the Hilbert--Schmidt inner product) is given by
\begin{equation}\label{rhocortho}
    X_{\mathcal{V}} = \sum_{j} \frac{ \tr( X \Lambda_{j} ) }{ \tr[(\Lambda_{j})^2] } \Lambda_{j}.
\end{equation}
The orthogonal complement $X_{\perp}$ is then given by $X_{\perp} \coloneqq X - X_{\mathcal{V}}$.

For the decomposition of the uncontrollable Hamiltonian, $H_{\mathrm{u}}(t) = H_{\mathrm{u},\mathcal{V}}(t) + H_{\mathrm{u},\perp}(t)$, we can show that
\begin{equation}
    \forall t \in [0,T],\quad H_{\mathrm{u},\mathcal{V}}(t) \in \mathcal{V},\quad \comm*{ H_{\mathrm{u},\perp}(t) }{ \mathcal{V} } \equiv 0,
    \label{eqA_ComponentsOfHu}
\end{equation}
where $\comm*{ H_{\mathrm{u},\perp}(t) }{ \mathcal{V} } \equiv 0$ means that $\comm*{ H_{\mathrm{u},\perp}(t) }{ X } \equiv 0$ for all $X \in \mathcal{V}$.
Indeed, assumption (ii) of Lie-algebraic control implies that
\begin{align}
    i \comm*{ H_{\mathrm{u},\perp}(t) }{ \mathcal{V} }
    = i \comm*{ H_{\mathrm{u}}(t) }{ \mathcal{V} } - i \comm*{ H_{\mathrm{u},\mathcal{V}}(t) }{ \mathcal{V} } \subseteq \mathcal{V}.
\end{align}
On the other hand, by assumption (i) of Lie-algebraic control and the cyclic property of the trace, we obtain
\begin{equation}
    \tr\ab( i\comm*{ H_{\mathrm{u},\perp}(t) }{ \mathcal{V} } \mathcal{V} ) = \tr\ab( H_{\mathrm{u},\perp}(t)\, i\comm*{ \mathcal{V} }{ \mathcal{V} } ) = 0.
\end{equation}
Therefore, $i \comm*{ H_{\mathrm{u},\perp}(t) }{ \mathcal{V} }$ lies in both $\mathcal{V}$ and its orthogonal complement, and thus vanishes.

As stated in Sec.~\ref{SectionIV.A}, assumption~(ii) of Lie-algebraic control implies that the time-evolution operator factorizes as $U = U_{\mathrm{u}} U_{\mathrm{c}}$, where $U_{\mathrm{u}}^{\dagger} \mathcal{V} U_{\mathrm{u}} = \mathcal{V}$ and $U_{\mathrm{c}} \in e^{i \mathcal{V}}.$
Applying Eq.~\eqref{eqA_ComponentsOfHu} to the final Hamiltonian $H^{\mathrm{f}} = H_{\mathrm{u}}(T)$, we obtain
\begin{align}
    U^{\dagger} H^{\mathrm{f}} U
    &= U_{\mathrm{c}}^{\dagger} (U_{\mathrm{u}}^{\dagger} H_{\mathcal{V}}^{\mathrm{f}} U_{\mathrm{u}}) U_{\mathrm{c}} + U_{\mathrm{c}}^{\dagger} (U_{\mathrm{u}}^{\dagger} H_{\perp}^{\mathrm{f}} U_{\mathrm{u}}) U_{\mathrm{c}}
    \nonumber \\
    &= U_{\mathrm{c}}^{\dagger} (U_{\mathrm{u}}^{\dagger} H_{\mathcal{V}}^{\mathrm{f}} U_{\mathrm{u}}) U_{\mathrm{c}} + U_{\mathrm{u}}^{\dagger} H_{\perp}^{\mathrm{f}} U_{\mathrm{u}},
    \nonumber \\
    &= \ab[ U_{\mathrm{c}}^{\dagger} (U_{\mathrm{u}}^{\dagger} H_{\mathcal{V}}^{\mathrm{f}} U_{\mathrm{u}}) U_{\mathrm{c}} - (U_{\mathrm{u}}^{\dagger} H_{\mathcal{V}}^{\mathrm{f}} U_{\mathrm{u}}) ] + U_{\mathrm{u}}^{\dagger} H^{\mathrm{f}} U_{\mathrm{u}},
\end{align}
where we use
\begin{equation}
    \comm*{ U_{\mathrm{u}}^{\dagger} H_{\perp}^{\mathrm{f}} U_{\mathrm{u}} }{ \mathcal{V} } = U_{\mathrm{u}}^{\dagger} \comm*{ H_{\perp}^{\mathrm{f}} }{ U_{\mathrm{u}} \mathcal{V} U_{\mathrm{u}}^{\dagger} } U_{\mathrm{u}} = U_{\mathrm{u}}^{\dagger} \comm*{ H_{\perp}^{\mathrm{f}} }{ \mathcal{V} } U_{\mathrm{u}} = 0
\end{equation}
from Eq.~\eqref{eqA_ComponentsOfHu}.
Substituting this decomposition into the definition of the work~\eqref{defM_Work}, we obtain
\begin{align}
    W(U) 
    &= -\tr[U_{\mathrm{c}}^{\dagger} (U_{\mathrm{u}}^{\dagger} H_{\mathcal{V}}^{\mathrm{f}} U_{\mathrm{u}}) U_{\mathrm{c}} \rho^{\mathrm{i}}] +\tr[U_{\mathrm{u}}^{\dagger} H_{\mathcal{V}}^{\mathrm{f}} U_{\mathrm{u}} \rho^{\mathrm{i}}]
    \nonumber \\
    &\qquad -\tr[U_{\mathrm{u}}^{\dagger} H^{\mathrm{f}} U_{\mathrm{u}} \rho^{\mathrm{i}}] +\tr[H^{\mathrm{i}} \rho^{\mathrm{i}}]
    \nonumber \\
    &\eqqcolon W_{\mathrm{c}}(U_{\mathrm{c}}, U_{\mathrm{u}}) + W(U_{\mathrm{u}}),
\end{align}
where
\begin{multline}
    W_{\mathrm{c}}(U_{\mathrm{c}}, U_{\mathrm{u}}) \\
    \coloneqq -\tr[U_{\mathrm{c}}^{\dagger} (U_{\mathrm{u}}^{\dagger} H_{\mathcal{V}}^{\mathrm{f}} U_{\mathrm{u}}) U_{\mathrm{c}} \rho^{\mathrm{i}}] +\tr[U_{\mathrm{u}}^{\dagger} H_{\mathcal{V}}^{\mathrm{f}} U_{\mathrm{u}} \rho^{\mathrm{i}}].
\end{multline}
Since $U_{\mathrm{c}}^{\dagger} (U_{\mathrm{u}}^{\dagger} H_{\mathcal{V}}^{\mathrm{f}} U_{\mathrm{u}}) U_{\mathrm{c}} \in \mathcal{V}$ and $\rho^{\mathrm{i}}_{\perp}$ is orthogonal to $\mathcal{V}$ by definition, we can equivalently write 
\begin{multline}
    W_{\mathrm{c}}(U_{\mathrm{c}}, U_{\mathrm{u}}) \\
    = -\tr[U_{\mathrm{c}}^{\dagger} (U_{\mathrm{u}}^{\dagger} H_{\mathcal{V}}^{\mathrm{f}} U_{\mathrm{u}}) U_{\mathrm{c}} \rho_{\mathcal{V}}^{\mathrm{i}}] +\tr[U_{\mathrm{u}}^{\dagger} H_{\mathcal{V}}^{\mathrm{f}} U_{\mathrm{u}} \rho_{\mathcal{V}}^{\mathrm{i}}].
\end{multline}

\section{Proof of Theorem~\ref{thmM_MainResult}}
\label{AppendixC}
\subsection{Optimal protocol}
We employ the Lagrange multiplier method with inequality constraints~\cite{Andrei_Other2022y} to maximize the extracted work under the norm constraint. 
The corresponding Lagrangian is given by
\begin{equation}
    \mathcal{L}[H_{\mathrm{c}}]
    \coloneqq W_{\mathrm{c}}(U_{\mathrm{c}}(T)) -\int_{0}^{T} \alpha(t) \ab( \Norm{ H_{\mathrm{c}}(t) }^2 - \omega^2 ) \d{t}.
\end{equation}
Here, $\alpha(t)$ is a Lagrange multiplier, which is required to satisfy $\alpha(t) \geq 0$ for the maximization problem~\cite{Andrei_Other2022y}.

To obtain the stationary condition $\fdif{\mathcal{L}} = 0$, we first consider the variation of the time-evolution operator with respect to the control Hamiltonian, which is given by
\begin{equation}
    \fdif{U_{\mathrm{c}}(T)} = -i U_{\mathrm{c}}(T) \int_{0}^{T} U_{\mathrm{c}}(t)^{\dagger} \fdif{H_{\mathrm{c}}^{I}\!(t)} U_{\mathrm{c}}(t) \d{t}. 
\end{equation}
Then, by using $\fdif{U_{\mathrm{c}}^{\dagger}} = - U_{\mathrm{c}}^{\dagger} \fdif{U_{\mathrm{c}}} U_{\mathrm{c}}^{\dagger}$ (obtained from the variation $\delta(U_{\mathrm{c}}^\dag U_{\mathrm{c}})=0$), the variation of the work is given by
\begin{align}
    \fdif{W_{\mathrm{c}}(U_{\mathrm{c}}(T))} 
    &= \tr\ab\Big( U_{\mathrm{c}}(T)^{\dagger} \fdif{U_{\mathrm{c}}(T)} \comm{ U_{\mathrm{c}}(T)^{\dagger} H_{\mathcal{V}}^{\mathrm{f},I}\!(T) U_{\mathrm{c}}(T) }{ \rho_{\mathcal{V}}^{\mathrm{i}} } )
    \nonumber \\
    &= \tr[i U_{\mathrm{c}}(T)^{\dagger} \fdif{U_{\mathrm{c}}(T)} M ]
    \nonumber \\
    &= \int_{0}^{T} \tr\ab[ \fdif{H_{\mathrm{c}}^{I}\!(t)} U_{\mathrm{c}}(t) M U_{\mathrm{c}}(t)^{\dagger} ] \d{t},
\end{align}
where we introduce a time-independent operator $M$ by
\begin{equation}
    M \coloneqq -i \comm*{ U_{\mathrm{c}}(T)^{\dagger} H_{\mathcal{V}}^{\mathrm{f},I}\!(T) U_{\mathrm{c}}(T) }{ \rho_{\mathcal{V}}^{\mathrm{i}} }.
    \label{eqC_defM}
\end{equation}
The variation of the Lagrangian is then given by
\begin{equation}
    \fdif{\mathcal{L}}
    = \int_{0}^{T} \tr\ab[ \fdif{H_{\mathrm{c}}^{I}\!(t)} \ab( U_{\mathrm{c}}(t) M U_{\mathrm{c}}(t)^{\dagger} -\frac{2\alpha(t)}{\tr I} H_{\mathrm{c}}^{I}(t) ) ] \d{t}.
    \label{eqM_VariationOfL}
\end{equation}

Here, assumption~(i) of Lie-algebraic control, that the controllable subspace $\mathcal{V}$ is closed under the commutator, implies that $U_{\mathrm{c}}(t) M U_{\mathrm{c}}(t)^{\dagger}$ belongs to $\mathcal{V}$.
Since the variation $\fdif{H_{\mathrm{c}}^{I}\!(t)}$ is taken over all $\mathcal{V}$ under the norm constraint due to assumptions~(i) and (ii), we may in particular choose 
\begin{equation*}
    \fdif{H_{\mathrm{c}}^{I}\!(t)} \propto \ab( U_{\mathrm{c}}(t) M U_{\mathrm{c}}(t)^{\dagger} -\frac{2\alpha(t)}{\tr I} H_{\mathrm{c}}^{I}(t) ),\  \forall t \in (0,T).
\end{equation*}
Substituting this choice into Eq.~\eqref{eqM_VariationOfL}, we obtain
\begin{equation*}
    \fdif{ \mathcal{L} } \propto  \int_{0}^{T} \Norm{ U_{\mathrm{c}}(t) M U_{\mathrm{c}}(t)^{\dagger} -\frac{2\alpha(t)}{\tr I} H_{\mathrm{c}}^{I}(t) }^{2} \d{t}.
\end{equation*}
Therefore, the stationary condition, $\fdif{\mathcal{L}} = 0$ for all $\fdif{H_{\mathrm{c}}^{I}(t)} \in \mathcal{V}$, is equivalent to the equation
\begin{equation}
    \frac{2\alpha(t)}{\tr I} H_{\mathrm{c}}^{I}(t) = U_{\mathrm{c}}(t) M U_{\mathrm{c}}(t)^{\dagger}.
\end{equation}

First, consider the case $M \neq 0$. 
Then, we have $\alpha(t) \neq 0$ for all $t \in (0,T)$.
According to the theory of Lagrange multiplier method with inequality constraints, this implies that the norm constraint must be saturated for all $t \in (0, T)$~\cite{Andrei_Other2022y}.
Therefore, we obtain
\begin{equation}
    H_{\mathrm{c}}^{I}(t) = \omega U_{\mathrm{c}}(t) \mathsf{H}  U_{\mathrm{c}}(t)^{\dagger},\quad
    \mathsf{H} \coloneqq \frac{ M }{ \Norm*{M} }.
\end{equation}
Solving this equation in conjunction with the Schrödinger equation, $i\dot{U}_{\mathrm{c}}(t) = H_{\mathrm{c}}^{I}(t) U_{\mathrm{c}}(t)$, we obtain
\begin{equation}
    H_{\mathrm{c}}^{I}(t) = \omega \mathsf{H},\quad 
    U_{\mathrm{c}}(T) = e^{ - i \omega T \mathsf{H} }.
    \label{eqM_OptimalSolutionNonzeroC}
\end{equation}
Substituting this solution back into the definition of the operator $M$ in Eq.~\eqref{eqC_defM}, we obtain the self-consistent equation that determines the optimal control direction as
\begin{equation}
    C \mathsf{H} = -i \comm*{ H_{\mathcal{V}}^{\mathrm{f},I}\!(T) }{ e^{ -i \omega T \mathsf{H} } \rho_{\mathcal{V}}^{\mathrm{i}} e^{ i \omega T \mathsf{H} } },\quad
    \Norm*{\mathsf{H}} = 1,
    \label{eqM_SelfConsistentEqNonzeroC}
\end{equation}
where $C \coloneqq \Norm*{M} > 0$, and we use $\comm*{ \mathsf{H} }{ e^{ - i \omega T \mathsf{H} } } = 0$.


Next, consider $M = 0$, which corresponds to $\alpha(t) \equiv 0$. 
Then, the optimal time-evolution operator satisfies $\comm*{ H_{\mathcal{V}}^{\mathrm{f},I}\!(T) }{ U_{\mathrm{c}}(T) \rho_{\mathcal{V}}^{\mathrm{i}} U_{\mathrm{c}}(T)^{\dagger} } = 0$ from Eq.~\eqref{eqC_defM}.
Now, the minimum time required to implement $U_{\mathrm{c}}(T)$ is given by $T_{U} = \omega^{-1} \Norm*{ \Log U_{\mathrm{c}}(T) }$, where $\Log$ is the inverse of the exponential map $X\mapsto e^{X}$ in the vicinity of the identity.
Indeed, from the norm constraint and the Schrödinger equation, we obtain
\begin{equation}
    \omega T \geq \int_{0}^{T} \Norm*{ H_{\mathrm{c}}^{I}(t) } \d{t} 
    = \int_{0}^{T} \Norm{ \odv{U_{\mathrm{c}}}{t} U_{\mathrm{c}}^{\dagger} } \d{t}. 
    \label{eq_IntegratedNormConstraint}
\end{equation}
The quantity on the right-hand side represents the length of the path $U_{\mathrm{c}}(t)\ (t \in [0,T])$ on $e^{i \mathcal{V}}$.
Then, the standard characterization of geodesics on compact Lie groups~\cite{Lee_Other2019c,Lee_Other1997d} shows that the minimum length is given by $\Norm*{ \Log U_{\mathrm{c}}(T) }$, which provides the lower bound for the right-hand side of \eqref{eq_IntegratedNormConstraint}.

Therefore, we have $T_U=\omega^{-1} \Norm*{ \Log U_{\mathrm{c}}(T) } \leq T$ by assumption that $U_{\mathrm{c}}(T)$ can be implemented within time $T$. 
When $T \leq \ell_{T}/\omega$, this inequality must in fact be saturated: $T_U=\omega^{-1} \Norm*{ \Log U_{\mathrm{c}}(T) } = T$~(see \Appendix{G} for the proof).
Now, if we take the control Hamiltonian as $H_{\mathrm{c}}^{I}(t) = \omega \mathsf{H}$ with $\mathsf{H} \coloneqq i \Log U_{\mathrm{c}}(T) / \omega T$,
it indeed implements the optimal unitary $U_{\mathrm{c}}(T) = e^{-i \omega T \mathsf{H}}$ for time $T$.
Moreover, it satisfies our norm constraint $\Norm*{\mathsf{H}} = 1$.
Finally, we can see that $\mathsf{H}$ satisfies the self-consistent equation~\eqref{eqM_SelfConsistentEq} with $C = 0$ due to the assumption $M = 0$.

\subsection{Relation between optimal work and $C$}
For the second part [Eq.~\eqref{workEq}], we write the optimal work as 
\begin{align}
    \mathcal{W}_{\mathrm{c}}(T;\omega) = \max_{X \in \mathcal{V},\, \Norm*{X} \leq 1} f_{T}(X,\omega),
\end{align}
where we define $f_{T}(X,\omega) \coloneqq W_{\mathrm{c}}(e^{ -i \omega T X})$.
Here, we explicitly write the $\omega$-dependence of the optimal work since we here vary $\omega$ with fixed $T$.
Let $\mathsf{H}_{\omega}$ be the maximizer for the parameter $\omega$, i.e., $f_{T}(\mathsf{H}_{\omega}, \omega) = \mathcal{W}_{\mathrm{c}}(T; \omega)$.
Note that $\mathsf{H}_{\omega}$ was simply denoted as $\mathsf{H}$ in the main text and the former part of the proof of Theorem~\ref{thmM_MainResult} for brevity.

The partial derivative of $f_{T}$ with respect to $\omega$ is given by
\begin{align}
    \pdv{f_{T}}{\omega}(X,\omega)
    &= -\tr\ab[ H_{\mathcal{V}}^{\mathrm{f},I}\!(T) \pdv*{ \ab( e^{-i\omega T X} \rho_{\mathcal{V}}^{\mathrm{i}} e^{+i\omega T X} ) }{\omega}  ]
    \nonumber \\
    &= i T \tr\ab( H_{\mathcal{V}}^{\mathrm{f},I}\!(T) \comm{ X }{ e^{-i\omega T X} \rho_{\mathcal{V}}^{\mathrm{i}} e^{+i\omega T X} } )
    \nonumber \\
    &= -i T \tr\ab( X \comm{ H_{\mathcal{V}}^{\mathrm{f},I}\!(T) }{ e^{-i\omega T X} \rho_{\mathcal{V}}^{\mathrm{i}} e^{+i\omega T X} } ).
    \label{eqM_DerivativeOfF}
\end{align}
Then, the Cauchy--Schwarz inequality yields
\begin{align}
    \abs{ \pdv{f_{T}}{\omega}(X,\omega) }
    &= T \abs{ \tr\ab\Big( \comm{ H_{\mathcal{V}}^{\mathrm{f},I}\!(T) }{ X } e^{-i\omega T X} \rho_{\mathcal{V}}^{\mathrm{i}} e^{+i\omega T X} ) }
    \nonumber \\
    &\leq T D \Norm{ \comm{ H_{\mathcal{V}}^{\mathrm{f},I}\!(T) }{ X } } \Norm*{ \rho_{\mathcal{V}}^{\mathrm{i}} }
    \nonumber \\
    &\leq 2 T D \Norm*{ H_{\mathcal{V}}^{\mathrm{f}} },
    \label{eqM_DerivativeOfFBound}
\end{align}
where we use $\Norm*{X} \leq 1$ and $\Norm*{ \rho_{\mathcal{V}}^{\mathrm{i}} } \leq \Norm*{ \rho^{\mathrm{i}} } \leq 1/\sqrt{D}$ in deriving the last inequality.
This shows that $f_{T}(X,\omega)$ is Lipschitz continuous with respect to $\omega$ for any $X$.
Therefore, we can apply Theorem~2 in Ref.~\cite{Milgrom_Econometrica2002f} to obtain (see below)
\begin{equation}
    \mathcal{W}_{\mathrm{c}}(T;\omega) = T \int_{0}^{\omega} \pdv{f_{T}}{\omega}(\mathsf{H}_{\omega'}, \omega') \d{\omega'}.
    \label{eqM_IntegralFormulaOfW}
\end{equation}

Setting $X = \mathsf{H}_{\omega}$ in Eq.~\eqref{eqM_DerivativeOfF}, substituting the self-consistent equation~\eqref{eqM_SelfConsistentEq}, and using $\Norm*{\mathsf{H}_{\omega}} = 1$, we obtain
\begin{equation}
    \pdv{f_{T}}{\omega}(\mathsf{H}_{\omega}, \omega) = T D C(\omega),\quad D \coloneqq \dim \mathcal{H}.
\end{equation}
Substituting this result into Eq.~\eqref{eqM_IntegralFormulaOfW} completes the proof of the second part of the theorem.

For completeness, we include the proof of Eq.~\eqref{eqM_IntegralFormulaOfW} in our setting.
From the bound~\eqref{eqM_DerivativeOfFBound}, we obtain
\begin{align}
    \abs{ \mathcal{W}_{\mathrm{c}}(T;\omega') - \mathcal{W}_{\mathrm{c}}(T;\omega) }
    &\leq \max_{\substack{X \in \mathcal{V} \\ \Norm{X} \leq 1}} \abs*{ f_{T}(X,\omega') - f_{T}(X,\omega) }
    \nonumber \\
    &\leq \max_{\substack{X \in \mathcal{V} \\ \Norm{X} \leq 1}} \abs{ \int_{\omega}^{\omega'} \pdv{f_{T}}{\omega}(X,\lambda) \d{\lambda} }
    \nonumber \\
    &\leq 2 T D \Norm*{ H_{\mathcal{V}}^{\mathrm{f}} }\ \abs{ \omega' - \omega },
\end{align}
where the first inequality follows from the bound $\abs*{ \max_{x \in X} g(x) - \max_{x \in X} h(x) } \leq \max_{x \in X} \abs*{ g(x) - h(x) }$ for any functions $g$ and $h$ on a common domain $X$.
This implies that $\mathcal{W}_{\mathrm{c}}$ is Lipschitz continuous with respect to $\omega$ and hence differentiable with respect to $\omega$ almost everywhere by Rademacher's theorem~\cite{Villani2009-hr}.
Then, the same argument leading to the inequality~\eqref{eqM_OneSidedDerivatives} implies
\begin{equation}
    \pdv{ \mathcal{W}_{\mathrm{c}}(T;\omega) }{\omega} = \pdv{f_{T}}{\omega}(\mathsf{H}_{\omega}, \omega)\ \text{almost everywhere.}
\end{equation}
Integrating this equation with $\mathcal{W}_\mathrm{c}(T;0) = 0$ yields Eq.~\eqref{eqM_IntegralFormulaOfW}.

\section{Optimal control operator in $\mathfrak{su}(2)$ control}
\label{AppendixD}
Here, we derive the optimal control operator 
\begin{align}
    \mathsf{H} = -\frac{i \comm*{ H_{\mathcal{V}}^{\mathrm{f},I}\!(T) }{ \rho_{\mathcal{V}}^{\mathrm{i}} } }{ \Norm*{ \comm*{ H_{\mathcal{V}}^{\mathrm{f},I}\!(T) }{ \rho_{\mathcal{V}}^{\mathrm{i}} }} }
    \label{eqA_optimalSU2}
\end{align}
in $\mathfrak{su}(2)$ control.
To this end, we first choose, without loss of generality, a standard basis $\{S_{1}, S_{2}, S_{3}\}$ of $\mathcal{V} \cong \mathfrak{su}(2)$ with commutation relations $\comm*{S_{\alpha}}{S_{\beta}} = i \sum_{\gamma} \epsilon_{\alpha\beta\gamma} S_{\gamma}$ and
\begin{equation}
    H_{\mathcal{V}}^{\mathrm{f},I}\!(T) = C_{H} S_{1},\quad
    \rho_{\mathcal{V}}^{\mathrm{i}} = C_{\rho} \ab( S_{1} \cos\phi_{T} + S_{2} \sin\phi_{T} ),
    \label{eqM_BasisForSU2}
\end{equation}
where $\phi_{T} \in [0,\pi]$ denotes the angle between $H_{\mathcal{V}}^{\mathrm{f},I}\!(T)$ and $\rho_{\mathcal{V}}^{\mathrm{i}}$ in the Hilbert--Schmidt inner product.
The coefficients are given by $C_{H} = \Norm*{ H_{\mathcal{V}}^{\mathrm{f}} } / c$ and $C_{\rho} = \Norm*{ \rho_{\mathcal{V}}^{\mathrm{i}} } / c$ with $c \coloneqq \Norm*{ S_{\alpha} }$.

As discussed in Sec.~\ref{SectionIV.C}, the self-consistent equation~\eqref{eqM_SelfConsistentEq} implies that the optimal $\mathsf{H}$ must be orthogonal to both $H_{\mathcal{V}}^{\mathrm{f},I}\!(T)$ and $\rho_{\mathcal{V}}^{\mathrm{i}}$ with respect to the Hilbert--Schmidt inner product.
Thus, when $H_{\mathcal{V}}^{\mathrm{f},I}\!(T)$ and $\rho_{\mathcal{V}}^{\mathrm{i}}$ do not commute, the optimal direction satisfies 
\begin{align}
    \mathsf{H} = \pm S_{3} / c,
    \label{eqM_su2OptimalControlOp}
\end{align}
which follows from Eq.~\eqref{eqM_BasisForSU2}.
Equivalently, $\mathsf{H}$ can be written as $\mathsf{H} = \mp i \comm*{ H_{\mathcal{V}}^{\mathrm{f},I}\!(T) }{ \rho_{\mathcal{V}}^{\mathrm{i}} } / \Norm*{ \comm*{ H_{\mathcal{V}}^{\mathrm{f},I}\!(T) }{ \rho_{\mathcal{V}}^{\mathrm{i}} } }$.
When $H_{\mathcal{V}}^{\mathrm{f},I}\!(T)$ and $\rho_{\mathcal{V}}^{\mathrm{i}}$ commute with each other, the optimal direction $\mathsf{H}$ lies in the plane perpendicular to $S_1$, and there remains a degree of freedom in choosing a basis in that plane.
We may therefore again choose $\mathsf{H}$ as in Eq.~\eqref{eqM_su2OptimalControlOp} without loss of generality.

With Eq.~\eqref{eqM_su2OptimalControlOp}, the time-evolution operator $e^{-i\omega T \mathsf{H}}$ generates the rotation around the $S_{3}$-axis by an angle $\pm \omega T/c$, and the final state becomes
\begin{multline*}
    e^{ -i\omega T \mathsf{H} } \rho_{\mathcal{V}}^{\mathrm{i}} e^{ i\omega T \mathsf{H} }
    \\
    = C_{\rho} \ab[ S_{1} \cos\ab(\frac{\pm\omega T}{c} + \phi_{T}) + S_{2} \sin\ab(\frac{\pm\omega T}{c} + \phi_{T}) ].
\end{multline*}
Therefore, the right-hand side of the self-consistent equation~\eqref{eqM_SelfConsistentEq} evaluates to
\begin{multline}
    -i \comm*{ H_{\mathcal{V}}^{\mathrm{f},I}\!(T) }{ e^{ -i\omega T \mathsf{H} } \rho_{\mathcal{V}}^{\mathrm{i}} e^{ i\omega T \mathsf{H} } } \\
    = \pm c\, C_{H} C_{\rho} \sin\ab(\pm \frac{\omega T}{c} + \phi_{T}) \mathsf{H}.
\end{multline}
Since the scalar $C$ on the left-hand side of Eq.~\eqref{eqM_SelfConsistentEq} must be nonnegative, the sign in Eq.~\eqref{eqM_su2OptimalControlOp} is chosen so that $\pm \sin\ab(\pm \omega T/c + \phi_{T}) \geq 0$.
For that purpose, below we summarize the sign of $\pm \sin\ab(\pm x + \phi_{T})$ as a function of $x$:
\begin{align*}
    &\begin{array}{|c||W{c}{2.5em}|W{c}{2.5em}|W{c}{2.5em}|} \hline
         \multicolumn{1}{|c}{x} & \multicolumn{3}{l|}{\hspace{-0.6em}0\hspace{2.5em}\phi_{T}\hspace{1.1em}\pi-\phi_{T}} \\ \hline
         +\sin(+x+\phi_{T}) & + & + & - \\
         -\sin(-x+\phi_{T}) & - & + & - \\ \hline
    \end{array}
    \quad \phi_{T} \in [0,\pi/2], \\
    &\begin{array}{|c||W{c}{2.5em}|W{c}{2.5em}|W{c}{2.5em}|} \hline
         \multicolumn{1}{|c}{x} & \multicolumn{3}{l|}{\hspace{-0.6em}0\hspace{1.7em}\pi-\phi_{T}\hspace{1.0em}\phi_{T}}  \\ \hline
         +\sin(+x+\phi_{T}) & + & - & - \\
         -\sin(-x+\phi_{T}) & - & - & + \\ \hline
    \end{array}
    \quad \phi_{T} \in (\pi/2,\pi].
\end{align*}
From this result, we should choose the $+$ sign in Eq.~\eqref{eqM_su2OptimalControlOp} when $0 \leq \omega T/c \leq \min\{\phi_{T}, \pi-\phi_{T}\}$.
When $\phi_{T} < \omega T/c < \pi - \phi_{T}$, both signs satisfy the self-consistent equation.
However, a direct evaluation of the extractable work shows that the $+$ sign always yields a larger value.
Therefore, the $+$ sign should be chosen when $0 \leq \omega T/c \leq \pi - \phi_{T}$. 

The optimal control operator is thus given by
\begin{equation}
    \mathsf{H} = \frac{S_3}{\Norm*{S_{\alpha}}},
\end{equation}
which coincides with Eq.~\eqref{eqA_optimalSU2} when $\comm*{ H_{\mathcal{V}}^{\mathrm{f},I}\!(T) }{ \rho_{\mathcal{V}}^{\mathrm{i}} } \neq 0$.
Then, corresponding extractable work is given by
\begin{multline}
    \label{extsu2}
    \mathcal{W}_{\mathrm{c}}(T) = D\, \Norm*{ H_{\mathcal{V}}^{\mathrm{f}} } \Norm*{ \rho_{\mathcal{V}}^{\mathrm{i}} } \\ 
    \times\ab[ \cos\phi_{T} -  \cos\ab(\frac{\omega T}{\Norm*{S_{\alpha}}} + \phi_{T}) ]
\end{multline}
when $0 \leq \omega T/c \leq \pi - \phi_{T}$.

Finally, we show that the complementary case $\omega T/c > \pi - \phi_{T}$ corresponds to the trivial case where an optimal control Hamiltonian does not saturate the norm constraint.
We first note that the extractable work is bounded from above as
\begin{equation}
    \label{eqM_SU2_CauchySchwartz}
    W_{\mathrm{c}}(U_{\mathrm{c}}) \leq D \Norm*{ H_{\mathcal{V}}^{\mathrm{f}} } \Norm*{ \rho_{\mathcal{V}}^{\mathrm{i}} } +\tr[H_{\mathcal{V}}^{\mathrm{f},I}(T) \rho_{\mathcal{V}}^{\mathrm{i}}],
\end{equation}
which is a consequence of the Cauchy--Schwarz inequality.
This upper bound is attained by the following protocol:
\begin{equation}
    H_{\mathrm{c}}^{I}(t) = \frac{c (\pi - \phi_{T} )}{T} \frac{ S_{3} }{ \Norm*{S_{3}} }\quad \text{with} \quad \Norm*{ H_{\mathrm{c}}^{I}(t) } < \omega,
\end{equation}
which realizes an optimal control unitary $\tilde{U}_{\mathrm{c}} = e^{ -i (\pi - \phi_{T}) S_{3} }$ that saturates the upper bound~\eqref{eqM_SU2_CauchySchwartz}.
This implies that $\ell_{T} \leq c (\pi - \phi_{T})$.
Therefore, the case $\omega T/c > \pi - \phi_{T}$ corresponds to the trivial case $\omega T > \ell_{T}$ where the maximum work extraction is possible by Eq.~\eqref{eqM_MaximumWorkExtractionProtocol}.
(Although it is not necessary for the proof, we remark that the argument here combined with Eq.~\eqref{extsu2} implies that $\ell_{T} = c (\pi - \phi_{T})$.)

\section{Numerical solution to the self-consistent equation}
\label{AppendixE}
As mentioned in Sec.~\ref{SectionIV.D}, when $\dim\mathcal{V} > 3$, an analytical solution to the self-consistent equation is generally infeasible.
Nevertheless, we can solve the equation numerically using a gradient-based method, with the gradient of the cost function in Eq.~\eqref{eq_CostFunction} obtained analytically as given in Eq.~\eqref{eqM_Gradient}.
Before presenting the result, we recall that $F(X) \coloneqq -i \comm*{ H_{\mathcal{V}}^{\mathrm{f},I}\!(T) }{ e^{ -i\omega T X } \rho_{\mathcal{V}}^{\mathrm{i}} e^{ i\omega T X } }$ is the right-hand side of the self-consistent equation~\eqref{eqM_SelfConsistentEq}.
\begin{proposition}[Gradient of the cost function]
    The gradient of the cost function $g(X)$ in Eq.~\eqref{eq_CostFunction} is given by
    \begin{multline}
        \nabla g(X) 
        = 2 \Bigg\{ \mathcal{K}_{X}\circ\mathcal{J}_{X}\ab(F(X) - \frac{ \tr[X F(X)]^{+} }{D} X) \\
        - \frac{ \tr[X F(X)]^{+} }{D} F(X) \Bigg\},
        \label{eqM_Gradient}
    \end{multline}
    where the linear maps $\mathcal{K}_{X}$ and $\mathcal{J}_{X}$ are defined by
    \begin{align*}
        \mathcal{K}_{X} \colon Z
        &\mapsto \omega \int_{0}^{T} e^{ i\omega t X } Z e^{ -i\omega t X } \d{t},\quad 
        \\
        \mathcal{J}_{X} \colon Z
        &\mapsto \comm{ e^{ -i\omega T X } \rho_{\mathcal{V}}^{\mathrm{i}} e^{ i\omega T X } }{ \comm*{ H_{\mathcal{V}}^{\mathrm{f},I}\!(T) }{ Z } }.
    \end{align*}
\end{proposition}
\begin{proof}
    The proof proceeds by direct calculation.
    We begin by recalling that the gradient $\nabla g(X)$ of the cost function is defined through the first variation:
    \begin{equation}
        \odif{g(X)} = \frac{1}{D} \tr[ \d{X}\, \nabla g(X) ]
    \end{equation}
    for any variation $\d{X} \in \mathcal{V}$.
    Using the explicit form of $g(X)$ given in Eq.~\eqref{eq_CostFunction}, its differential can be written as
    \begin{multline}
        \label{eqM_DerivativeOfG}
        \odif{g(X)} = \frac{2}{D} \tr\ab[ \ab( F(X) - \frac{ \tr[X F(X)]^{+} }{D} X ) \odif{F(X)} ] \\
        -\frac{2}{D} \tr\ab( \d{X}\, \frac{ \tr[X F(X)]^{+} }{D} F(X) ).
    \end{multline}
    Therefore, the task is to calculate $\tr[A \odif{F(X)}]$.

    For that purpose, we use the following identity:
    \begin{multline}
        \odif{ \ab\Big( e^{-i\omega T X} \rho_{\mathcal{V}}^{\mathrm{i}} e^{i\omega T X} ) } \\
        = -i \comm{ \mathcal{K}_{X}^{\dagger}(\d{X}) }{ e^{-i\omega T X} \rho_{\mathcal{V}}^{\mathrm{i}} e^{i\omega T X} },
    \end{multline}
    where $\mathcal{K}_{X}^{\dagger}$ denotes the adjoint of $\mathcal{K}_{X}$ defined via $\tr[A\, \mathcal{K}_{X}(B)] = \tr[\mathcal{K}_{X}^{\dagger}(A) B]$.
    By employing this equation, we calculate
    \begin{align}
        &\tr[A \odif{F(X)}] 
        \nonumber \\
        &\qquad= -i \tr\ab\{ A \comm{ H_{\mathcal{V}}^{\mathrm{f},I}(T) }{ \odif{\ab( e^{-i\omega T X} \rho_{\mathcal{V}}^{\mathrm{i}} e^{+i\omega T X} )} } \}
        \nonumber \\
        &\qquad= i \tr\ab\{ \odif{\ab( e^{-i\omega T X} \rho_{\mathcal{V}}^{\mathrm{i}} e^{+i\omega T X} )} \comm*{ H_{\mathcal{V}}^{\mathrm{f},I}(T) }{ A } \}
        \nonumber \\
        &\qquad= \tr\ab\{ \mathcal{K}_{X}^{\dagger}(\d{X}) \comm{ { e^{-i\omega T X} \rho_{\mathcal{V}}^{\mathrm{i}} e^{i\omega T X} } }{ \comm*{ H_{\mathcal{V}}^{\mathrm{f},I}(T) }{ A } } \}
        \nonumber \\
        &\qquad= \tr\ab\big[ \d{X}\; \mathcal{K}_{X}\circ\mathcal{J}_{X}(A) ].
    \end{align}
    Substituting this result into Eq.~\eqref{eqM_DerivativeOfG} completes the proof of Eq.~\eqref{eqM_Gradient}.
\end{proof}

The linear maps $\mathcal{K}_{X}$ and $\mathcal{J}_{X}$ in the proposition can be evaluated using the spectral decomposition of $X$.  
Let $X = \sum_{j} x_{j} \Pi_{j}$ be the spectral decomposition of $X$, where $\Pi_{j}$ denotes the projection operator onto the eigenspace with an eigenvalue $x_{j}$.  
Then, the map $\mathcal{K}_{X}$ is given by
\begin{equation}
    \mathcal{K}_{X}(Z) = \omega T \sum_{j} \Pi_{j} Z \Pi_{j} + \sum_{j\neq k} \tfrac{ e^{ -i \omega T (x_{j} - x_{k}) } - 1 }{ -i (x_{j} - x_{k}) } \Pi_{j} Z \Pi_{k}.
\end{equation}
Similarly, the final state under the evolution generated by $X$ reads 
\begin{equation}
    e^{-i\omega T X} \rho_{\mathcal{V}}^{\mathrm{i}} e^{i\omega T X} 
    = \sum_{jk} e^{ -i\omega T (x_{j} - x_{k}) } \Pi_{j} \rho_{\mathcal{V}}^{\mathrm{i}} \Pi_{k},
\end{equation}
which enables the evaluation of both $\mathcal{J}_{X}$ and $F(X)$.
Since the above procedure involves diagonalizing the matrix $X$, the computational cost for evaluating the gradient scales as $\order{d^3}$ when $X$ is a $d \times d$ matrix.
As we have seen in Sec.~\ref{SectionV}, we have $d=D$ when solving with the original representation, and $d=n$ when utilizing the standard representation.

Being nonlinear, the self-consistent equation~\eqref{eqM_SelfConsistentEq} may admit multiple solutions.
In fact, as the operational time $T$ increases, we find several critical times at which the number of solutions increases. 
Since the solutions correspond to the stationary points of the Lagrangian $\mathcal{L}$, this phenomenon is regarded as a transition in the control landscape, which attracts recent attention~\cite{Ldq7O_topological_beato2024, h1RY4_towards_beato2024}.
In this case, the choice of the initial guess is crucial to obtain a solution that maximizes work extraction within time~$T$.
To determine the best initial guess, let us begin by considering the small-$T$ regime.

For small $T$, the self-consistent equation~\eqref{eqM_SelfConsistentEq} can be expanded as
\begin{multline}
    C \mathsf{H} = 
    -i \comm*{ H_{\mathcal{V}}^{\mathrm{f}} }{ \rho_{\mathcal{V}}^{\mathrm{i}} } + T \comm{ H_{\mathcal{V}}^{\mathrm{f}} }{ \comm*{ \rho_{\mathcal{V}}^{\mathrm{i}} }{ \omega \mathsf{H} +H_{\mathrm{u}}(0) } } + \order{T^2}.
\end{multline}
Accordingly, when $\comm*{ H_{\mathcal{V}}^{\mathrm{f}} }{ \rho_{\mathcal{V}}^{\mathrm{i}} } \neq 0$, the solution in the small-$T$ regime is obtained as
\begin{equation}
    C \mathsf{H} \simeq -i \comm*{ H_{\mathcal{V}}^{\mathrm{f}} }{ \rho_{\mathcal{V}}^{\mathrm{i}} }.
\end{equation}
On the other hand, when $\comm*{ H_{\mathcal{V}}^{\mathrm{f}} }{ \rho_{\mathcal{V}}^{\mathrm{i}} } = 0$, 
any solution of the linear equation
\begin{equation}
    \ab(\operatorname{ad}(H_{\mathcal{V}}^{\mathrm{f}}) \circ\operatorname{ad}(\rho_{\mathcal{V}}^{\mathrm{i}}) - \mu)(\mathsf{H})
    = - \frac{1}{\omega}\comm{ H_{\mathcal{V}}^{\mathrm{f}} }{ \comm*{\rho_{\mathcal{V}}^{\mathrm{i}}}{H_{\mathrm{u}}(0)} }
    \label{eqA_LinearEq}
\end{equation}
satisfies the self-consistent equation up to first order in $T$ with $C = \mu \omega T$.
Here, we introduce the linear map $\operatorname{ad}(H_{\mathcal{V}}^{\mathrm{f}}) \circ\operatorname{ad}(\rho_{\mathcal{V}}^{\mathrm{i}}) \colon Z \mapsto \comm{ H_{\mathcal{V}}^{\mathrm{f}} }{ \comm*{ \rho_{\mathcal{V}}^{\mathrm{i}} }{ Z } }$.
Therefore, one should select the one that achieves the optimal work extraction.

In general, solutions of Eq.~\eqref{eqA_LinearEq} may be found by working with the simultaneous eigenbasis of $H_{\mathcal{V}}^{\mathrm{f}}$ and $\rho_{\mathcal{V}}^{\mathrm{i}}$ as follows.
(Recall that we are considering the case where they commute.)
Let $\{ \ket*|j> \}_{j}$ be a simultaneous eigenbasis of $H_{\mathcal{V}}^{\mathrm{f}}$ and $\rho_{\mathcal{V}}^{\mathrm{i}}$ satisfying $H_{\mathcal{V}}^{\mathrm{f}} \ket*|j> = E_{j}$ and $\rho_{\mathcal{V}}^{\mathrm{i}} \ket*|j> = q_{j} \ket*|j>$.
Then, we have
\begin{equation}
    \operatorname{ad}(H_{\mathcal{V}}^{\mathrm{f}}) \circ\operatorname{ad}\ab(\rho_{\mathcal{V}}^{\mathrm{i}})( \ketbra*|j><k| ) = \lambda_{jk} \ketbra*|j><k|
    \label{eqA_EigenValueEq}
\end{equation}
with $\lambda_{jk} \coloneqq (E_j - E_k)(q_j - q_k)$.
If the right-hand side of Eq.~\eqref{eqA_LinearEq} is nonzero, we can solve it for $\mathsf{H}$ by expanding both $\mathsf{H}$ and $H_{\mathrm{u}}(0)$ in the operator basis $\{ \ketbra*|j><k| \}_{jk}$, obtaining
\begin{align}
    \mathsf{H} = -\frac{1}{\omega} \sum_{jk} \frac{ \lambda_{jk} }{ \lambda_{jk} - \mu } \ket*|j> \braket<j|H_{\mathrm{u}}(0)|k> \bra*<k|.
\end{align}
Here, the parameter $\mu$ is determined from the condition $\Norm*{ \mathsf{H} } = 1$, which results in the following equation for the parameter $\mu$:
\begin{equation}
    \sum_{jk} \frac{ \lambda_{jk}^2 }{ (\lambda_{jk} - \mu)^2 } \ab|{ \braket<j|H_{\mathrm{u}}(0)|k> }|^2 = D \omega^2.
\end{equation}
This equation has at most $D^2$ solutions and can be solved numerically by, e.g., bisection search or Newton's method.

If the right-hand side of Eq.~\eqref{eqA_LinearEq} vanishes, Eq.~\eqref{eqA_LinearEq} is just the eigenvalue equation for the linear map $\operatorname{ad}(H_{\mathcal{V}}^{\mathrm{f}}) \circ\operatorname{ad}(\rho_{\mathcal{V}}^{\mathrm{i}})$, which is solved in Eq.~\eqref{eqA_EigenValueEq}.
In this case, we can also explicitly obtain the work $W_{\mathrm{c}}(e^{-i\omega T \mathsf{H}})$ extracted by a solution $\mathsf{H}$ as follows (note that we assume  $\comm*{ H_{\mathcal{V}}^{\mathrm{f}} }{ \rho_{\mathcal{V}}^{\mathrm{i}} } = 0$):
\begin{align}
    W_{\mathrm{c}}(e^{-i\omega T \mathsf{H}}) 
    &= \frac{\omega^2 T^2}{2} \tr\ab\Big( \mathsf{H} \comm{ H_{\mathcal{V}}^{\mathrm{f}} }{ \comm*{ \rho_{\mathcal{V}}^{\mathrm{i}} }{\mathsf{H}} } ) +\order{T^3} \nonumber \\
    &= \frac{\omega^2 T^2}{2} D \mu +\order{T^3}.
\end{align}
Therefore, among the candidates, choosing $\mathsf{H}$ to be a normalized eigenvector of the composite map $\operatorname{ad}(H_{\mathcal{V}}^{\mathrm{f}}) \circ\operatorname{ad}(\rho_{\mathcal{V}}^{\mathrm{i}})$ belonging to the largest eigenvalue $\lambda_{\max}$ yields the optimal work extraction up to $\order{T^2}$, with the corresponding parameter $\mu=\lambda_{\max}$.

Based on the above analysis in the small-$T$ regime, we adopt the following initial guess at $T = 0$:
\begin{equation}
    \mathsf{H} \propto 
    \begin{cases}
        -i \comm*{ H_{\mathcal{V}}^{\mathrm{f}} }{ \rho_{\mathcal{V}}^{\mathrm{i}} } & \comm*{ H_{\mathcal{V}}^{\mathrm{f}} }{ \rho_{\mathcal{V}}^{\mathrm{i}} } \neq 0; \\
        \Lambda_{\mathrm{max}} & \comm*{ H_{\mathcal{V}}^{\mathrm{f}} }{ \rho_{\mathcal{V}}^{\mathrm{i}} } = 0,
    \end{cases}
\end{equation}
where $\Lambda_{\max} \in \mathcal{V}$ denotes a solution of Eq.~\eqref{eqA_LinearEq} that yields the largest work extraction up to $\order{T^2}$.
Given the optimal solution $\mathsf{H}$ at time $T$, it is used as the initial guess to compute a solution at $T + \d{T}$.
If this guess fails to yield a larger extractable work than in the previous step, we switch to random initial guesses, chosen to be orthogonal to both $H_{\mathcal{V}}^{\mathrm{f},I}(T+ \d{T})$ and $\rho_{\mathcal{V}}^{\mathrm{i}}$ and normalized as $\Norm*{X} = 1$, until a solution yielding larger work than $W_{\mathrm{c}}(e^{-i\omega (T+\d{T}) \mathsf{H}}, U_{\mathrm{u}}(T+\d{T}))$ is found.

\section{Derivation of Eqs.~\eqref{eq_SUnRhoProjection} and \eqref{eq_SUnInnerProducts} in the $\mathrm{SU}(n)$-Hubbard model}
\label{AppendixF}
Here, we provide a derivation of Eqs.~\eqref{eq_SUnRhoProjection} and \eqref{eq_SUnInnerProducts}.
Since both equations can be derived straightforwardly using an orthonormal basis of $\mathcal{V} \cong \mathfrak{su}(n)$, we first construct such a basis, and then use it to derive Eqs.~\eqref{eq_SUnRhoProjection} and \eqref{eq_SUnInnerProducts}.

We first take an orthonormal basis of the Hilbert space, which will be used to calculate the Hilbert–Schmidt inner product.
For the $\mathrm{SU}(n)$-Hubbard model on a $V$-site lattice $\Lambda$, the Hilbert space is given by
\begin{equation*}
    \mathcal{H} \coloneqq \operatorname{span}_{\mathbb{C}}\ab\Big\{ c_{x_{N}\alpha_{N}}^{\dagger} \cdots c_{x_{1}\alpha_{1}}^{\dagger} \ket|0> \colon x_{j}\in \Lambda,\ \alpha_{j} \in \{1,\dots,n\} \},
\end{equation*}
where $\ket|0>$ is the vacuum state, and $c_{x \alpha}^{\dagger}$ denotes the fermionic creation operator of flavor $\alpha$ at site $x$, satisfying $\acomm*{ c_{x \alpha}^{\dagger} }{ c_{y \beta} } = \delta_{xy} \delta_{\alpha\beta}$ and $\acomm*{ c_{x \alpha} }{ c_{y \beta} } = 0$.
Let $\bm{\xi} = (\xi_{x\alpha}) \in \{0,1\}^{n V}$ be a $(n V)$-dimensional vector satisfying $N = \sum_{x,\alpha} \xi_{x\alpha}$, where $N$ is the total particle number.
Then, a basis of $\mathcal{H}$ can be written as $\{ \ket|\bm{\xi}> \coloneqq \prod_{x \in \Lambda} \prod_{\alpha=1}^{n} (c_{x\alpha}^{\dagger})^{\xi_{x\alpha}} \ket|0> \}$, where the products of non-commuting operators $\{c_{x\alpha}^{\dagger}\}$ are arranged in an arbitrary but fixed order.

We then calculate the Hilbert--Schmidt inner products for the generators $\{ E_{\mu\nu} \coloneqq \sum_{x \in \Lambda} c_{x\mu}^{\dagger} c_{x\nu} \}$ of $\mathcal{V} \cong \mathfrak{su}(n)$.
For this purpose, we note that the operator $E_{\mu\nu}$ with $\mu \neq \nu$ annihilates a fermion with flavor $\nu$ and creates a fermion with flavor $\mu$.
Consequently, we have $\braket<\bm{\xi}| E_{\alpha\beta} E_{\mu\nu} |\bm{\xi}> = 0$ for any $\bm{\xi}$, unless $(\beta,\nu) = (\alpha, \mu)$ or $(\mu,\alpha)$.
Therefore, we have
\begin{equation}
    \tr[E_{\alpha\beta} E_{\mu\nu}] = 0\quad\text{unless $(\beta,\nu) = (\alpha, \mu)$ or $(\mu,\alpha)$.}
    \label{eq_SUnInnerProduct_Calc1}
\end{equation}
Let us consider the case $(\beta,\nu) = (\mu,\alpha)$ and $\mu \neq \nu$.
In this case, the inner product can be written as
\begin{equation}
    \tr[E_{\nu\mu} E_{\mu\nu}]
    = \sum_{\bm{\xi}} \sum_{y,z \in \Lambda} \braket*<\bm{\xi}| c_{y\nu}^{\dagger} c_{y\mu} c_{z\mu}^{\dagger} c_{z\nu} |\bm{\xi}>.
\end{equation}
Here, we have
\begin{align}
    \braket*<\bm{\xi}| c_{y\nu}^{\dagger} c_{y\mu} c_{z\mu}^{\dagger} c_{z\nu} |\bm{\xi}>
    &= 
    \begin{cases}
        0 & (y \neq z); \\
        (1 - \xi_{y\mu}) \xi_{y\nu} & (y = z).
    \end{cases}
\end{align}
Therefore, we obtain
\begin{align}
    \tr[E_{\nu\mu} E_{\mu\nu}] 
    &= \sum_{\bm{\xi}} \sum_{y \in \Lambda} (1 - \xi_{y\mu}) \xi_{y\nu}
    = \sum_{y \in \Lambda} \binom{ nV - 2 }{N - 1}
    \nonumber \\
    &= V \binom{ nV - 2 }{N - 1}.
    \label{eq_SUnInnerProduct_Calc2}
\end{align}
Since $E_{\nu\mu} = E_{\mu\nu}^{\dagger}$, this result shows that $\{ E_{\mu\nu} \}_{\mu\neq\nu}$ are mutually orthonormal and normalized.
Since the Hilbert--Schmidt inner product is invariant under any unitary transformation, we conclude that Hermitian operators
\begin{equation*}
    X_{\mu\nu} \coloneqq \frac{ E_{\mu\nu} + E_{\mu\nu}^{\dagger} }{ \sqrt{2} },\ Y_{\mu\nu} \coloneqq \frac{ i(E_{\mu\nu} - E_{\mu\nu}^{\dagger}) }{ \sqrt{2} },\ (\mu < \nu)
\end{equation*}
form an orthonormal set and satisfy $\tr[X_{\mu\nu}^2] = \tr[Y_{\mu\nu}^2] = V \binom{nV - 2}{N-1}$.

We next consider the other case $(\beta,\nu) = (\alpha,\mu)$, including the case with $\alpha = \mu$.
In this case, the action of the operator $E_{\mu\mu}$ can be easily calculated as
\begin{equation}
    E_{\mu\mu} \ket|\bm{\xi}> = N_{\mu}(\bm{\xi}) \ket|\bm{\xi}>,
\end{equation}
where $N_{\mu}(\bm{\xi}) \coloneqq \sum_{x \in \Lambda} \xi_{x\mu}$ denotes the number of fermions of the $\mu$th flavor in the state $\ket|\bm{\xi}>$.
Then, we obtain
\begin{align}
    \tr[ E_{\mu\mu} E_{\nu\nu} ] 
    &= \sum_{\bm{\xi}} N_{\mu}(\bm{\xi}) N_{\nu}(\bm{\xi})
    = \sum_{x,y\in\Lambda} \sum_{\bm{\xi}} \xi_{x\mu} \xi_{y\nu}.
\end{align}
Here, we have
\begin{align}
    \sum_{\bm{\xi}} \xi_{x\mu} \xi_{y\nu}
    &=
    \begin{cases}
        \binom{nV - 2}{N-2} & (x,\mu) \neq (y,\nu); \\
        \binom{nV - 1}{N-1} & (x,\mu) = (y,\nu)
    \end{cases}
    \nonumber \\
    &= \delta_{xy} \delta_{\mu\nu} \binom{nV - 2}{N-1} + \binom{nV - 2}{N-2}.
\end{align}
Therefore, we obtain
\begin{align}
    \tr[ E_{\mu\mu} E_{\nu\nu} ]
    &= \delta_{\mu\nu} V \binom{nV - 2}{N-1} + V^2 \binom{nV - 2}{N-2}.
\end{align}
This result shows that the operators $\{ E_{\mu\mu} \}_{\mu=1}^{n}$ are not orthogonal to each other.
However, their Fourier transforms 
\begin{equation}
    F_{k} \coloneqq \frac{1}{\sqrt{n}} \sum_{\mu=1}^{n} e^{-2\pi ik\mu/n} E_{\mu\mu}
\end{equation}
can be verified to be orthogonal to each other.
Indeed, we have
\begin{align}
    \tr[F_{k}^{\dagger} F_{p}]
    &= \frac{1}{n} \sum_{\mu,\nu=1}^{n} e^{2\pi i (k\mu -p\nu) / n} \tr[ E_{\mu\mu} E_{\nu\nu} ]
    \nonumber \\
    &= \delta_{kp} V \binom{nV - 2}{N-1} + \delta_{k0} \delta_{p0}\: n V^2 \binom{nV - 2}{N-2}.
    \label{eq_SUnInnerProduct_Calc3}
\end{align}
This result further shows that the operators $\{ F_{k} \}_{k=1}^{n-1}$ are orthonormal to each other, unlike $F_{0}$, which is proportional to the particle number operator and thus proportional to the identity operator: $F_{0} = N/\sqrt{n}$.
Again, since the Hilbert--Schmidt inner product is invariant under any unitary transformation, we conclude that Hermitian operators
\begin{alignat*}{3}
    Z_{2m} &\coloneqq \sqrt{\frac{2}{n}}\sum_{\mu=1}^{n} E_{\mu\mu} \cos\ab(\frac{2\pi m\mu}{n}) &\quad &(1 \leq m \leq \lfloor \frac{n}{2} \rfloor) \\
    Z_{2m+1} &\coloneqq \sqrt{\frac{2}{n}}\sum_{\mu=1}^{n} E_{\mu\mu} \sin\ab(\frac{2\pi m\mu}{n}) &\quad &(1 \leq m \leq \lfloor \frac{n-1}{2} \rfloor)
\end{alignat*}
are orthonormal to each other and satisfy $\tr[Z_{k}^2] = V \binom{nV - 2}{N-1}$ for any $k \in \{1,\cdots,n-1\}$.

From these results we construct an orthonormal basis 
\begin{equation}
    \mathcal{B} \coloneqq \ab\{ X_{\mu\nu}, Y_{\mu\nu} \}_{\mu < \nu} \cup \{ Z_{k} \}_{k=1}^{n-1}
    \label{eq_SUnOrthonormalBasis}
\end{equation}
satisfying, for all $\Lambda_{a}, \Lambda_{b} \in \mathcal{B}$, the orthonormality condition
\begin{equation}
    \tr[ \Lambda_{a} \Lambda_{b} ] = \delta_{ab} V \binom{nV - 2}{N - 1} \eqqcolon \delta_{ab} C_{V,N}^{(n)}.
\end{equation}

Now, we derive Eq.~\eqref{eq_SUnRhoProjection}. 
We first note that 
\begin{equation}
    \tr[ \rho^{\mathrm{i}} X_{\mu\nu} ] = \tr[ \rho^{\mathrm{i}} Y_{\mu\nu} ] = 0
\end{equation}
for $\mu \neq \nu$, since $E_{\mu\nu}$ alters the particle numbers of the $\mu$th and $\nu$th flavors.
For $E_{\mu\mu}$, we trivially have $\tr[ \rho^{\mathrm{i}} E_{\mu\mu} ] = N_{\mu}$.
Therefore, the orthogonal projection of $\rho^{\mathrm{i}}$ onto $\mathcal{V}$ is obtained from Eq.~\eqref{rhocortho} as
\begin{align}
    \rho_{\mathcal{V}}^{\mathrm{i}}
    &= \frac{1}{C_{V,N}^{(n)}} \sum_{k=1}^{n-1} \tr[\rho^{\mathrm{i}} Z_{k}] Z_{k}
    \nonumber \\
    &= \frac{1}{C_{V,N}^{(n)}} \ab( \sum_{k=0}^{n-1} \tr[\rho^{\mathrm{i}} Z_{k}] Z_{k} - \tr[\rho^{\mathrm{i}} Z_{0}] Z_{0} )
    \nonumber \\
    &= \frac{1}{C_{V,N}^{(n)}} \ab( \sum_{\mu=1}^{n} \tr[\rho^{\mathrm{i}} E_{\mu\mu}] E_{\mu\mu} - \frac{N^2}{n} )
    \nonumber \\
    &= \frac{1}{C_{V,N}^{(n)}} \sum_{\mu=1}^{n} \ab( N_{\mu} - \frac{N}{n} ) E_{\mu\mu},
\end{align}
which reproduces Eq.~\eqref{eq_SUnRhoProjection}.
Here, we write $Z_{0} \coloneqq F_{0} = N/\sqrt{n}$, and the third equality follows because $\{E_{\mu\mu}\}_{\mu=1}^{n}$ and $\{Z_{k}\}_{k=0}^{n-1}$ are related by a unitary transformation preserving the Hilbert–Schmidt inner product.

To derive Eq.~\eqref{eq_SUnInnerProducts}, we begin by noting that
\begin{equation}
    \tr[ \pi(E_{\alpha\beta}) \pi(E_{\mu\nu}) ] = \tr\ab\Big[ \dyad*{\alpha}{\beta}\, \dyad*{\mu}{\nu} ]
    = \delta_{\alpha\nu} \delta_{\beta\mu}.
\end{equation}
Comparing this equation with Eqs.~\eqref{eq_SUnInnerProduct_Calc1} and \eqref{eq_SUnInnerProduct_Calc2}, for $\alpha \neq \beta$ and $\mu \neq \nu$, we obtain
\begin{equation}
    \tr[ \pi(E_{\alpha\beta}) \pi(E_{\mu\nu}) ] = \frac{1}{ C_{V,N}^{(n)} } \tr[ E_{\alpha\beta} E_{\mu\nu} ].
    \label{eq_InnerProductSUn1}
\end{equation}
We also have
\begin{align}
    \tr[ \pi(F_{k}^{\dagger}) \pi(F_{p}) ]
    &= \frac{1}{n} \sum_{\mu,\nu=1}^{n} e^{ 2\pi i \frac{ k \mu - p \nu }{n} } \tr[\pi(E_{\mu\mu}) \pi(E_{\nu\nu})]
    \nonumber \\
    &= \frac{1}{n} \sum_{\mu,\nu=1}^{n} e^{ 2\pi i \frac{ k \mu - p \nu }{n} } \delta_{\mu\nu}
    \nonumber \\
    &= \frac{1}{n} \sum_{\mu=1}^{n} e^{ 2\pi i \frac{ (k - p) \mu }{n} }
    \nonumber \\
    &= \delta_{kp}.
\end{align}
Comparing this equation with Eq.~\eqref{eq_SUnInnerProduct_Calc3}, we obtain for $k,p \in \{1,\cdots, n-1\}$,
\begin{equation}
    \tr[ \pi(F_{k}^{\dagger}) \pi(F_{p}) ] = \frac{1}{ C_{V,N}^{(n)} } \tr[ F_{k}^{\dagger} F_{p} ].
    \label{eq_InnerProductSUn2}
\end{equation}

Combined together, the equations \eqref{eq_InnerProductSUn1} and \eqref{eq_InnerProductSUn2} imply that
\begin{equation}
    \tr[ \tilde{\Lambda}_{a} \tilde{\Lambda}_{b} ] = C_{V,N}^{(n)} \tr[ \pi(\tilde{\Lambda}_{a}) \pi(\tilde{\Lambda}_{b}) ]
    \label{eq_InnerProductSUn3}
\end{equation}
for arbitrary $\tilde{\Lambda}_{a}, \tilde{\Lambda}_{b} \in \{ E_{\mu\nu} \}_{\mu \neq \nu} \cup \{ F_{k} \}_{k=1}^{n-1}$.
Since these operators are related to the orthonormal basis $\mathcal{B}$ of $\mathcal{V}$ given in Eq.~\eqref{eq_SUnOrthonormalBasis} via a unitary transformation, Eq.~\eqref{eq_InnerProductSUn3} implies
\begin{equation}
    \tr[ \Lambda_{a} \Lambda_{b} ] = C_{V,N}^{(n)} \tr[ \pi(\Lambda_{a}) \pi(\Lambda_{b}) ]
    \label{eq_InnerProductSUn4}
\end{equation}
for any $\Lambda_{a}, \Lambda_{b} \in \mathcal{B}$.
Finally, any traceless operators in $\mathcal{V}$ can be written as a linear combination of elements in $\mathcal{B}$.
Therefore, Eq.~\eqref{eq_InnerProductSUn4} implies Eq.~\eqref{eq_SUnInnerProducts} for any traceless operators in $\mathcal{V}$.

\section{Proof of $\Norm*{ \Log U_{\mathrm{c}}(T) } = \omega T$ when $\omega T \leq \ell_{T}$}
\label{AppendixG}
Here, we provide a proof of $\Norm*{ \Log U_{\mathrm{c}}(T) } = \omega T$ when $\omega T \leq \ell_{T}$, which is required to complete the proof of Theorem~\ref{thmM_MainResult} in \Appendix{C}.
We recall that $U_{\mathrm{c}}(T)$ is assumed to achieve the optimal work extraction for the operational time $T$ so that $\mathcal{W}_{\mathrm{c}}(T) = W_{\mathrm{c}}(U_{\mathrm{c}}(T))$.
Throughout the section, the parameter $T$ is fixed, and we simply write $H_{\mathcal{V}} = H_{\mathcal{V}}^{\mathrm{f},I}(T)$ for brevity.

The claim follows from Lemma~\ref{lemma1} below, which formalizes the intuition that, for fixed $T$, tightening the constraint on control strength yields strictly less extractable work.
In the main text, the optimal work $\mathcal{W}_{\mathrm{c}}(T;\omega')$ was mainly written as $\mathcal{W}_{\mathrm{c}}(T)$ to emphasize its dependence on $T$, while the dependence on the norm constraint $\omega'$ was kept implicit.
Here, since $T$ is fixed and we vary the norm constraint $\omega'$, we instead write $\mathcal{W}_{\mathrm{c}}(\omega') \equiv \mathcal{W}_{\mathrm{c}}(T;\omega')$.

\begin{lemma}
    \label{lemma1}
    Suppose that $\omega T \leq \ell_{T}$.
    Then, $\mathcal{W}_{\mathrm{c}}(\omega')$ is strictly increasing in $\omega'$ for $0 < \omega' \leq \omega$, i.e., we have
    \begin{equation}
        \mathcal{W}_{\mathrm{c}}(\omega'') < \mathcal{W}_{\mathrm{c}}(\omega')
        \label{eqM_StrictMonotonicity}
    \end{equation}
    whenever $\omega'' < \omega' \leq \omega$.
\end{lemma}

\subsection{Proof of $\Norm*{ \Log U_{\mathrm{c}}(T) } = \omega T$ when $\omega T \leq \ell_{T}$ \\ given Lemma~\ref{lemma1}}
Suppose by contradiction that we had $\omega' T \coloneqq \Norm*{ \Log U_{\mathrm{c}}(T) } < \omega T$.
Then, the same work extraction $\mathcal{W}_{\mathrm{c}}(\omega)$ could be achieved under a tighter norm constraint $\Norm*{ H_{\mathrm{c}}(t) } \leq \omega'$ by the control Hamiltonian
\begin{equation}
    H_{\mathrm{c}}^{I}(t) = \omega' \frac{ i \Log U_{\mathrm{c}}(T) }{ \Norm*{ \Log U_{\mathrm{c}}(T) } },\quad t \in (0,T).
\end{equation}
That is, we would have
\begin{equation}
    \mathcal{W}_{\mathrm{c}}(\omega') \geq \mathcal{W}_{\mathrm{c}}(\omega)
\end{equation}
However, since $\omega' < \omega$ by assumption, this contradicts the strict monotonicity~\eqref{eqM_StrictMonotonicity} from Lemma~\ref{lemma1}.
Hence, we must have $\Norm*{ \Log U_{\mathrm{c}}(T) } = \omega T$.
(Note that we have already shown $\Norm*{ \Log U_{\mathrm{c}}(T) } \leq \omega T$ in \Appendix{C}.)
\qed

\subsection{Proof of Lemma~\ref{lemma1}}
Unlike the other parts of the proof of Theorem~\ref{thmM_MainResult}, this part requires concepts and results from the theory of Lie algebras, such as Cartan subalgebras and roots.

\subsubsection{Preliminaries on Lie algebras for the proof of Lemma~\ref{lemma1}}
In the following we assume that a Lie algebra $\mathcal{V}$ is finite-dimensional, as is the case in the main text.
\begin{definition}[Simple Lie algebra]
    A Lie algebra $\mathcal{V}$ is called simple if there exists no subalgebra $\mathfrak{h} \subset \mathcal{V}$ such that
    \begin{equation}
        \mathfrak{h} \neq \{0\}, \mathcal{V}
        \quad\text{and}\quad
        i \comm*{ X }{ \mathfrak{h} } \subseteq \mathfrak{h}\text{ for all } X \in \mathcal{V}.
    \end{equation}
\end{definition}

\begin{definition}[Semisimple Lie algebra]
    A Lie algebra $\mathcal{V}$ is called semisimple if there exists no subalgebra $\mathfrak{h} \subset \mathcal{V}$ satisfying the following two conditions:
    \begin{enumerate}
        \item $\mathfrak{h} \neq \{0\}, \mathcal{V}
        \quad\text{and}\quad
        i \comm*{ X }{ \mathfrak{h} } \subseteq \mathfrak{h}\text{ for all } X \in \mathcal{V}$;
        \item $[\underbrace{\mathfrak{h}, \cdots, [\mathfrak{h}}_{n}, \mathfrak{h}] ]= 0$ for some $n \geq 2$.
    \end{enumerate}
    This is equivalent to the standard definition of semisimple Lie algebras as those having no nontrivial solvable ideals.
\end{definition}

As mentioned in the main text~(Section~\ref{SectionVB}), every semisimple Lie algebra can be decomposed into a direct sum of simple Lie algebras~\cite{Humphreys_Other1972o, Knapp_Other1996h, Fulton1999-ai}.

\begin{definition}[Cartan subalgebra]
    A Cartan subalgebra of a Lie algebra $\mathcal{V}$ is a maximal abelian subalgebra $\mathfrak{h}$ of $\mathcal{V}$ such that $\operatorname{ad}(X) \colon Z \in \mathcal{V} \mapsto \comm*{X}{Z}$ is diagonalizable over $\mathbb{C}$ for any $X \in \mathfrak{h}$.
\end{definition}

\begin{definition}[Compact Lie algebra]
    A Lie algebra $\mathcal{V}$ is called compact if a corresponding Lie group $e^{i\mathcal{V}}$ is compact.
    Note that a compact Lie algebra is a Lie algebra over $\mathbb{R}$ but not over $\mathbb{C}$.
\end{definition}

\begin{definition}[Root space decomposition for compact Lie algebras]
Let $\mathfrak{h}$ be a Cartan subalgebra of a compact Lie algebra $\mathcal{V}$.
Then, the operators $i\operatorname{ad}(X)$ with $X \in \mathfrak{h}$ are anti-symmetric and commute with each other.
Therefore, they can be simultaneously block-diagonalized into two-dimensional invariant subspaces and the kernel.
This implies that there exists a linearly independent set $\{X_{\alpha}, Y_{\alpha}\}$ such that
\begin{equation}
   \forall A \in \mathfrak{h},\quad i[A,X_\alpha] = \alpha(A) Y_\alpha,\ i[A,Y_\alpha] = - \alpha(A) X_\alpha
   \label{eqM_CartanWeylBasis}
\end{equation}
for some nonzero linear functional $\alpha \colon \mathfrak{h} \to \mathbb{R}$ called a root.
Then, $\mathcal{V}$ can be decomposed as
\begin{equation}
   \mathcal{V} = \mathfrak{h} \oplus 
   \bigoplus_{\alpha \neq 0} \mathcal{V}_{\alpha},\quad \mathcal{V}_{\alpha} \coloneqq \operatorname{span}_{\mathbb{R}}\{ X_{\alpha}, Y_{\alpha} \}.
\end{equation}
\end{definition}

As an example, consider the $\mathfrak{su}(2)$ algebra with $[S_{i}, S_{j}] = i\sum_{k} \epsilon_{ijk} S_{k}$.
For a Cartan subalgebra $\mathfrak{h} = \operatorname{span} S_{3}$, we have
\begin{equation}
    i[S_{3}, S_{2}] = S_{1},\quad i[S_{3}, S_{1}] = -S_{2}.
\end{equation}
Therefore, the $\mathfrak{su}(2)$ algebra has one root (up to sign) satisfying $\alpha(S_{3}) = 1$ with $X_{\alpha} = S_{2}$ and $Y_{\alpha} = S_{1}$.

As another example, consider the $\mathfrak{su}(3)$ algebra consisting of $3\times 3$ traceless Hermitian matrices.
For a Cartan subalgebra $\mathfrak{h} \coloneqq \{ \sum_{j=1}^{3} a_{j} \ketbra*|j><j| \colon \sum_{j=1}^{3} a_{j} = 0 \}$ consisting of diagonal matrices, the operators $\{ X_{\alpha}, Y_{\alpha} \}$ are given by
\begin{gather}
    X_{\alpha_{1}} \coloneqq \frac{ \ketbra*|1><2| + \ketbra*|2><1| }{\sqrt{2}},\quad Y_{\alpha_{1}} \coloneqq \frac{ i\ab( \ketbra*|1><2| - \ketbra*|2><1| ) }{\sqrt{2}},  \nonumber\\
    X_{\alpha_{2}} \coloneqq \frac{ \ketbra*|2><3| + \ketbra*|3><2| }{\sqrt{2}},\quad Y_{\alpha_{2}} \coloneqq \frac{ i\ab( \ketbra*|2><3| - \ketbra*|3><2| ) }{\sqrt{2}},  \nonumber\\
    X_{\alpha_{3}} \coloneqq \frac{ \ketbra*|1><3| + \ketbra*|3><1| }{\sqrt{2}},\quad Y_{\alpha_{3}} \coloneqq \frac{ i\ab( \ketbra*|1><3| - \ketbra*|3><1| ) }{\sqrt{2}}. \nonumber
\end{gather}
Indeed, one can directly verify, e.g.,
\begin{equation*}
    i \comm*{A}{ X_{\alpha_{1}} } = (a_{1} - a_{2}) Y_{\alpha_{1}},\ i \comm*{A}{ Y_{\alpha_{1}} } = -(a_{1} - a_{2}) X_{\alpha_{1}}
\end{equation*}
for arbitrary $A = \sum_{j} a_{j} \ketbra*|j><j| \in \mathfrak{h}$, so that 
\begin{equation*}
    \alpha_{1}(A) = a_{1} - a_{2},\ \
    \alpha_{2}(A) = a_{2} - a_{3},\ \
    \alpha_{3}(A) = a_{1} - a_{3}.
\end{equation*}

\subsubsection{Proof of Lemma~\ref{lemma1}}
We first show that $\mathcal{W}_{\mathrm{c}}(\omega')$ is nondecreasing in $\omega'$.
This is trivial because any control Hamiltonian $H_{\mathrm{c}}(t)$ satisfying a tighter norm constraint automatically satisfies any looser one.
Then, the maximum in the definition of the optimal work~\eqref{optimize} implies that $\mathcal{W}_{\mathrm{c}}(\omega')$ is nondecreasing in $\omega'$.

To elevate this monotonicity to the strict monotonicity, we use the facts from the theory of Lie algebras.
Let $\mathcal{U}_{\mathrm{c}}(\omega'')$ be an optimal time-evolution operator such that $\mathcal{W}_{\mathrm{c}}(\omega'') = W_{\mathrm{c}}(\mathcal{U}_{\mathrm{c}}(\omega''))$.
In this proof, we use the notation $\mathcal{U}_{\mathrm{c}}(\omega'')$ instead of $U_{\mathrm{c}}(\omega'')$ to emphasize that we do not use any information about the optimal unitary that depends on Lemma~\ref{lemma1}, including the relation $\Norm*{ \Log U_{\mathrm{c}}(T) } = \omega T$.
Let $\tilde{H}_{\mathrm{c}}^{I}(t)$ be the optimal control Hamiltonian in the interaction picture corresponding to $\mathcal{U}_{\mathrm{c}}(\omega'')$,
and write $\rho_{\mathcal{V}}^{\mathrm{f}}(\omega'') \coloneqq \mathcal{U}_{\mathrm{c}}(\omega'') \rho_{\mathcal{V}}^{\mathrm{i}} \mathcal{U}_{\mathrm{c}}(\omega'')^{\dagger}$.

Consider $V_{\varepsilon} \coloneqq e^{ -i \varepsilon T Z} \mathcal{U}_{\mathrm{c}}(\omega'')$ for some $Z \in \mathcal{V}$ with $\Norm*{Z} \leq 1$, which can be realized under the looser norm constraint $\Norm*{ H_{\mathrm{c}}^{I}(t) } \leq \omega'' + \varepsilon$
by the following control Hamiltonian:
\begin{equation}
    H_{\mathrm{c}}^{I}(t)
    =
    \begin{cases}
        \frac{\omega'' + \varepsilon}{\omega''} \tilde{H}_{\mathrm{c}}^{I}\ab( \frac{\omega'' + \varepsilon}{\omega''} t ) & (0 < t \leq \frac{\omega''}{\omega'' + \varepsilon} T); \\
        (\omega'' + \varepsilon) Z & (\frac{\omega''}{\omega'' + \varepsilon} T < t < T).
    \end{cases}
    \label{eqM_BelongedProtocol}
\end{equation}
Then, the work extractable by the unitary $V_{\varepsilon}$ is evaluated to be
\begin{align}
    W_{\mathrm{c}}(V_{\varepsilon})
    &= \mathcal{W}_{\mathrm{c}}(\omega'')
    \nonumber \\
    &\quad + \varepsilon T \tr\ab\Big( Z (-i) \comm{ H_{\mathcal{V}} }{ \rho_{\mathcal{V}}^{\mathrm{f}}(\omega'') } ) 
    \nonumber \\
    &\quad +\frac{\varepsilon^2 T^2}{2} \tr\ab\Big( Z \comm{ H_{\mathcal{V}} }{ \comm*{ \rho_{\mathcal{V}}^{\mathrm{f}}(\omega'') }{Z} } ) +\order{\varepsilon^3}.
    \label{eqM_WorkIncrement}
\end{align}

Below we prove that there exists a constant $\delta > 0$ and $Z \in \mathcal{V}$ such that $W_{\mathrm{c}}(V_{\varepsilon}) > \mathcal{W}_{\mathrm{c}}(\omega'')$ for all $\varepsilon \in (0,\delta)$.
If $\comm*{ H_{\mathcal{V}} }{ \rho_{\mathcal{V}}^{\mathrm{f}}(\omega'') } \neq 0$, this holds trivially: 
it suffices to choose
\begin{equation}
    Z = \frac{ -i \comm{ H_{\mathcal{V}} }{ \rho_{\mathcal{V}}^{\mathrm{f}}(\omega'') } }{ \Norm*{ \comm{ H_{\mathcal{V}} }{ \rho_{\mathcal{V}}^{\mathrm{f}}(\omega'') } } }\in\mathcal{V}.
    \label{eqM_FirstOrderInclZ}
\end{equation}

If $\comm{ H_{\mathcal{V}} }{ \rho_{\mathcal{V}}^{\mathrm{f}}(\omega'') } = 0$, we use the following lemma to find a suitable $Z \in \mathcal{V}$:
\begin{lemma}[Condition for the maximum work extraction]
    \label{lemma2}
    Suppose that a unitary $V_{\mathrm{c}} \in e^{i\mathcal{V}}$ satisfies $\comm*{ H_{\mathcal{V}} }{ V_{\mathrm{c}} \rho_{\mathcal{V}}^{\mathrm{i}} V_{\mathrm{c}}^{\dagger} } = 0$.
    Let $\mathfrak{h}$ be a Cartan subalgebra containing both $H_{\mathcal{V}}$ and $V_{\mathrm{c}} \rho_{\mathcal{V}}^{\mathrm{i}} V_{\mathrm{c}}^{\dagger}$, and let $\Phi \coloneqq \{\alpha\}$ be the associated root system.
    Then, it holds that
    \begin{multline}
        \forall \alpha \in \Phi,\ \alpha\ab\big(H_{\mathcal{V}}) \alpha\ab\big(V_{\mathrm{c}} \rho_{\mathcal{V}}^{\mathrm{i}} V_{\mathrm{c}}^{\dagger}) \leq 0 \\
        \Longrightarrow
        W_{\mathrm{c}}(V_{\mathrm{c}}) = \max_{V'_{\mathrm{c}} \in e^{i\mathcal{V}}} W_{\mathrm{c}}(V'_{\mathrm{c}}).
        \label{eqM_ConditionFormaximumWork}
    \end{multline}
\end{lemma}

Here, the assumptions $\omega'' < \omega' \leq \omega$ and $\omega T \leq \ell_{T}$ imply that the unitary $\mathcal{U}_{\mathrm{c}}(\omega'')$ satisfies $\Norm*{ \Log \mathcal{U}_{\mathrm{c}}(\omega'') } \leq \omega'' T < \ell_{T}$~(see Eq.~\eqref{eq_IntegratedNormConstraint} and the surrounding discussion).
Then, the minimality of $\ell_{T}$ implies that $W_{\mathrm{c}}(\mathcal{U}_{\mathrm{c}}(\omega'')) < \max_{V'_{\mathrm{c}} \in e^{i\mathcal{V}}} W_{\mathrm{c}}(V'_{\mathrm{c}})$.
Now, Lemma~\ref{lemma2} combined with $\comm*{ H_{\mathcal{V}} }{ \rho_{\mathcal{V}}^{\mathrm{f}}(\omega'') } = 0$ implies
that there exists a root $\beta$ such that $\beta\ab\big(H_{\mathcal{V}}) \beta\ab\big( \rho_{\mathcal{V}}^{\mathrm{f}}(\omega'') ) > 0$.

We choose $Z$ to be an element of the root space $\mathcal{V}_{\beta}$ associated with the root $\beta$, such as
\begin{equation}
    Z = \frac{ X_{\beta} }{ \Norm*{ X_{\beta} } },
    \label{eqM_SecondOrderInclZ}
\end{equation}
where $X_{\beta}$ is that in Eq.~\eqref{eqM_CartanWeylBasis}.
Using Eqs.~\eqref{eqM_CartanWeylBasis} and \eqref{eqM_WorkIncrement}, we obtain 
\begin{equation}
    W_{\mathrm{c}}(V_{\varepsilon}) \\
    = \mathcal{W}_{\mathrm{c}}(\omega'') + \underbrace{\frac{D \varepsilon^2 T^2}{2} \beta\ab\big(H_{\mathcal{V}}) \beta\ab\big( \rho_{\mathcal{V}}^{\mathrm{f}}(\omega'') ) }_{> 0} + \order{\varepsilon^3}.
    \label{eqM_SecondOrderInclWc}
\end{equation}
This implies that there exists a constant $\delta > 0$ and $Z \in \mathcal{V}$ such that $W_{\mathrm{c}}(V_{\varepsilon}) > \mathcal{W}_{\mathrm{c}}(\omega'')$ for all $\varepsilon \in (0,\delta)$.

Then, we can choose $\varepsilon'$ such that $0 < \varepsilon' < \min\{\delta, \omega' - \omega''\}$, and obtain the following chain of inequalities:
\begin{equation}
    \mathcal{W}_{\mathrm{c}}(\omega'') < W_{\mathrm{c}}(V_{\varepsilon'})
    \leq \mathcal{W}_{\mathrm{c}}(\omega'' + \varepsilon') \leq \mathcal{W}_{\mathrm{c}}(\omega'),
\end{equation}
where the third inequality follows because $\mathcal{W}_{\mathrm{c}}(\omega')$ is nondecreasing.
This completes the proof of Eq.~\eqref{eqM_StrictMonotonicity}.
\qed

\subsection{Proof of Lemma~\ref{lemma2}}
\subsubsection{Additional preliminaries on Lie algebras for the proof of Lemma~\ref{lemma2}}
For the proof of Lemma~\ref{lemma2}, we employ the following additional concepts and facts about Lie algebras~\cite{Humphreys_Other1972o, Knapp_Other1996h, Fulton1999-ai, Sepanski_Other2006f}.

\begin{proposition}
    \label{prop_ConjugacyOfCartanSubalgebras}
    Any two Cartan subalgebras $\mathfrak{h}$ and $\mathfrak{h}'$ of a compact Lie algebra $\mathcal{V}$ are conjugate to each other, meaning that there exists $V \in e^{i \mathcal{V}}$ such that $\mathfrak{h}' = V \mathfrak{h} V^{\dagger}$~\cite[Theorem~4.34]{Knapp_Other1996h}.
\end{proposition}

\begin{definition}[Weyl group]
    The Weyl group $\mathfrak{W}$ of a Cartan subalgebra $\mathfrak{h}$ is the group of automorphisms of $\mathfrak{h}$ generated by the following reflections:
    \begin{equation}
        s_{\alpha}(X) \coloneqq X - 2 \alpha(X) \frac{ T_{\alpha} }{\tr T_{\alpha}^2},\quad \forall X \in \mathfrak{h},
        \label{eqM_WeylGenerators}
    \end{equation}
    where $T_{\alpha}$ is the dual of the root $\alpha$ defined by
    \begin{equation}
        \tr[ X T_{\alpha} ] = \alpha(X),\quad \forall X \in \mathfrak{h}.
    \end{equation}
\end{definition}

\begin{proposition}
    \label{prop_WeylGroupNormalizers}
    For any $\sigma \in \mathfrak{W}$, there exists $V_{\sigma} \in e^{i\mathcal{V}}$ such that $\sigma(X) = V_{\sigma} X V_{\sigma}^{\dagger}$ for all $X \in \mathfrak{h}$~\cite[Theorem~4.54]{Knapp_Other1996h}.
\end{proposition}

\begin{definition}[Weyl chamber]
    Let $\mathfrak{h}$ be a Cartan subalgebra and $\Phi \coloneqq \{\alpha\}$ be the associated root system.
    Then, $\mathfrak{h}$ is decomposed into disjoint regions, called the Weyl chambers, by hyperplanes determined by the equations $\alpha(X) = 0$ for each $\alpha \in \Phi$.
    In other words, any $A$ and $B$ in $\mathfrak{h}$ lie in the closure of the same Weyl chamber if and only if  $\alpha(A) \alpha(B) \geq 0$ for all $\alpha \in \Phi$.
\end{definition}

\begin{proposition}
    \label{prop_SimpleTransitivityOfWeylGroup}
    Each element of $\mathfrak{W}$ maps a Weyl chamber to another.
    Moreover, for any two Weyl chambers $W_{1}$ and $W_{2}$, there exists exactly one element $\sigma \in \mathfrak{W}$ such that $\sigma(W_{1}) = W_{2}$~\cite[Theorem 6.43]{Sepanski_Other2006f};\cite[Proposition 8.23]{Hall_Other2015c}.
\end{proposition}

\begin{proposition}
    \label{prop_WeylChamberIntersection}
    Let $\mathfrak{h}$ be a Cartan subalgebra and $\mathfrak{W}$ be its Weyl group.
    For arbitrary $X \in \mathfrak{h}$, the intersection of $\mathfrak{h}$ and the adjoint orbit of $X$ coincides with the Weyl group orbit of $X$~\cite[Theorem 6.36]{Sepanski_Other2006f}:
    \begin{equation*}
        \{ V X V^{\dagger} \colon V \in e^{i \mathcal{V}} \} \cap \mathfrak{h} = \{ \sigma(X) \colon \sigma \in \mathfrak{W} \}\ (\eqqcolon \mathfrak{W}\cdot X).
    \end{equation*}
    Moreover, the closure of each Weyl chamber contains exactly one element of $\mathfrak{W}\cdot X$~\cite[Proposition 8.29]{Hall_Other2015c}.
\end{proposition}

\begin{figure}
    \centering
    \includegraphics[width=0.9\linewidth]{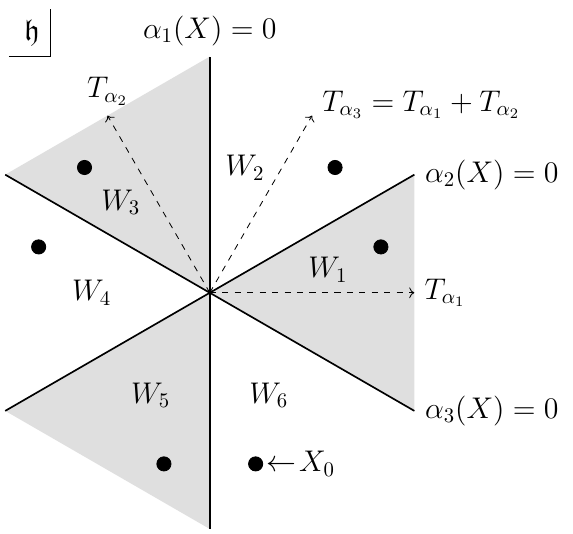}
    \caption{Weyl chambers for the $\mathfrak{su}(3)$ algebra, labeled $W_{1}, \dots, W_{6}$.
    The algebra has three roots up to sign, $\alpha_{1}, \alpha_{2}, \alpha_{3}$, with the relation $\alpha_{3} = \alpha_{1} + \alpha_{2}$, and their dual elements denoted by $T_{\alpha_{j}}$.
    The Weyl group is generated by reflections about the hyperplanes $\alpha_{j}(X) = 0$ $(j=1,2,3)$ defined in Eq.~\eqref{eqM_WeylGenerators}.
    (One can confirm Proposition~\ref{prop_SimpleTransitivityOfWeylGroup} by inspection.)
    Black dots show the Weyl group orbit of an element $X_{0} \in \mathfrak{h}$, which coincides with the intersection of the adjoint orbit of $X_{0}$ with $\mathfrak{h}$ (Proposition~\ref{prop_WeylChamberIntersection}).
    }
    \label{fig_WeylChambersSU3}
\end{figure}

We show the Weyl chambers for $\mathfrak{su}(3)$ algebra in Fig.~\ref{fig_WeylChambersSU3}.

\vspace{-0.75cm}
\subsubsection{Proof of Lemma~\ref{lemma2}}
Let $\tilde{U}_{\mathrm{c}} \in e^{i \mathcal{V}}$ be a maximizer: $W_{\mathrm{c}}(\tilde{U}_{\mathrm{c}}) = \max_{V'_{\mathrm{c}} \in e^{i \mathcal{V}}} W_{\mathrm{c}}(V'_{\mathrm{c}})$.
From the argument around Eqs.~\eqref{eqM_BelongedProtocol}--\eqref{eqM_FirstOrderInclZ}, $\tilde{U}_{\mathrm{c}}$ must satisfy $\comm*{ H_{\mathcal{V}} }{ \tilde{U}_{\mathrm{c}} \rho_{\mathcal{V}}^{\mathrm{i}} \tilde{U}_{\mathrm{c}}^{\dagger} } = 0$; otherwise we could find a unitary that leads to the work larger than $W_\mathrm{c}(\tilde{U}_\mathrm{c})$.
Here, we can assume that $\tilde{U}_{\mathrm{c}} \rho_{\mathcal{V}}^{\mathrm{i}} \tilde{U}_{\mathrm{c}}^{\dagger} \in \mathfrak{h}$ without loss of generality.
To see this, consider the centralizer algebra of $H_{\mathcal{V}}$ defined by
\begin{equation}
    \mathfrak{c}[H_{\mathcal{V}}] \coloneqq \ab\{ X \in \mathcal{V} \colon \comm*{ H_{\mathcal{V}} }{ X } = 0 \},
\end{equation}
which is compact as it is a subalgebra of a compact algebra $\mathcal{V}$.
The subalgebra $\mathfrak{h}$ is a Cartan subalgebra of $\mathfrak{c}[H_{\mathcal{V}}]$ as well because it is a \emph{maximal} abelian subalgebra containing $H_{\mathcal{V}}$.
Then, from Proposition~\ref{prop_ConjugacyOfCartanSubalgebras}, there exists an element $V \in e^{ i \mathfrak{c}[H_{\mathcal{V}}] }$ which maps a Cartan subalgebra containing both $-H_{\mathcal{V}}$ and $\tilde{U}_{\mathrm{c}} \rho_{\mathcal{V}}^{\mathrm{i}} \tilde{U}_{\mathrm{c}}^{\dagger}$ to $\mathfrak{h}$:
in particular we have
\begin{equation}
    (V \tilde{U}_{\mathrm{c}}) \rho_{\mathcal{V}}^{\mathrm{i}} (V \tilde{U}_{\mathrm{c}})^{\dagger} \in \mathfrak{h}.
\end{equation}
Since $V \in e^{ i \mathfrak{c}[H_{\mathcal{V}}] }$ commutes with $H_{\mathcal{V}}$, it does not alter the extractable work:
\begin{equation}
    \tr[ H_{\mathcal{V}} (V \tilde{U}_{\mathrm{c}}) \rho_{\mathcal{V}}^{\mathrm{i}} (V \tilde{U}_{\mathrm{c}})^{\dagger}] = \tr[ H_{\mathcal{V}} \tilde{U}_{\mathrm{c}} \rho_{\mathcal{V}}^{\mathrm{i}} \tilde{U}_{\mathrm{c}}^{\dagger} ].
\end{equation}
Rewriting $V \tilde{U}_{\mathrm{c}}$ as $\tilde{U}_{\mathrm{c}}$, we can assume without loss of generality that $\tilde{U}_{\mathrm{c}} \rho_{\mathcal{V}}^{\mathrm{i}} \tilde{U}_{\mathrm{c}}^{\dagger}$ also belongs to $\mathfrak{h}$.

If $V_{\mathrm{c}} \rho_{\mathcal{V}}^{\mathrm{i}} V_{\mathrm{c}}^{\dagger}$ and $\tilde{U}_{\mathrm{c}} \rho_{\mathcal{V}}^{\mathrm{i}} \tilde{U}_{\mathrm{c}}^{\dagger}$ belong to the closure of the same Weyl chamber, the claim of Lemma 2 follows immediately:
Proposition~\ref{prop_WeylChamberIntersection} implies $V_{\mathrm{c}} \rho_{\mathcal{V}}^{\mathrm{i}} V_{\mathrm{c}}^{\dagger} = \tilde{U}_{\mathrm{c}} \rho_{\mathcal{V}}^{\mathrm{i}} \tilde{U}_{\mathrm{c}}^{\dagger}$, and hence we obtain $W_{\mathrm{c}}(V_{\mathrm{c}}) = W_{\mathrm{c}}(\tilde{U}_{\mathrm{c}})$.

Consider the other case where $V_{\mathrm{c}} \rho_{\mathcal{V}}^{\mathrm{i}} V_{\mathrm{c}}^{\dagger}$ and $\tilde{U}_{\mathrm{c}} \rho_{\mathcal{V}}^{\mathrm{i}} \tilde{U}_{\mathrm{c}}^{\dagger}$ belong to the closure of different Weyl chambers, say $\overline{W}_{1}$ and $\overline{W}_{2}$, respectively.
Here, we note that $\tilde{U}_{\mathrm{c}}$ must satisfy $\alpha(H_{\mathcal{V}}) \alpha(\tilde{U}_{\mathrm{c}} \rho_{\mathcal{V}}^{\mathrm{i}} \tilde{U}_{\mathrm{c}}^{\dagger}) \leq 0$ for all $\alpha \in \Phi$.
Otherwise, if there existed $\beta \in \Phi$ such that $\beta(H_{\mathcal{V}}) \beta(\tilde{U}_{\mathrm{c}} \rho_{\mathcal{V}}^{\mathrm{i}} \tilde{U}_{\mathrm{c}}^{\dagger}) > 0$, then we could extract more work by an argument similar to that around Eq.~\eqref{eqM_SecondOrderInclWc}, which contradicts the maximality of $W_{\mathrm{c}}(\tilde{U}_{\mathrm{c}})$.
From the definition of the Weyl chamber and the relation $\alpha(X)=-\alpha(-X)$, this fact and the assumption that $\alpha(H_{\mathcal{V}}) \alpha(V_{\mathrm{c}} \rho_{\mathcal{V}}^{\mathrm{i}} V_{\mathrm{c}}^{\dagger}) \leq 0$ for all $\alpha \in \Phi$ imply that $-H_{\mathcal{V}}$ belongs to both $\overline{W}_{1}$ and $\overline{W}_{2}$.
(Note that $H_{\mathcal{V}}$ can belong to the closure of multiple Weyl chambers, if $H_{\mathcal{V}}$ lies on their boundaries.)
Let $\sigma \in \mathfrak{W}$ satisfy $\sigma(\overline{W}_{1}) = \overline{W}_{2}$, which exists from Proposition~\ref{prop_SimpleTransitivityOfWeylGroup}.
Then, Propositions~\ref{prop_WeylGroupNormalizers} and \ref{prop_WeylChamberIntersection} imply that 
\begin{equation}
    \sigma(H_{\mathcal{V}}) = H_{\mathcal{V}}\quad\text{and}\quad \sigma(V_{\mathrm{c}} \rho_{\mathcal{V}}^{\mathrm{i}} V_{\mathrm{c}}^{\dagger}) = \tilde{U}_{\mathrm{c}} \rho_{\mathcal{V}}^{\mathrm{i}} \tilde{U}_{\mathrm{c}}^{\dagger}.
\end{equation}
Here, the first relation follows because both $-H_{\mathcal{V}}$ and $\sigma(-H_{\mathcal{V}})$ belong to the closure $\overline{W}_{2}$.
Since $\sigma \in \mathfrak{W}$ preserves the inner product due to Proposition~\ref{prop_WeylGroupNormalizers}, these relations imply $W_{\mathrm{c}}(V_{\mathrm{c}}) = W_{\mathrm{c}}(\tilde{U}_{\mathrm{c}})$, completing the proof.
\qed

\end{document}